\documentclass[a4paper,fleqn]{cas-dc}
\PassOptionsToPackage{hyphens}{url}
\usepackage[english]{babel} 
\usepackage[utf8]{inputenc}
\usepackage{soul}
\usepackage{xcolor}
\usepackage{graphicx} 
\usepackage{longtable}
\usepackage{tabularx}
\usepackage{booktabs} 
\usepackage{array}
\usepackage{afterpage}
\usepackage{placeins}
\usepackage{xtab} 

\usepackage{multicol}
\usepackage{comment}
\usepackage{adjustbox}
\usepackage[most]{tcolorbox}
\usepackage{amssymb}
\usepackage{pifont}
\usepackage[hyphens]{url}
\usepackage[numbers]{natbib}
\newtcolorbox{promptbox}[1][]{
    colback=white,          
    colframe=black,         
    fonttitle=\bfseries,    
    title=#1,               
    boxrule=0.5mm,          
    sharp corners,          
    width=\linewidth        
}

\def\tsc#1{\csdef{#1}{\textsc{\lowercase{#1}}\xspace}}
\tsc{WGM}
\tsc{QE}
\tsc{EP}
\tsc{PMS}
\tsc{BEC}
\tsc{DE}

\begin{document}
\let\WriteBookmarks\relax
\def\floatpagepagefraction{1}
\def\textpagefraction{.001}
\shorttitle{SLR for transformer-based Software Vulnerability detection}
\shortauthors{Naseer et~al.}

\title [mode = title]{A systematic literature Review for Transformer-based Software Vulnerability detection}                      



\author[1]{Fiza Naseer}[]

\ead{f.naseer@herts.ac.uk}

\credit{Data curation, Conceptualization of this study, Methodology, Analaysis,  Writing - Original draft preparation}

\affiliation[1]{organization={
Department of Computer Science, Cybersecurity and Computing Systems Research Group, University of Hertfordshire, Hertfordshire},
                addressline={College Lane}, 
                city={Hatfield},
                postcode={AL10 9AB}, 
                state={Hertfordshire},
                country={UK}}

\author[1]{Javed Ali Khan}[style=chinese]
\cormark[1]
\credit{ Methodology, Analaysis, Revising - Original draft preparation, Supervision}
\ead{j.a.khan@herts.ac.uk}
\author[1]{Muhammad Yaqoob}[%
   ]
\ead{m.yaqoob3@herts.ac.uk}
\credit{ Methodology, Analaysis, Revising - Original draft preparation, Supervision}

\author[1]{Alexios Mylonas}
\credit{ Revising - Original draft preparation, Supervision}
\ead{a.mylonas@herts.ac.uk}

\author[2]{Ishaya Gambo }
\credit{ Revising - Original draft preparation, Supervision}
\ead{ipgambo@oauife.edu.ng}

\affiliation[2]{organization={Department of Software Engineering, Obafemi Awolowo University Ile-Ife},
                postcodesep={220005}, 
                city={Ile-Ife},
                country={Nigeria}}


\cortext[cor1]{Javed Ali Khan}


\begin{abstract}
\textbf{Context:} Software vulnerabilities pose significant security threats to software systems, especially as software is increasingly used across many areas of daily life, including health, government, and finance. Recently, transformer-based models have demonstrated promising results in automatic software vulnerability identification due to their robust contextual modelling and representation learning capabilities. \textbf{Objectives:} While numerous systematic literature reviews (SLRs) have examined machine learning and deep learning methods for identifying vulnerabilities, a more transformer-centric analysis remains to be explored. This SLR critically analysed 80 studies published between 2021 and 2025 that utilised transformer models to identify software vulnerabilities. \textbf{Methods:} Using Kitchenham’s SLR guidelines, we methodically evaluate current research from various perspectives, encompassing study trends, datasets and sources, programming languages, transformer frameworks, detection detail levels, assessment metrics, reference models, types of vulnerabilities, and experimental configurations. \textbf{Results:} We classify transformer models into encoder, decoder, and combined architectures and analyse both pre-trained and fine-tuned versions utilized on source code, logs, and smart contracts. The results emphasise prevailing research trends, frequently utilised benchmarks, and main baselines. It also uncovers crucial technical issues like data imbalance, interpretability, scalability, and generalization across programming languages. \textbf{Conclusion:} By integrating current evidence and recognising unaddressed research areas, this SLR provides a consolidated resource for researchers and professionals seeking to develop more reliable, precise, and interpretable transformer-based vulnerability identification systems.

\end{abstract}

\begin{graphicalabstract}
\includegraphics{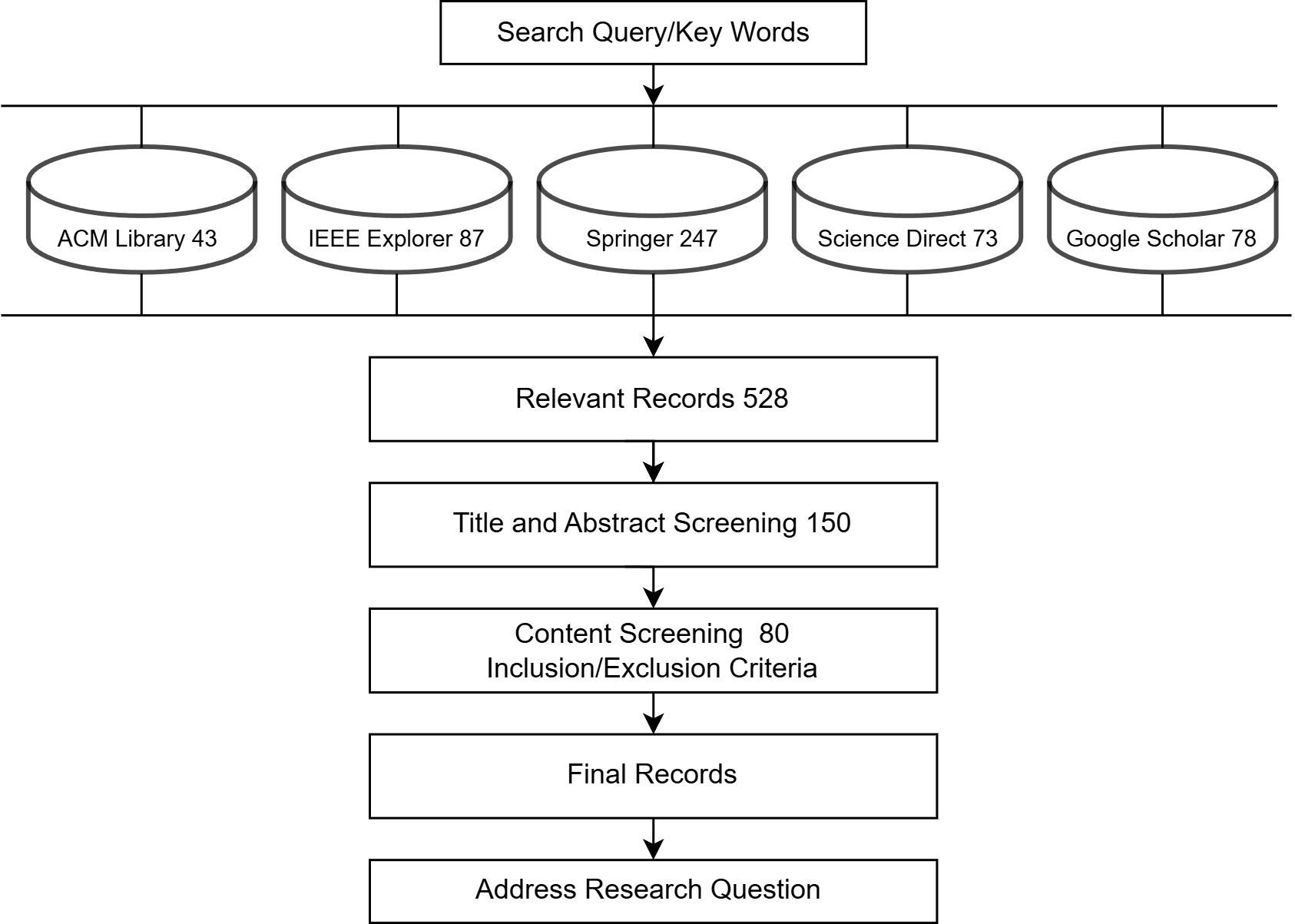}
\end{graphicalabstract}

\begin{highlights}
\item A detailed critical analysis of 80 transformer-based software vulnerability detection approaches to identify insightful information.
\item Identify frequently used transformer architectures, evaluation matrices, and datasets, providing a holistic overview to the software vendors and researchers about the state-of-the-art.
\item Analysing existing approaches to identify different vulnerability types explored to date and mapping them with the CWE.
\item Categorising existing transformer-based software vulnerability-based approaches to different levels of granularity.
\item Identifying state-of-the-art research on multi-lingual software vulnerability detection using transformers.
\end{highlights}

\begin{keywords}
Software Vulnerability \sep Transformer \sep SLR \sep CodeBERT \sep Multi-lingual Software Vulnerability
\end{keywords}
\maketitle
\section{Introduction}
Software vulnerabilities are potential loopholes in software that attackers can exploit to cause harm \cite{chernis2018machine}. Code vulnerability detection plays an important role in software security. Its main goal is to find and fix weaknesses in software code to reduce the risk of cyberattacks and system failures \cite{lin2020software}. These vulnerabilities may include coding errors, design flaws or insecure programming practices that can lead to issues such as data breaches, denial-of-service attacks, and information leakage. As software systems become larger and more complex, security risks and vulnerabilities have become more serious concerns. Detecting these vulnerabilities is a challenging task for developers as it requires analysing large volumes of code \cite{liu2023software}. On the other hand, increased reliance on software applications in critical domains, such as health \cite{ameh2025c3, matloob2025healthcare} and finance \cite{gao2024sguard}, software vulnerabilities pose significant threats, including privacy breaches, service disruptions, and unauthorized access \cite{harzevili2025systematic}. As reported on the CWE site, the complete CWE list contains 943 weaknesses as of version 4.17 \cite{CWESite}. The majority of security flaws today arise from insecure coding practices \cite{bahaa2024db}, which can lead to significant financial losses. For example, an unpatched Apache Struts vulnerability in 2017 caused a major Equifax data breach, exposing the personal information of 147 million users \cite{Example}.

To address these issues, researchers have developed a variety of approaches to detect code vulnerabilities and their types. Methods include conventional static, dynamic, and hybrid code analysis approaches, as well as ML \cite{shiri2024systematic} and DL \cite{harzevili2025systematic} based approaches that can automatically identify potential vulnerabilities. The conventional approaches can be divided into three categories. In the Static Analysis category, a program is analysed from its source code without executing it \cite{ghaffarian2017software}. In the Dynamic Analysis category, a given program is analysed by executing it with specific input data and monitoring its runtime behaviour \cite{alaoui2022deep}. In the Hybrid Analysis category, a given program is analysed with a mixture of static analysis and dynamic analysis techniques\cite{senanayake2023android}
However, static analysis techniques often suffer from high false positives. Dynamic analysis techniques suffer from low code coverage, and hybrid analysis techniques suffer from the limitations of both approaches and are inefficient to operate in practice \cite{ghaffarian2017software, lin2020software}. 

To improve the effectiveness and efficiency of vulnerability detection and reduce manual effort, many learning-based vulnerability detection methods that use ML \cite{ghaffarian2017software,wu2022code} and DL \cite{nong2022open} have been proposed recently. However, ML-based detection methods are limited by the quality of feature engineering and the ability to extract deep features, and they often exhibit high false alarm rates in practice, making them challenging to meet the needs of practical applications. DL has the advantage of processing large amounts of data and mining deep features, and is thus increasingly used for vulnerability detection tasks \cite{zhang2023vuld, perl2015vccfinder}. DL-based approaches still focus on coarse-grained vulnerability prediction, where models only point out vulnerabilities at the file or function level, which remains coarse-grained \cite{fu2022linevul}.
Self-attention-based automated vulnerability models (transformers) are particularly effective. Transformer-based pre-trained language models extract features from long sequences and perform well on natural language tasks, making them suitable for programming language analysis \cite{zhang2023vuld, devlin2019bert, feng2020codebert}.  Several experimental studies highlight their effectiveness at capturing complex code patterns \cite{bahaa2024db, cao2024vulnerability, chan2023transformer, curto2024multivd} . For example, in a research experiment, the transformer model consistently outperformed graph neural network models with improvements in average F1, precision and recall scores. The reason for this gap is the transformer classifier, which generates multiple attention patterns at each layer, yielding contextualised vectors that combine multiple weighted graph structures. This argument supports the use of transformer-based models, such as DetectBERT, for classifying vulnerable statements, particularly when many encoder layers are needed to capture complex data patterns \cite{gujar2024detectbert}.

To highlight the importance of transformers in detecting and classifying software vulnerabilities in software code, it is essential to present a state-of-the-art systematic literature review (SLR) that provides opportunities for researchers and software practitioners to improve existing approaches. Considering its importance, researchers have developed several SLRs for automatic software vulnerability detection; however, they mainly discuss ML/DL for vulnerability detection \cite{senanayake2023android, le2022survey, shiri2024systematic}, they do not specifically address transformer-based models and their interpretation, as summarised in Table \ref{tab:Comaprision}), leaving a research gap. This study analysed 80 research articles covering a wide range of transformer architectures (encoder-based, decoder-based, hybrid) and pre-trained/fine-tuned variants applied to source code analysis. We investigate the following research questions:

\begin{itemize}
    \item RQ1: What types of studies are currently gaining attention in transformer-based software vulnerability detection?
    \item RQ2: What are the commonly used datasets, their sources, and the programming languages involved?
    \item RQ3: Which transformer models are frequently used for software vulnerability detection, and what are their reported accuracies?
    \item RQ4: What evaluation metrics are commonly used to evaluate transformer-based approaches' performances?
    \item RQ5: What types of software vulnerabilities are most frequently detected?
    \item RQ6: What hyperparameters and environment settings are commonly used for transformer-based experiments?
    \item RQ7: At what level of granularity is vulnerability detection performed?
    \item RQ8: What baseline models are used for comparing transformer-based approaches' performances?
    \item RQ9: How many existing approaches focus on multi-lingual software vulnerability detection using transformers?
\end{itemize}

The structure of the paper is as follows: Section 2 elaborates on the related work. Section 3 describes the methodology adopted for the proposed systematic literature review (SLR). Section 4 presents the answers to the research questions. Section 5 discusses the threats to validity. Section 6 outlines open challenges and potential future work on software vulnerability detection, and Section 7 presents the conclusion.

\section{Related Work}
In this section, we elaborate on existing SLRs on software vulnerabilities and explain how the proposed SLR differs from them. The comparative study of the proposed SLR with the existing SLRs is depicted in analyze existing reviews in Table \ref{tab:Comaprision}. Whereas Table \ref{tab:Comaprision} compares existing SLRs in terms of their coverage of key aspects related to transformer-based software vulnerability detection. Specifically, it evaluates whether each SLR addresses transformer models, hyper-parameter analysis, detection granularity, vulnerability types, baseline models, evaluation metrics, and data sources.

Senanayake et al. \cite{senanayake2023android} examine 118 research articles focused on detecting and preventing vulnerabilities in Android source code. It explores both ML-based and conventional methods, showing the relative frequency with which each approach is used. The study reviews various analysis techniques, including static, dynamic, and hybrid analyses. It also discusses various tools and repositories that support vulnerability detection and compares different tools and frameworks used in Android application analysis, highlighting their strengths and limitations.  Similarly, Harzevili et al. \cite{shiri2024systematic} analysed 138 research papers and explored ML techniques for detecting vulnerabilities in automated software. It provides a detailed review of benchmark datasets, repositories, and data types used in these studies. The survey also examines how datasets are represented and embedded, the types of models applied, trends over time, the top 18 vulnerability types, and the tools commonly used for vulnerability detection in the reviewed articles. While Le et al. \cite{le2022survey} primarily focus on software vulnerability assessment, they highlight the characteristics of vulnerabilities identified during the discovery phase and their prioritisation by severity. It reviews 84 research articles and provides a detailed analysis of data sources and data-driven approaches, particularly those involving Natural Language Processing (NLP), ML and DL techniques. The survey also summarises commonly used data sources, features, models, evaluation methods, and metrics for software vulnerability assessment and prioritisation. 

Moreover, Croft et al. \cite{croft2022data} SLR mainly focuses on data preparation processes and their role in software vulnerability prediction. They review 61 relevant studies and highlight the most commonly used programming languages in this research area. The survey also discusses the main types of data sources and compares their frequency of use. Additionally, it outlines common data labelling methods and data cleaning techniques. The study presents a taxonomy of data-related challenges in software vulnerability research, identified from the reviewed papers. Additionally, Ghaffarian et al. \cite{ghaffarian2017software} explore the use of ML and data mining techniques for detecting software vulnerabilities. It reviews two main categories of research: one focusing on the analysis of program syntax and semantics, and the other on software metrics-based approaches. The survey summarizes recent work on vulnerability prediction models, anomaly detection methods, recognition of vulnerable code patterns, and other miscellaneous techniques.
On the other hand, Eberebdu et al. \cite{eberendu2022systematic} reviewed 55 articles related to software vulnerability detection. It discusses trends in detection methods, various detection approaches, topics commonly addressed, and the characteristics and causes of software vulnerabilities. The study covers techniques such as neural networks, machine learning, code clone detection, and static and dynamic analysis methods. Bassi et al. \cite{bassi2023systematic} reviewed 77 articles focused on software vulnerability prediction. It explores DL and ML techniques, tools, and feature types used in prediction models. The survey also covers data balancing techniques, cross-validation methods, feature extraction methods and datasets, evaluation metrics, and parameter tuning approaches. Finally, Sohan et al. \cite{sohan2020systematic} focus specifically on malware detection in JavaScript. It reviews 32 articles and addresses research questions about trends in the field, dataset types and sizes, data analysis methods, detection techniques, performance metrics, and common challenges. The survey also examines the machine learning and data analysis methods employed in these studies.
\begin{table*}
\caption{Comparison Between Contribution of Our Survey and Existing Vulnerability Detection SLRs}
\label{tab:Comaprision}
\begin{tabularx}{\textwidth}{@{} LCCCCCCCC @{}}
\toprule
\textbf{SLR} & \textbf{Transformers} & \textbf{Hyper-parameter} & \textbf{Granularity} & \textbf{Vul.Types} & \textbf{Baseline} & \textbf{Eval.Metric} & \textbf{Data Source} & \textbf{Multi-language} \\
\midrule
Senanayke et al.\cite{senanayake2023android} & \ding{55} & \ding{55} & \ding{55} & \ding{55} & \ding{55} & \ding{55} & \ding{51} & \ding{55} \\
H.M. Le et al.\cite{le2022survey} & \ding{55} & \ding{55} & \ding{55} & \ding{55} & \ding{55} & \ding{55} & \ding{51} & \ding{55} \\
Harzevili et al.\cite{shiri2024systematic} & Partial & \ding{55} & \ding{55} & \ding{51} & \ding{55} & \ding{55} & \ding{51} & \ding{55} \\
Croft et al.\cite{croft2022data} & \ding{55} & \ding{55} & \ding{51} & Partial & \ding{55} & \ding{55} & \ding{55} & \ding{55} \\
Ghaffarian et al.\cite{ghaffarian2017software} & \ding{55} & \ding{55} & \ding{51} & \ding{55} & \ding{55} & \ding{51} & Partial & \ding{55} \\
Eberendu et al.\cite{eberendu2022systematic} & \ding{55} & \ding{55} & \ding{55} & Partial & \ding{55} & \ding{55} & Partial & \ding{55} \\
Bassi and Singh \cite{bassi2023systematic} & \ding{55} & \ding{51} & \ding{55} & \ding{55} & \ding{55} & \ding{51} & \ding{51} & \ding{55} \\
Sohan and Basalamah\cite{sohan2020systematic} & \ding{55} & \ding{55} & \ding{55} & \ding{55} & \ding{51} & \ding{51} & Partial & \ding{55} \\
Our Survey & \ding{51} & \ding{51} & \ding{51} & \ding{51} & \ding{51} & \ding{51} & \ding{51} & \ding{51} \\
\bottomrule
\end{tabularx}
\end{table*}
\vspace{2pt}

Most existing SLRs focus on general vulnerability-detection techniques and provide limited or no coverage of transformer-based approaches, as shown in Table \ref{tab:Comaprision}. For example, Senanayake et al. \cite{senanayake2023android}  and H. M. Le et al. \cite{le2022survey} cover data sources but do not address transformers, hyperparameters, granularity, vulnerability types, baseline models, or evaluation metrics in detail. Harzevili et al. \cite{harzevili2025systematic} partially discuss transformer-based approaches and vulnerability types, but do not examine hyperparameters, granularity, baseline models, or evaluation metrics comprehensively. Similarly, Croft et al. \cite{croft2022data}, and Ghaffarian et al. \cite{ghaffarian2017software} address certain aspects such as granularity and evaluation metrics. However, their coverage remains incomplete or partial across other dimensions. On the other hand, Eberendu et al. \cite{eberendu2022systematic} and Bassi and Singh \cite{bassi2023systematic} provide partial or selective coverage of vulnerability types, evaluation metrics, and data sources, yet lack a focused analysis of transformer models. Sohan and Basalamah \cite{sohan2020systematic} address baseline models and evaluation metrics but do not discuss transformers or detection granularity in detail. In contrast, the proposed SLR approach provides comprehensive coverage across all evaluated dimensions, including transformer models, hyperparameters, detection granularity, vulnerability types, baseline comparisons, evaluation metrics, and data sources. This highlights the novelty and contribution of the proposed work, as it offers the first systematic and in-depth review dedicated to transformer-based vulnerability detection, to the best of our knowledge. The proposed SLR provides opportunities for software developers, researchers, and vendors to equip themselves with state-of-the-art transformer-based vulnerability approaches to further improve the performance of existing vulnerability-based approaches.

In contrast, the proposed research explores the existing research focused on vulnerability detection and prediction. The review emphasizes studies that use transformer-based approaches for identifying software vulnerabilities. Harzevili et al. \cite{shiri2024systematic} briefly discuss transformer-based approaches, but the study’s scope is limited and does not provide an in-depth analysis.  We examine the use of pre-trained models, their fine-tuning strategies, and key hyperparameter configurations. The proposed SLR also unfolded the reported performance evaluation methods, results, commonly addressed vulnerability types and the diversity of data sources and datasets employed in the literature. In addition, we consider articles that use multiple programming languages and assess how different techniques contribute to effective software vulnerability detection and prediction. 

\section{Methodology}
\subsection{Studies Source}

\begin{figure*}[!ht]
    \centering
    \includegraphics[width=0.95\textwidth]{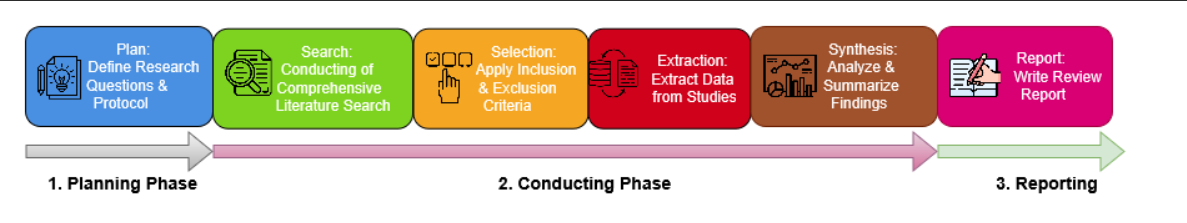}
    \caption{SLR methodology stages following the Kitchenham guidelines \cite{kitchenham2007guidelines}}
    \label{fig:Kitchenham}
\end{figure*}

In this article, we conduct a systematic literature review following Kitchenham’s guidelines \cite{kitchenham2007guidelines}. Figure \ref{fig:Kitchenham} shows the Kitchenham methodology stages, which define a structured methodology for conducting systematic literature reviews in software engineering through three phases: planning phase, conducting phase that includes search, selection, extraction and Reporting phase that includes results report writing and discussion, to ensure rigour, transparency, and reproducibility. We collect and examine research published between 2021 and 2025 that focuses on software vulnerability detection using Transformer-based models. We examine articles from 2021 because Transformer models \cite{vaswani2017attention} were introduced in 2017, and researchers began actively exploring their applications shortly thereafter; however, a substantial body of mature and promising research has been available since 2021, based on knowledge and search. The overall workflow of the systematic methodology is illustrated in Figure \ref{fig:SLR}. To ensure comprehensive coverage, we draw the SLR data from several widely used and reputable digital libraries, including the ACM Digital Library, ScienceDirect, IEEE Xplore, Springer, and Google Scholar. We included Google Scholar to ensure that any related and important transformer-based vulnerability detection approaches are not missed in the proposed SLR, as it searches through various research databases to identify potential papers for further analysis.  

\subsection{Search String}
For the proposed SLR, we developed the following research string to identify transformer-based vulnerability detection approaches:

(“software vulnerability detection” OR “code vulnerability detection” OR “software vulnerability prediction” OR “software flaw detection” OR “security vulnerability detection” OR “software anomaly detection”)
AND
(“transformer model” OR “transformer-based model”)

\subsubsection{Inclusion Criteria}
The following criteria were used to include studies in the proposed systematic literature review:
\begin{itemize}
\item Studies that address software vulnerability detection, classification, or prediction.
\item Studies that employ transformer-based approaches.
\item When multiple versions of a study exist, only the most recent and comprehensive version was included.
\item Studies published in the English language.
\item Studies published between 2021 and 2025.
\item Articles focusing on the significance, strategies, techniques, application domains, or challenges of transformer-based models for software vulnerability detection.

\end{itemize}

\subsubsection{ Exclusion Criteria}

The following criteria were used to exclude studies from this review:
\begin{itemize}
\item Studies that do not specifically address software vulnerabilities. For example, works on network security.
\item Duplicate publications.
\item Pre-print articles.
\item Studies not published in English.
\item Literature that is not directly related to transformer-based methods.
\end{itemize}
\subsection{Data Extraction}

\begin{figure*}
    \centering
 \includegraphics[width=0.95\textwidth]{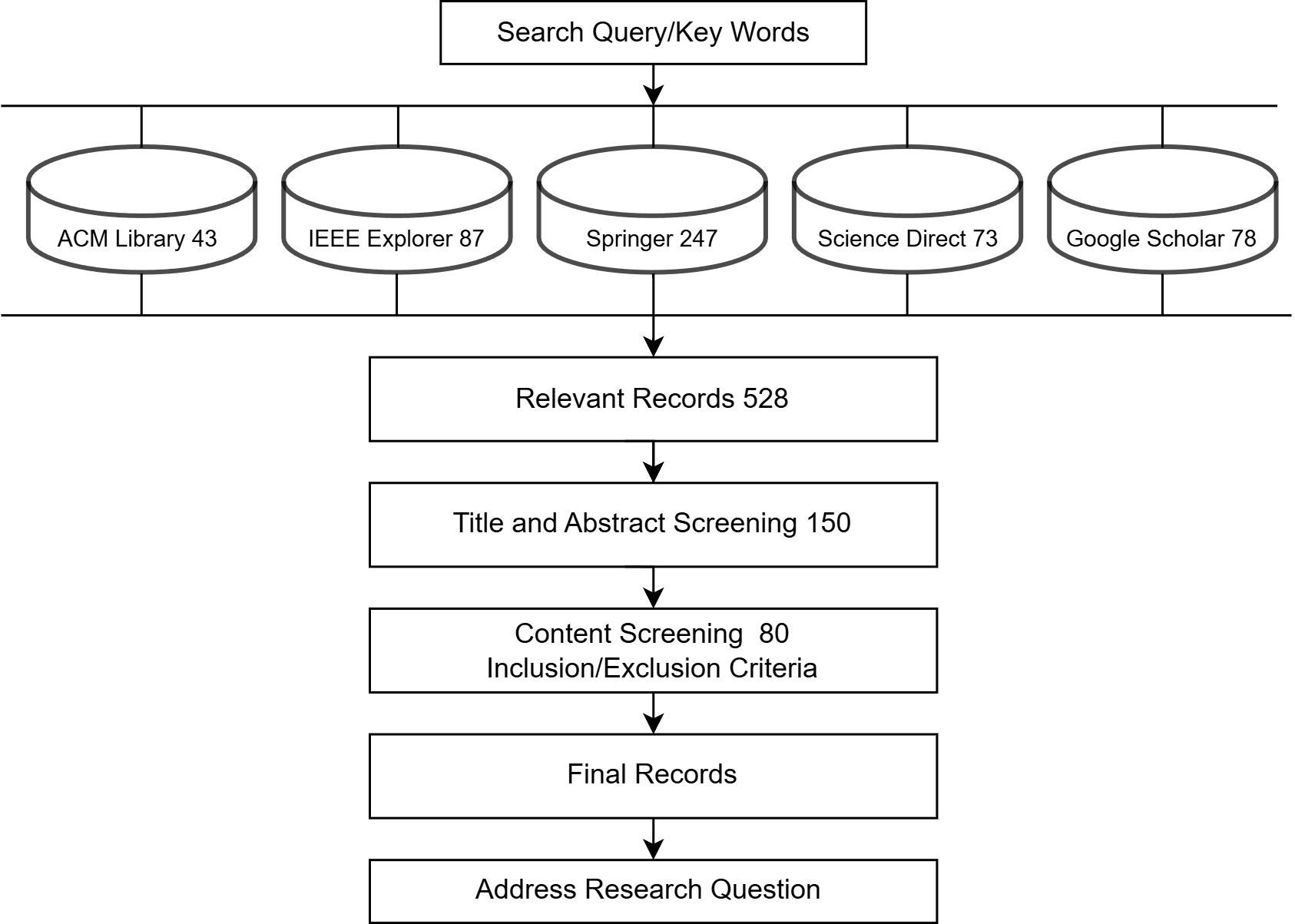}
    \caption{Overall workflow of our systematic survey}
    \label{fig:SLR}
\end{figure*}

Figure \ref{fig:SLR} shows the publication collection process for the proposed SLR. After conducting a search in the ACM Digital Library using relevant keywords, we found 43 articles. From these, we selected 15 articles that satisfy the inclusion criteria defined for the proposed SLR. The remaining articles were excluded based on titles and abstracts, as they mainly addressed general topics in machine learning or artificial intelligence rather than software vulnerability detection using a transformer. Additionally, when we run the search query on the ACM Digital Library, it returns names of various conference proceedings that were not relevant to the search query. When we explored further, we found that the papers published under these themes or conference titles were not relevant, so we excluded them from the final count.  After conducting a search on Springer, a total of 247 articles were retrieved. From these, only 6 articles were found to be relevant and selected for inclusion in the proposed SLR. A total of 203 articles were excluded based on their titles because they primarily focused on unrelated domains, such as medical applications for disease detection, computer networking, time-series forecasting, and business-related software. The remaining 53 articles were rejected after reviewing their abstracts, which did not align with the objectives of the proposed study.

\begin{figure}
    \centering
    \includegraphics[width=.9\columnwidth]{"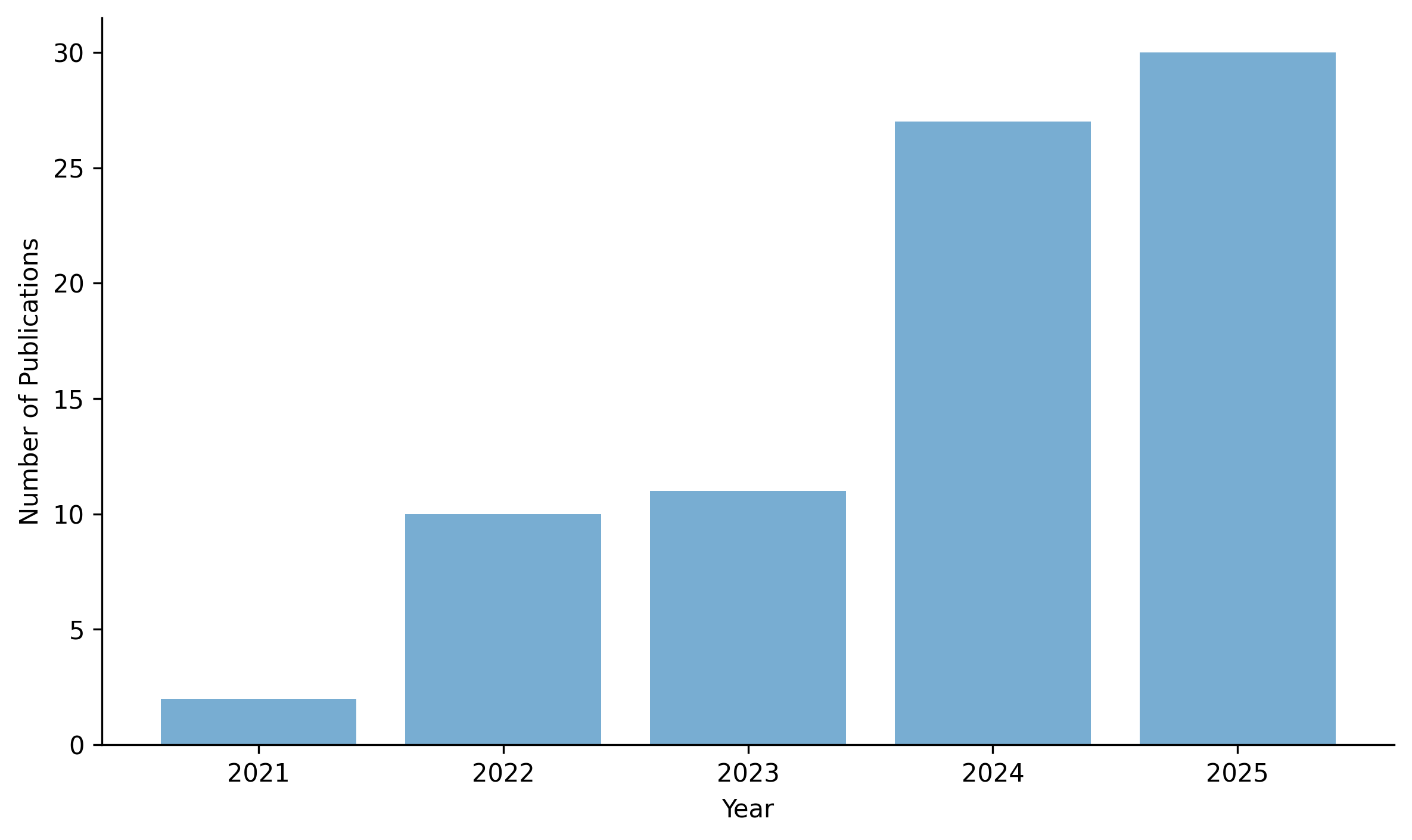"}
    \caption{Number of Publications per Year }
    \label{fig:Publication}
\end{figure}

We searched IEEE Xplore using the keywords "Software vulnerability detection using transformers" because they yielded the most relevant results compared to the main search query. We found 87 papers in total. Of these, we selected 36 papers for the proposed SLR and excluded 51. We removed 33 papers because their titles were unrelated to our topic, and 18 more were excluded after reading their abstracts, as they did not align with the research objectives. We applied a search query on ScienceDirect and initially retrieved 73 articles. We then filtered the results to include only research articles within the computer science domain, reducing the set to 50 articles for further screening. After a detailed review of titles and abstracts, we excluded 33 additional articles and selected only 17 articles relevant to the proposed study based on the inclusion criteria. We conducted a search on Google Scholar and initially found 78 articles. Following a screening process, we shortlisted 16 articles for detailed evaluation. Out of the remaining, 44 articles were excluded based on irrelevant titles, and 18 were rejected based on abstract content. After removing duplicates from the shortlisted set, we finalized 5 articles for inclusion in the proposed survey. Precisely, we had a total of 80 articles that are included in this review paper. Figure \ref{fig:Publication} illustrates the number of research articles published during the period of 2021 to 2025. It shows an increasing trend of transformer-based approaches for software vulnerability detection, with the highest number of papers published in 2025. It highlights the importance of a detailed SLR on transformer-based approaches to software vulnerability detection by identifying the key findings to date. Additionally, before submitting the SLR paper, we quickly reviewed the latest studies published in 2026 \cite{khan2026leveraging,reza2026empirical,do2026novel} and included them for the SLR, but we did not examine them in detail.
\subsection{Quality Assessment}
Before performing a detailed analysis of the research articles, we evaluate the relevance and rigor of the selected studies. We establish specific quality criteria for assessment and formulate a set of questions. These questions help determine whether a study should be included for further analysis. Only studies that receive positive answers to these questions are considered relevant.
\begin{itemize}
\item Q1: Does the paper clearly state a research goal related to software vulnerability detection in its introduction?
\item Q2: Does the proposed approach employ specifically transformer/attention-based models?
\item Q3: Is the vulnerability detection technique clearly defined and described in a way that makes it repeatable or reproducible?
\item Q4: Does the study present an explicit contribution to the field of software vulnerability detection?
\item Q5: Is there a clear and well defined methodology for validating the proposed transformer based approach?
\item Q6: Are the subject projects or software systems used for validation appropriate and aligned with the stated research objectives?
\item Q7: Are the datasets used in the study relevant to software vulnerability detection?
\item Q8: Does the study include baseline models or control techniques to demonstrate the effectiveness of the proposed approach?
\item Q9: Are the evaluation metrics appropriate and relevant to measuring vulnerability detection performance?
\item Q10: Do the reported results align with the stated research objectives, and are they presented clearly and in a meaningful manner?
\end{itemize}

\section{Results and Discussion}
In this section, we present the detailed analysis of the selected research articles. The findings are discussed with respect to all research questions and are supported by Tables and Figures to better illustrate them for potential readers.
\subsection{What are the current research foci in software vulnerabilities using transformers?}
To answer this research question, we analysed each research article included in the proposed SLR to identify the current foci in software vulnerability detection. The holistic foci of current research trends are shown in Table \ref{tab:Trends}. By analysing recent research studies on software vulnerability detection, we can observe a clear study framework and methodological diversity. Among the reviewed works, 39 papers primarily address binary vulnerability detection, aiming to distinguish between vulnerable and non-vulnerable code segments. In contrast, a group of 38 papers focuses on multi-class detection, identifying distinct categories of vulnerabilities such as buffer overflows, injection flaws, and access control issues, reflecting a growing trend toward more fine-grained vulnerability classification.  Moreover, only 3 studies explore vulnerability prediction, there aims to forecast the likelihood of vulnerabilities before they occur. For example, Liu et al. \cite{liu2024making} methodology starts by extracting multidimensional code representations, including plain text, flattened control/data-flow sequences, and structural program graphs. In the second step, they use a combination of CodeBERT and bidirectional LSTM (BLSTM) models in a multi-model training regime to capture semantic and syntactic patterns. At the final step, they evaluate the approach across multiple real-world datasets.
\begin{table*}[!h]
\caption{Current Research Trends of Studies}
\label{tab:Trends}
\begin{tabularx}{\textwidth}{@{} p{5cm} p{1cm} p{10cm} @{}}
\toprule
\textbf{Category} & \textbf{Count} & \textbf{Articles} \\
\midrule
Binary Vulnerability Classification & 39 & \cite{li2022software,chen2022hlt,mahyari2024harnessing,bui2023detecting,zhao2025vulnerability,hin2022linevd,gupta2024dl,liang2024source,wu2021self,liu2023software,zhang2023vuld,jianjie2023code,zhang2024twlog,almakayeel2024deep,myllari2025ladle,he2025vultr,wang2024scl,cao2025multi,jiang2024haformer,wu2025information,yang2024large,xuan2025large,ni2025abundant,do2024optimizing,liu2024pre,rusinova2024explaining,smaili2025transformer,kalouptsoglou2025transfer,wang2025m2cvd,wang2025line,sun2025hgtjit,gunda2025transformer,li2025macd,vanam2025software,cui2025vulgtda,tian2025efvd,oladokun2025effective,perera2025codebert} \\
\hline
Vulnerability Detection on Specific Types / Multi-class Type Classification & 38 & \cite{saimbhi2024vulnerai,bahaa2024db,purba2023software,ferrag2025securefalcon,lu2023assessing,sun2024enhancing,kim2022vuldebert,gujar2024detectbert,peng2023ptlvd,zhao2024python,kim2024robust,nguyen2023mando,wu2021peculiar,tanko2025approach,thapa2022transformer,cao2024vulnerability,mamede2022transformer,hanif2022vulberta,hou2022vulnerability,zahid2025detectbert,shiaeles2023vuldetect,kaanan2024llm,islam2023unbiased,sun2024gptscan,curto2024multivd,alqarni2022low,gong2023gratdet,liu2024automatic,mechri2025secureqwen,ehrenberg2024python,wang2024extensive,jeon2024design,katz2025siexvults,shir2025robust,ferretti2025detecting,sultan2025codevul,shang2025cegt,tao2025transformer} \\
\hline
Vulnerability Prediction & 3 & \cite{le2024software,liu2024making,fu2022linevul} \\
\bottomrule
\end{tabularx}
\end{table*}
Similarly, Le et al. \cite{le2024software} predict vulnerability by constructing vulnerability datasets using CVEfixes and fine‑tuning a CodeBERT model for both function‑level and line‑level vulnerability prediction with and without data sampling techniques to evaluate performance under data scarcity, and additionally explore using ChatGPT as an alternative predictive model, finding that CodeBERT performance deteriorates in low resource settings while ChatGPT yields substantial gains in prediction accuracy. On the other hand, Fu and Tantithamthavorn \cite{fu2022linevul} apply a  CodeBERT with token embeddings to source code to learn contextual semantic representations and then uses a classifier on top of those embeddings to perform vulnerability prediction at both function and fine grained line levels, leveraging self attention mechanisms to identify vulnerable code patterns and rank likely vulnerable lines via attention scores, resulting in improved line level vulnerability detection over prior sequence or graph based models. 

\begin{tcolorbox}[ title={RQ1 Research Finding}]
Researchers have shown interest in software binary vulnerability detection,
multi-class software vulnerability detection and software vulnerability prediction. Among these, 
Binary vulnerability detection is comparatively dominant. However, the trends show that
fine-grained software vulnerability analysis is receiving equal attention from the 
research community to better understand frequently occurring software vulnerabilities. Moreover, we identify only three research papers focusing on software vulnerability prediction, indicating an alternative research domain that researchers can further explore for early software vulnerability detection.
\end{tcolorbox}

\subsection{What are the commonly used datasets, their sources, and the programming languages involved?}
Software vulnerability detection and classification are emerging issue in software development that needs efforts from software researchers and vendors to minimise their effects on existing software systems. For this purpose, researchers and software vendors have begun developing various datasets and baseline approaches to improve the detection and classification of software vulnerabilities. In this research question, we investigate the datasets used in transformer-based software vulnerability detection research articles, their frequency of use, the programming languages included in each dataset, and their corresponding links, aiming to provide a holistic overview of these resources for the research community and software vendors. 

\begin{table*}
\footnotesize
\caption{The datasets used in the research articles, along with their links and the programming languages. }
\label{tab:dataset} 
\begin{tabularx}{\textwidth}{@{} p{3cm} p{2cm} p{7.5cm} p{3.5cm} @{}}
\toprule
 \textbf{Dataset} &  \textbf{Language} &  \textbf{URL} &  \textbf{Articles}\\
 \midrule

BigVul & C/C++ & \url{https://www.kaggle.com/datasets/kaggler10240/msr-data} &
\cite{lu2023assessing,peng2023ptlvd,liang2024source,rusinova2024explaining,curto2024multivd,he2025vultr,wang2024scl,cao2025multi,liu2024making,xuan2025large,ni2025abundant,li2025macd,kalouptsoglou2025transfer,sultan2025codevul} \\
\hline
SARD & C, C++, Java, PHP, and C\# & \url{https://samate.nist.gov/SARD} &
\cite{bahaa2024db,mahyari2024harnessing,kim2022vuldebert,liang2024source,hou2022vulnerability,shiaeles2023vuldetect,he2025vultr,cui2025vulgtda,tao2025transformer} \\
\hline
VulDeePecker & C/C++ & \url{https://github.com/CGCL-codes/VulDeePecker} &
\cite{purba2023software,liu2024pre,thapa2022transformer,hanif2022vulberta} \\
\hline
FormAI & C & \url{https://github.com/FormAI-Dataset/FormAI} &
\cite{ferrag2023securefalcon} \\
\hline
Devign & C/C++ & \url{https://github.com/epicosy/devign} &
\cite{lu2023assessing,bui2023detecting,sun2024enhancing,liu2023software,zhang2023vuld,islam2023unbiased,he2025vultr,wang2024scl,cao2025multi,liu2024making,xuan2025large,do2024optimizing,wang2025line,wang2025m2cvd} \\
\hline
Reveal & C/C++ & \url{https://huggingface.co/datasets/claudios/ReVeal} &
\cite{lu2023assessing,liu2024pre,zhang2023vuld,hanif2022vulberta,islam2023unbiased,wang2024scl,xuan2025large,tian2025efvd,wang2025line,wang2025m2cvd,kalouptsoglou2025transfer,sultan2025codevul} \\
\hline
LVDAndro & Android & \url{https://github.com/softwaresec-labs/LVDAndro} &
\cite{oladokun2025effective} \\
\hline
FFmpeg & C/C++ & \url{https://github.com/ffmpeg/ffmpeg} &
\cite{tao2025transformer,smaili2025transformer} \\
\hline
FFmpeg+QEMU & C/C++ & \url{https://github.com/ffmpeg/ffmpeg}; \url{https://github.com/qemu/qemu} &
\cite{kalouptsoglou2025transfer,smaili2025transformer} \\
\hline
OpenSSL 1.0.1e & C, C++, \& assembly & \url{https://github.com/openssl/openssl} &
\cite{tao2025transformer} \\
\hline
PostgreSQL 9.2.4 & as above & \url{https://github.com/postgres/postgres} &
\cite{tao2025transformer} \\
\hline
Apache Subversion 1.8.3 & as above & \url{https://github.com/apache/subversion} &
\cite{tao2025transformer} \\
\hline
DiverseVul & C/C++ & \url{https://github.com/wagner-group/diversevul} &
\cite{sultan2025codevul} \\
\hline
NVD & Multi & \url{https://github.com/fkie-cad/nvd-json-data-feeds} &
\cite{mahyari2024harnessing,kim2022vuldebert,liu2024automatic,ehrenberg2024python} \\
\hline
Debian & Multi & \url{https://www.debian.org/download} &
\cite{zhao2025vulnerability} \\
\hline
CVEfixes & Multi & \url{https://github.com/secureIT-project/CVEfixes} &
\cite{gujar2024detectbert,zahid2025detectbert} \\
\hline
VUDENC & Python & \url{https://huggingface.co/datasets/DetectVul/Vudenc} &
\cite{gujar2024detectbert,zhao2024python,zahid2025detectbert} \\
\hline
LineVD & C/C++ & \url{https://github.com/davidhin/linevd} &
\cite{hin2022linevd} \\
\hline
SB Curated & Solidity & \url{https://github.com/smartbugs/smartbugs-curated} &
\cite{kim2024robust,nguyen2023mando} \\
\hline
GitHub & Multi & \url{https://github.com} &
\cite{mechri2025secureqwen,ehrenberg2024python,alqarni2022low} \\
\hline
CodeXGLUE & Multi & \url{https://github.com/microsoft/CodeXGLUE} &
\cite{liu2024pre,hanif2022vulberta,jianjie2023code} \\
\hline
D2A & C/C++ & \url{https://github.com/IBM/D2A} &
\cite{liu2024pre,hanif2022vulberta,islam2023unbiased} \\
\hline
SolidiFI-Benchmark & Solidity & \url{https://github.com/DependableSystemsLab/SolidiFI-benchmark} &
\cite{nguyen2023mando} \\
\hline
SmartBugs Wild & Solidity & \url{https://github.com/smartbugs/smartbugs-wild} &
\cite{nguyen2023mando,wu2021peculiar} \\
\hline
OWASP WebGoat & Java Web & \url{https://github.com/WebGoat/WebGoat} &
\cite{tanko2025approach} \\
\hline
SeVC dataset & C/C++ & \url{https://github.com/SySeVR/SySeVR} &
\cite{thapa2022transformer,shiaeles2023vuldetect} \\
\hline
OverflowGen & Solidity & \url{https://figshare.com/s/fbaf47e3ac2a9581dbd7} &
\cite{cao2024vulnerability} \\
\hline
CodeSearchNet & Multi & \url{https://github.com/github/CodeSearchNet} &
\cite{fu2022linevul} \\
\hline
Draper VDISC & C/C++ & \url{https://osf.io/d45bw/} &
\cite{chen2022hlt,liu2024pre,hanif2022vulberta,kaanan2024llm,sultan2025codevul} \\
\hline
Juliet Test Suite v1.3 & C/C++ & \url{https://samate.nist.gov/SARD/test-suites/112} &
\cite{jianjie2023code} \\
\hline
MVD & C/C++/Python
/Java & \url{https://github.com/mvd-dataset/MVD} &
\cite{islam2023unbiased} \\
\hline
Top200 & Solidity & \url{https://github.com/MetaTrustLabs/GPTScan-Top200} &
\cite{sun2024gptscan} \\
\hline
Web3Bugs & Solidity & \url{https://github.com/MetaTrustLabs/GPTScan-Web3Bugs} &
\cite{sun2024gptscan} \\
\hline
DeFiHacks & Solidity & \url{https://github.com/MetaTrustLabs/GPTScan-DefiHacks} &
\cite{sun2024gptscan} \\
\hline
HDFS (project) & Java & \url{https://github.com/apache/hadoop} &
\cite{zhang2024twlog} \\
\hline
OpenStack & Python & \url{https://opendev.org/openstack} &
\cite{zhang2024twlog} \\
\hline
Kubernetes & Go & \url{https://github.com/kubernetes/kubernetes} &
\cite{zhang2024twlog} \\
\hline
Peculiar & Solidity & \url{https://github.com/wuhongjun15/Peculiar} &
\cite{gong2023gratdet} \\
\hline
MegaVul & C/C++ & \url{https://github.com/Icyrockton/MegaVul} &
\cite{cao2025multi} \\
\hline
Linux Kernel & C & \url{https://github.com/torvalds/linux} &
\cite{liu2024making} \\
\hline
BinKit & C & \url{https://github.com/SoftSec-KAIST/BinKit} &
\cite{jiang2024haformer} \\
\hline
HDFS (LogHub) & Log Text & \url{https://github.com/logpai/loghub} &
\cite{wu2025information} \\
\hline
Blue Gene/L (BGL) & Log Text & \url{https://github.com/logpai/loghub} &
\cite{wu2025information} \\
\hline
Defects4J & Java & \url{https://github.com/rjust/defects4j} &
\cite{yang2024large} \\
\hline
BugsInPy & Python & \url{https://github.com/soarsmu/BugsInPy} &
\cite{yang2024large} \\
\hline
SmartCon & Solidity & \url{https://github.com/s00ne/SmartConDetect} &
\cite{jeon2024design} \\
\hline
VDET-for-Java & Java & \url{https://github.com/TQRG/VDET-for-Java} &
\cite{mamede2022transformer} \\
\hline
CVE & Multi & \url{https://www.kaggle.com/datasets/andrewkronser/cve-common-vulnerabilities-and-exposures} &
\cite{katz2025siexvults} \\
\hline
Drebin & Android applications & \url{https://www.kaggle.com/datasets/shashwatwork/android-malware-dataset-for-machine-learning} &
\cite{almakayeel2024deep} \\
\bottomrule
\end{tabularx}
\end{table*}

Table \ref{tab:dataset} summarizes the datasets used in recent software vulnerability detection studies, including their programming languages, sources, and corresponding references. The table \ref{tab:dataset} shows that most research is based on C/C++ datasets, with BigVul, Devign, Reveal, and SARD being the most frequently used benchmarks. It also highlights the growing use of multi-language datasets such as CVEfixes and CodeXGLUE, as well as the increasing importance of Solidity datasets in smart contract vulnerability detection. Overall, the Table \ref{tab:dataset} indicates that current research is still centred on traditional C/C++ vulnerability detection, while Python, Java, Go, and Solidity are receiving increasing attention. It also shows that most vulnerability detection studies rely on publicly available datasets derived from real-world software repositories and vulnerability databases. Widely used datasets include Devign, Big-Vul, Reveal and SARD, which are constructed by mining source code from GitHub and linking it to vulnerability labels from the National Vulnerability Database (NVD) and Common Vulnerabilities and Exposures (CVE) records \cite{su2026source}. For Example, Saimbhi and Akpinar \cite{saimbhi2024vulnerai} use a custom dataset of PHP code snippets that cover 17 types of vulnerabilities. The process involved parsing PHP code to extract individual functions, which were then analysed using prompt engineering techniques. These prompts were submitted to the GPT-3.5 Turbo model through the OpenAI API to identify possible software vulnerabilities. Similarly, Li et al. \cite{li2022software} use a custom dataset consisting of C programs from publicly available vulnerability datasets, while the model itself is trained on unlabelled code to learn normal coding patterns. Each function or code snippet is encoded into a representation capturing syntactic and semantic features, and an anomaly attention mechanism is applied to highlight unusual patterns indicative of potential vulnerabilities.

\begin{tcolorbox}[ title={RQ2 Research Finding}]
To date, researchers use over 40 types of vulnerability datasets, most of 
which are publicly available, showing researchers growing interest. The BigVul (13) and 
Devign (13) is the most frequently used, followed by Reveal (12) and SARD (9), 
primarily for C/C++ code, while smart contract research relies on Solidity-based datasets. 
Also, there is a trend of developing a vulnerability dataset for multiple programming 
languages.
\end{tcolorbox}

\subsection{RQ3: Which transformer models are used in existing studies, and what are their reported results?}
Transformer-based approaches have been widely used for various natural language processing tasks, achieving comparatively better results than deep and machine learning-based approaches. In the literature, researchers have used various baseline transformers and transformer combinations to identify software vulnerabilities in different programming languages. Through this RQ, we examine the types and combinations of transformer-based models used by software researchers in their studies, along with the best reported results. With this, we aim to provide a holistic overview of the state-of-the-art in transformer-based vulnerability detection for software vendors and researchers to identify models that best suit the problem. Also, this will give opportunities to software researchers to identify areas that need further research. We consider only the best-performing results reported in each study, as determined by the proposed analysis. Table \ref{tab:Transformer} shows the details of the models used in the research articles along with the reported results of the evaluation metric.

 Table \ref{tab:Transformer} shows that transformer models have become the dominant paradigm in software vulnerability detection research. Among them, CodeBERT is the most frequently used backbone, often serving either as a stand-alone encoder or as the semantic foundation of hybrid architectures. A major trend is that researchers increasingly combine transformers with CNNs, BiLSTMs, GNNs, and graph-based reasoning mechanisms, indicating that plain token-sequence modeling is often considered insufficient for capturing the structural complexity of vulnerable code. Recent work also shows a shift toward specialized transformer frameworks, graph-aware models, and multi-model collaboration, while LLMs are increasingly explored as complementary components rather than complete replacements for task-specific code transformers. Researchers use baseline Transformer models and a combination of models to achieve the best results. For example, Alqarni and Azim \cite{alqarni2022low} proposed an improved version of BERT to detect vulnerabilities in source code by extending its architecture with deeper layers tailored for code analysis. Firstly, they preprocess the dataset and address the class imbalance by resampling to create a balanced training set. After that, the improved BERT algorithm was fine-tuned with the most effective hyperparameter to maximise accuracy. Bui and Do \cite{bui2023detecting} have proposed a Vulnerability detection approach for converting the source code into a code property graph (CPG), a unified representation that combines the syntax, control flow, and data flow of the program. Then, each node in this graph, such as statements, variables, or control structures, is converted into a vector using a pre-trained codeBERT model. The hybrid model Adaptive Transformer‑GCN (AT‑GCN) uses the Transformer's self‑attention to understand long-range relationships across the code and a Graph Convolutional Network (GCN) to capture local graph structure around each node. By learning from both global and local features, the model can more accurately identify the software vulnerabilities. In another research, Jian-Jie and Le \cite{jianjie2023code} cleaned the source code and converted it into a more concise form. In the first step, they embed code as tokens, AST tokens, and dependency graph nodes using BERT, convert tokens into high-dimensional vector representations, and then process them with the DL models TextCNN, BiLSTM, BiLSTM CNN, and GCN to extract different types of features. The outputs of all five models are combined via a stacking ensemble, and the resulting predictions are fed into a final classifier that determines whether a software defect is present in the code.

\begin{table*}[p]
\footnotesize
\centering
\caption{Transformer models used in articles along with reported results (Part 1)}
\label{tab:Transformer}

\begin{tabularx}{\textwidth}{@{} p{3cm} p{5cm} X @{}}
\hline
\textbf{Article} & \textbf{Models} & \textbf{Results} \\
\hline

Saimbhi and Akpinar \cite{saimbhi2024vulnerai} & VulnerAI, Transformer (GPT) & Accuracy 70.00\%, Precision 100.00\%, Recall 68.00\% \\
\hline
Li et al. \cite{li2022software} & Transformer with Anomaly Attention Mechanism & Accuracy 87.73\%, FPR 13.60\%, TPR 85.26\%, F1-score 93.23\% \\
\hline
Chen and Liu \cite{chen2022hlt} & LSTM and Transformer (HLT) & Accuracy 67.85\% and 67.00\%; F1-score 70.17\% and 72.03\% \\
\hline
Bahaa et al. \cite{bahaa2024db} & CNN and BiLSTM with BERT tokenizer (DB-CBIL) & Recall 100.00\%, Accuracy 99.81\%, Precision 99.51\%, AUC 99.84\%, F1-score 99.75\% \\
\hline
Purba et al. \cite{purba2023software} & GPT-3.5-Turbo, CodeGen (Transformer Ensemble) & FPR 74.22\%, FNR 3.96\%, TPR 96.04\%, Precision 57.40\%, F1-score 71.85\% \\
\hline
Ferrag et al. \cite{ferrag2023securefalcon} & Self-Attention + MLP (SecureFalcon) & Accuracy 94.00\% (binary), up to 92.00\% (multiclass) \\
\hline
Lu et al. \cite{lu2023assessing} & CodeBERT, CodeT5, and CodeGPT & Accuracy 97.98\%, Precision 91.55\%, Recall 70.50\%, F1-score 79.66\% (Big-Vul dataset) \\
\hline
Mahyari \cite{mahyari2024harnessing} & BERT and DistilBERT & Accuracy 98.25\% (BERT), 98.17\% (DistilBERT) \\
\hline
Bui and Do \cite{bui2023detecting} & GCN + Transformer & Accuracy 54.20\%, Precision 38.70\%, Recall 39.90\%, F1-score 41.10\% \\
\hline
Sun et al. \cite{sun2024enhancing} & CodeBERT (12-layer Transformer Encoder) & Accuracy 70.00\%, F1-score 68.00\% \\
\hline
Zhao and Liu \cite{zhao2025vulnerability} & SCDetect with improved self-attention mechanism (Si-AS) & Accuracy 96.00\%, Precision 98.00\%, Recall 91.00\%, F1-score 95.00\% \\
\hline
Kim et al. \cite{kim2022vuldebert} & BERT & CWE-119: FPR 0.30\%, FNR 4.00\%, TPR 96.00\%, Precision 99.90\%, F1-score 97.90\% \\
\hline
Gujar \cite{gujar2024detectbert} & DetectBERT & F1-score 64.88 $\pm$ 5.46\%, Precision 60.71 $\pm$ 8.08\%, Recall 73.08 $\pm$ 2.65\% \\
\hline
Peng et al. \cite{peng2023ptlvd} & CodeBERT Self-Attention (PTLVD) & Accuracy 60.05\%, Precision 54.90\%, Recall 80.99\%, F1-score 65.44\% \\
\hline
Zhao et al. \cite{zhao2024python} & CodeBERT & Accuracy 95.89--98.63\%, F1-score 94.74--98.33\% \\
\hline
Hin et al. \cite{hin2022linevd} & CodeBERT (LineVD) & F1-score 36.00\%, Recall 53.30\%, Precision 27.10\%, ROC-AUC 91.30\%, PR-AUC 64.20\% \\
\hline
Kim et al. \cite{kim2024robust} & Transformer-based LLMs (BERT, GPT-3, DistilBERT, PruneBERT, etc.) & Accuracy 97.00\%, Precision 97.00\%, Recall 97.00\%, F1-score 97.00\% (Dataset 3) \\
\hline
Liu et al. \cite{liu2024pre} & Custom BERT (PDBERT) & Accuracy 67.61\%, F1-score 59.41\% \\
\hline
Nguyen et al. \cite{nguyen2023mando} & MANDO-HGT (Heterogeneous Graph Transformer) & F1-score 95.40\% (Time Manipulation vulnerability) \\
\hline
Gupta et al. \cite{gupta2024dl} & DL-VulBERT (LSTM + Attention + BERT) & Accuracy 94.43\%, Precision 94.39\%, Recall 94.64\%, F1-score 94.33\% \\
\hline
Liang et al. \cite{liang2024source} & CodeBERT + Adaptive GNN & Accuracy 68.50\%, Precision 70.70\%, Recall 92.00\%, F1-score 82.90\% \\
\hline
Wu et al. \cite{wu2021peculiar} & Peculiar (GraphCodeBERT-based) & Precision 91.80\%, Recall 92.40\%, F1-score 92.10\% \\
\hline
Rusinova et al. \cite{rusinova2024explaining} & CodeBERT with LIME and SHAP & LIME 94.00\%, SHAP 99.00\% \\
\hline
Wu et al. \cite{wu2021self} & Transformer (Multi-Head Attention + FFN) & Accuracy 94.30\%, Precision 85.20\%, F1-score 90.20\%, FPR 6.20\%, FNR 4.20\% \\
\hline
Liu et al. \cite{liu2023software} & Vul-GPT & TF-IDF: Accuracy 49.82\%, Precision 46.47\%, Recall 71.90\%, F1-score 56.45\%; BM25: Accuracy 49.35\%, F1-score 56.18\% \\
\hline
Tanko et al. \cite{tanko2025approach} & Transformer Encoder + Transformer-based CNN + GGNN & CWE-78: Accuracy 74.07\%, Precision 79.59\%, Recall 75.58\%, F1-score 77.53\% \\
\hline
Zhang et al. \cite{zhang2023vuld} & VulD-Transformer + FastText & Accuracy improvement 1.42--6.70\%, Recall improvement 2.72--10.11\%, F1-score improvement 1.82--5.64\% \\
\hline
Thapa et al. \cite{thapa2022transformer} & CodeBERT, GraphCodeBERT, PLBART, CodeT5 & FPR 3.63\%, FNR 9.06\%, Precision 89.14\%, Recall 90.94\%, F1-score 90.03\% (BERT-base) \\
\hline
Cao \cite{cao2024vulnerability} & VDTransformer & Accuracy 92.80\%, Precision 93.90\%, Recall 89.00\%, F1-score 91.40\% \\
\hline
Fu and Tantithamthavorn \cite{fu2022linevul} & LineVul (BPE + CodeBERT + Attention) & F1-score 91.00\%, Precision 97.00\%, Recall 86.00\% \\
\hline
Mamede \cite{mamede2022transformer} & javaBERT & Accuracy 99.00\%, Precision 95.00\%, Recall 93.00\% \\
\hline
Hanif and Maffeis \cite{hanif2022vulberta} & VulBERTa-MLP/CNN (BPE) & VulBERTa-MLP: Precision 95.76\% \\
\hline
Hou et al. \cite{hou2022vulnerability} & Transformer (Attention + MLP) & Precision 95.04\%, Recall 88.89\%, F1-score 91.86\% \\
\hline
Zahid \cite{zahid2025detectbert} & DetectBERT, MiniLM, MPNet & Precision 60.71\%, Recall 73.08\%, F1-score 64.88\% \\
\hline
Omar and Shiaeles \cite{shiaeles2023vuldetect} & GPT-based VulDetect & F1-score 92.40\%, TPR 91.30\%, AUC 90.30\% (SARD dataset) \\
\hline
Kaanan et al. \cite{kaanan2024llm} & VulBERTDense (CodeBERT-based) & Accuracy 90.10\%, Precision 91.10\%, Recall 89.20\%, F1-score 89.70\% \\
\hline
Jianjie and Le \cite{jianjie2023code} & BERT + GCN + CNN + BiLSTM & CodeXGLUE: Accuracy 62.87\%, F1-score 60.51\%; Juliet: Accuracy 95.98\%, F1-score 87.80\% \\
\hline
Islam et al. \cite{islam2023unbiased} & RoBERTa-PFGCN & VulF: Accuracy 96.24\%, F1-score 95.85\%; MVB: Accuracy 98.23\%, F1-score 98.01\% \\
\hline
Sun et al. \cite{sun2024gptscan} & GPTScan (GPT + Static Analysis) & Top200: FP 13; Web3Bugs: TP 40, FP 30; DeFiHacks: TP 10, FN 4 \\
\hline

Curto et al. \cite{curto2024multivd} & Multitask CodeBERT & Accuracy 99.03\%, Precision 97.31\%, Recall 93.87\%, F1-score 95.51\% \\
\hline
Alqarni and Azim \cite{alqarni2022low} & BERT & Accuracy 99.30\% \\
\hline
Le et al. \cite{le2024software} & CodeBERT + ChatGPT (GPT-3.5) & F1-score 43.00\%, Precision 44.00\%, Recall 43.00\% (best on Rust) \\
\end{tabularx}
\end{table*}

\addtocounter{table}{-1} 

\begin{table*}[p]
\footnotesize
\centering
\caption{Transformer models used in articles along with reported results (Part 2)}
\label{tab:Transformer2}
\begin{tabularx}{\textwidth}{@{} p{3cm} p{5cm} X @{}}
\hline
Zhang et al. \cite{zhang2024twlog} & TWLog & HDFS F1-score 95.40\%; OpenStack F1-score 96.30\%; Kubernetes F1-score 90.90\% \\
\hline
Almakayeel \cite{almakayeel2024deep} & DLBITM-AMD (Transformer + RNN) & Accuracy 99.26\%, Precision 99.39\%, Recall 99.26\%, F1-score 99.32\% \\
\hline
Myllari et al. \cite{myllari2025ladle} & Ladle (Transformer-based) & Accuracy 99.76\% \\
\hline
Gong et al. \cite{gong2023gratdet} & GRATDet (Transformer-GP) & Accuracy 95.22\%, Precision 95.59\%, Recall 95.17\%, F1-score 95.16\% \\
\hline
Liu et al. \cite{liu2024automatic} & BiVulD (CodeBERT + LSTM) & CWE-119 F1-score 91.50\%; CWE-399 F1-score 90.40\% \\
\hline
He et al. \cite{he2025vultr} & VulTR (CodeBERT + BiLSTM) & SARD+NVD: Accuracy 95.71\%, F1-score 96.06\%, Recall 96.18\% \\
\hline
Mechri et al. \cite{mechri2025secureqwen} & SecureQwen (CodeQwen1.5, Qwen2) & Accuracy 95.00\%, Precision 99.00\%, Recall 99.00\% \\
\hline
Wang et al. \cite{wang2024scl} & SCL-CVD (GraphCodeBERT + MLP) & Accuracy 91.36\%, F1-score 46.59\% \\
\hline
Ehrenberg et al. \cite{ehrenberg2024python} & CodeBERT, RoBERTa, DistilBERT & Accuracy 97.53\%, Precision 92.65\%, Recall 92.60\%, F1-score 92.61\% \\
\hline
Cao and Dong \cite{cao2025multi} & MSVD (CodePTM + CodeBERT) & Accuracy 82.00\%, F1-score 71.62\%, AUC 86.78\% \\
\hline
Liu et al. \cite{liu2024making} & VulPCL (BiLSTM + CodeBERT) & CWE-264: Accuracy 100.00\%, F1-score 90.91\% \\
\hline
Jiang et al. \cite{jiang2024haformer} & HAformer & AUC 99.73\% \\
\hline
Wu et al. \cite{wu2025information} & Transformer-BERT & Precision 99.00\%, Recall 99.50\%, F1-score 99.30\% \\
\hline
Yang et al. \cite{yang2024large} & LLMAO (CodeGen-16B, GPT-2) & Top-1 22.30\%, Top-3 37.70\%, Top-5 46.30\% \\
\hline
Xuan et al. \cite{xuan2025large} & FG-CVD (BiSelf-Attention + MLP) & Accuracy 96.19\%, F1-score 57.34\% (BigVul) \\
\hline
Tian and Zhang \cite{tian2025efvd} & EFVD (CodeBERT + EA-GGNN + MLP + focal loss) & Maximum absolute improvement of 35.63\% in accuracy and 289.32\% in F1-score over baselines across three benchmark datasets \\
\hline
Katz et al. \cite{katz2025siexvults} & SIExVulTS (SentBERT + CodeQL + GraphCodeBERT) & Attack-surface detection: average F1-score 93.00\%; Exposure analysis: F1-score 85.71\%; Flow verification: Precision 87.23\%, Accuracy 95.50\% \\
\hline
Cui et al. \cite{cui2025vulgtda} & VulGTDA (GraphTransformer + domain adaptation) & QEMU: Accuracy 64.10\%, Precision 64.00\%, Recall 67.30\%, F1-score 65.60\%; FFmpeg: Accuracy 63.90\%, Precision 64.70\%, Recall 63.50\%, F1-score 64.10\%; SARD: Accuracy 83.60\%, Precision 88.90\%, Recall 80.10\%, F1-score 84.30\% \\
\hline
Vanam et al. \cite{vanam2025software} & OAssAI / SF-TransBiLSTM & Accuracy 98.20\% \\
\hline
Sun et al. \cite{sun2025hgtjit} & HgtJIT (Graph-transformer model) & Temporal split: Precision 61.00\%, Recall 51.00\%, F1-score 55.00\%, AUC 83.00\%; Cross-project split: Precision 62.00\%, Recall 64.00\%, F1-score 63.00\%, AUC 80.00\% \\
\hline
Oladokun and Rice \cite{oladokun2025effective} & CodeBERT, GraphCodeBERT & 99.00\% Accuracy \\

\hline
Wang et al. \cite{wang2025line} & CSLS (Transformer-based line-level semantic structure learning model) & Devign: Accuracy 70.57\%, Recall 59.36\%, Precision 71.70\%, F1-score 64.95\%; Reveal: Accuracy 91.86\%, Recall 39.91\%, Precision 65.46\%, F1-score 49.59\% \\

\hline
Shir et al. \cite{shir2025robust} & CodeBERT, Longformer, RNN, LSTM, GRU & LLVM-IR: Accuracy 93.60\% and 94.20\%; Assembly: Accuracy 91.00\% and 89.80\%; single-compilation LLVM-IR CodeBERT: 86.60\% multi-class accuracy \\
\hline
Wang et al. \cite{wang2025m2cvd} & M2CVD (transformer-based code models and LLMs) & Devign: Accuracy 69.25\%, Recall 61.51\%, Precision 68.38\%, F1-score 64.77\%; Reveal: Accuracy 91.78\%, Recall 45.18\%, Precision 62.42\%, F1-score 52.54\% \\
\hline
Ferretti et al. \cite{ferretti2025detecting} & BERT, DistilBERT, CodeBERT, Gemini, RF, G-NB, SVC, GBoost, DT & Best source-code single model: Accuracy 79.00\%, micro-F1 88.00\%; best byte-code single model: Accuracy 77.00\%, micro-F1 86.26\%; best meta-classifier: Accuracy 83.46\%, weighted F1 91.07\% \\
\hline
Perera et al. \cite{perera2025codebert} & CodeBERT-based embeddings + neural network classifier & Accuracy 97.37\%, Precision 96.77\%, Recall 98.36\%, F1-score 97.56\% \\
\hline
Sultan et al. \cite{sultan2025codevul} & CodeVul+ (GraphCodeBERT embeddings plus graph neural network reasoning) & Main multiclass result: AUC 89.60\%, Accuracy 67.10\%, Precision 66.90\%, Recall 66.10\%, F1-score 66.30\%; Juliet F1-score 89.10\% \\
\hline
Shang et al. \cite{shang2025cegt} & CEGT (GCN-Transformer) & Reentrancy F1-score 93.47\%; Timestamp dependence F1-score 89.33\%; Integer overflow F1-score 91.27\% \\
\hline
Kalouptsoglou et al. \cite{kalouptsoglou2025transfer} & CodeBERT fine-tuning / CodeBERT embeddings & Big-Vul: F1-score 91.60\% (fine-tuning), 91.40\% (word-level embeddings) \\
\hline
Tao et al. \cite{tao2025transformer} & CodeBERT-based bimodal Transformer + BiGRU & SARD coarse-grained: Accuracy 97.40\%, Precision 96.70\%, Recall 92.70\%, F1-score 94.70\%; fine-grained localization average IOU 84.50\% \\
\hline
Smaili et al. \cite{smaili2025transformer} & CodeGATNet (CodeBERT + CNN + gated attention) & FFmpeg: Accuracy 76.25\%, F1-score 75.63\%, MCC 52.64\%; QEMU: Accuracy 89.74\%, F1-score 87.48\%, MCC 78.82\%; FFmpeg+QEMU: Accuracy 78.66\%, F1-score 76.80\%, MCC 57.05\% \\
\hline
\end{tabularx}
\end{table*}


Similarly, Bahaa et al. \cite{bahaa2024db} introduce the DB-CBIL model. It is a hybrid deep learning approach developed to automatically detect vulnerabilities in software code. It combines DistilBERT, a compact version of the BERT transformer model, to generate contextual word embeddings from source code functions represented as Abstract Syntax Trees (ASTs). These embeddings capture both the syntax and semantics of the code. The model then employs two neural networks: a CNN to extract local features and a Bidirectional Long Short-Term Memory (BiLSTM) network to capture sequential dependencies in the code. By integrating these components, DB-CBIL effectively identifies vulnerable code patterns. While, Kaanan et al. \cite{kaanan2024llm} propose VulBERTDense, a vulnerability detection approach that combines multiple large language models  with a custom neural network layer. In the first step, they preprocess the data from the Draper VDISC dataset, labelling code snippets and applying padding and truncation. Then they use a dataloader to batch them for training. LLM is used as a feature extractor, turning code into meaningful embeddings, which are then refined by the added dense neural layer. The combined model predicts the vulnerability of CWE-120 as 'YES' or 'NO'. Furthermore, Sun et al. \cite{sun2024gptscan} use a hybrid model, GPTScan, to detect logic vulnerabilities in Solidity smart contracts by combining a GPT-based transformer with traditional static analysis. In the first step, they apply static reachability filtering to highlight candidate functions. Then they break down vulnerability types into scenarios and properties, which are matched using GPT prompts to assess semantic code patterns and identify key variables and statements. Finally, these GPT-informed findings are validated using static program analysis to eliminate false positives.


Mylläri et al. \cite{myllari2025ladle} proposed the Ladle technique that uses a sentence transformer, a pre-trained language model to embed overlapping short segments of log entries from multiple log types into a vector space. It calculates the anomaly score by comparing its embedding to a reference distribution for its log type. The system supports data drift adaptation by updating the reference collection with new log segments without retraining the model, allowing it to stay accurate as log behaviour evolves. Tested on a real-world dataset. Ladle showed high accuracy and outperformed traditional single-log anomaly detection approaches. Similarly, He et al.\cite{he2025vultr} introduce the VulTR model to detect software vulnerabilities by enhancing and refining key features across multiple layers of analysis. First, it extracts static features from source code functions, including code metrics and semantic representations. Then, a multi-layer enhancement module processes these features to better capture structural and contextual information. These enriched features are then fed into a deep learning classifier to predict whether a function is vulnerable or not. 

Furthermore, Wu et al. \cite{wu2025information} proposed a vulnerability detection methodology that employs a Transformer encoder-based model to detect software anomalies in system log sequences. Their designed model accepts log sequences of varying lengths, incorporating special tokens, each with temporal embeddings. These embeddings are combined with event representations and fed into Transformer encoder layers, with an attention mask ensuring the model ignores padded tokens. During supervised training, only the output token is used as a sequence-level representation and optimized with a binary classification objective using binary cross-entropy loss to distinguish between normal and anomalous sequences. Khan et al. \cite{khan2026leveraging} use a CodeBERT-based transformer model as the core architecture for vulnerability detection on code slices. Reza et al. \cite{reza2026empirical} explicitly evaluates four transformer-based models CodeBERT, CodeT5+, PLBART and UniXcoder. Among them, UniXcoder gives the best overall results in their JavaScript vulnerability detection experiments. Do et al. \cite{do2026novel} proposes a custom transformer/LLM-inspired ensemble architecture called RoS-Dex.

\begin{tcolorbox}[ title={RQ3 Research Finding}]
Transformer based models, especially CodeBERT and GraphCodeBERT variants, dominate 
recent software vulnerability detection research, with a clear trend toward hybrid, graph-aware, 
and task-specific architectures rather than plain transformer-only models.
\end{tcolorbox}

\begin{table}[t]
\centering
\footnotesize
\caption{Popular Transformer Model Categories Used in Articles}
\label{tab:transformer_backbone_frequency}

\begin{tabular}{@{} p{3cm} p{5cm} @{}}
\hline
\textbf{Transformer Backbone / Family} & \textbf{Articles} \\
\hline

\textbf{CodeBERT and CodeBERT-based variants} &
\cite{lu2023assessing,sun2024enhancing,peng2023ptlvd,zhao2024python,hin2022linevd,gupta2024dl,liang2024source,rusinova2024explaining,tanko2025approach,thapa2022transformer,
fu2022linevul,kaanan2024llm,curto2024multivd,le2024software,liu2024automatic,he2025vultr,tian2025efvd,katz2025siexvults,oladokun2025effective,shir2025robust,wang2025m2cvd,
ferretti2025detecting,perera2025codebert,kalouptsoglou2025transfer,tao2025transformer,smaili2025transformer,cao2025multi} \\
\hline

\textbf{BERT / RoBERTa / DistilBERT family} &
\cite{bahaa2024db,mahyari2024harnessing,zhao2025vulnerability,kim2022vuldebert,gujar2024detectbert,kim2024robust,liu2024pre,wu2021self,mamede2022transformer,hanif2022vulberta,
hou2022vulnerability,zahid2025detectbert,islam2023unbiased,alqarni2022low,ehrenberg2024python,wu2025information,mechri2025secureqwen,oladokun2025effective} \\
\hline

\textbf{GraphCodeBERT / Graph-aware pretrained code transformers} &
\cite{wu2021peculiar,thapa2022transformer,wang2024scl,katz2025siexvults,oladokun2025effective,sultan2025codevul,perera2025codebert,cui2025vulgtda,nguyen2023mando} \\
\hline

\textbf{GPT / LLM family} &
\cite{saimbhi2024vulnerai,purba2023software,kim2024robust,liu2023software,shiaeles2023vuldetect,sun2024gptscan,le2024software,yang2024large,wang2025m2cvd,ferretti2025detecting,mechri2025secureqwen,kalouptsoglou2025transfer} \\
\hline

\textbf{CodeT5 / PLBART / UniXcoder / other code LMs} &
\cite{lu2023assessing,thapa2022transformer,wang2025line,wang2025m2cvd,ferretti2025detecting,perera2025codebert,kalouptsoglou2025transfer,cao2025multi,sultan2025codevul,mechri2025secureqwen} \\
\hline

\textbf{Custom Transformer Architectures} &
\cite{li2022software,ferrag2023securefalcon,bui2023detecting,nguyen2023mando,zhang2023vuld,cao2024vulnerability,gong2023gratdet,jiang2024haformer,tian2025efvd,cui2025vulgtda,vanam2025software,sun2025hgtjit,wang2025line,shang2025cegt} \\
\hline


\hline
\end{tabular}
\end{table}

\begin{table*}[t]
\footnotesize
\caption{Evaluation metrics reported in the reviewed articles }
\label{tab:Evaluation} 
\begin{tabularx}{\textwidth}{@{} p{2.5cm} X @{} }
\toprule
 \textbf{Evaluation Metric} & \textbf{Articles}\\
 \midrule

Accuracy & \cite{saimbhi2024vulnerai,li2022software,chen2022hlt,bahaa2024db,ferrag2023securefalcon,lu2023assessing,mahyari2024harnessing,bui2023detecting,sun2024enhancing,zhao2025vulnerability,peng2023ptlvd,zhao2024python,liu2024pre,gupta2024dl,liang2024source,rusinova2024explaining,wu2021self,liu2023software,zhang2023vuld,jianjie2023code,zhang2024twlog,almakayeel2024deep,myllari2025ladle,he2025vultr,wang2024scl,cao2025multi,liu2024making,xuan2025large,ni2025abundant,do2024optimizing,ehrenberg2024python,mechri2025secureqwen,curto2024multivd,alqarni2022low,tanko2025approach,thapa2022transformer,cao2024vulnerability,kaanan2024llm,islam2023unbiased,jiang2024haformer,wu2025information,yang2024large,tian2025efvd,katz2025siexvults,cui2025vulgtda,vanam2025software,gunda2025transformer,li2025macd,oladokun2025effective,wang2025line,shir2025robust,wang2025m2cvd,ferretti2025detecting,perera2025codebert,sultan2025codevul,shang2025cegt,kalouptsoglou2025transfer,tao2025transformer,smaili2025transformer} \\
\hline

Precision & \cite{wang2024extensive,xuan2025large,wu2025information,cao2025multi,ehrenberg2024python,wang2024scl,mechri2025secureqwen,he2025vultr,liu2024automatic,gong2023gratdet,almakayeel2024deep,zhang2024twlog,le2024software,curto2024multivd,islam2023unbiased,jianjie2023code,kaanan2024llm,shiaeles2023vuldetect,hou2022vulnerability,mamede2022transformer,fu2022linevul,cao2024vulnerability,thapa2022transformer,tanko2025approach,liu2023software,wu2021self,wu2021peculiar,liang2024source,gupta2024dl,kim2024robust,hin2022linevd,zhao2024python,peng2023ptlvd,gujar2024detectbert,kim2022vuldebert,zhao2025vulnerability,bui2023detecting,lu2023assessing,ferrag2023securefalcon,purba2023software,bahaa2024db,saimbhi2024vulnerai,ni2025abundant,jeon2024design,tian2025efvd,katz2025siexvults,cui2025vulgtda,gunda2025transformer,li2025macd,sun2025hgtjit,oladokun2025effective,wang2025line,shir2025robust,wang2025m2cvd,ferretti2025detecting,perera2025codebert,sultan2025codevul,shang2025cegt,kalouptsoglou2025transfer,tao2025transformer,smaili2025transformer} \\
\hline

Recall & \cite{saimbhi2024vulnerai,bahaa2024db,ferrag2023securefalcon,lu2023assessing,bui2023detecting,sun2024enhancing,zhao2025vulnerability,kim2022vuldebert,gujar2024detectbert,peng2023ptlvd,zhao2024python,hin2022linevd,kim2024robust,gupta2024dl,liang2024source,wu2021peculiar,wu2021self,liu2023software,tanko2025approach,zhang2023vuld,thapa2022transformer,cao2024vulnerability,fu2022linevul,mamede2022transformer,hou2022vulnerability,zahid2025detectbert,shiaeles2023vuldetect,kaanan2024llm,jianjie2023code,islam2023unbiased,ehrenberg2024python,curto2024multivd,le2024software,zhang2024twlog,almakayeel2024deep,gong2023gratdet,he2025vultr,mechri2025secureqwen,wang2024scl,cao2025multi,wang2024extensive,xuan2025large,wu2025information,jiang2024haformer,ni2025abundant,jeon2024design,tian2025efvd,katz2025siexvults,cui2025vulgtda,gunda2025transformer,li2025macd,sun2025hgtjit,oladokun2025effective,wang2025line,shir2025robust,wang2025m2cvd,ferretti2025detecting,perera2025codebert,sultan2025codevul,shang2025cegt,kalouptsoglou2025transfer,tao2025transformer,smaili2025transformer} \\
\hline

F1-score & \cite{saimbhi2024vulnerai,li2022software,chen2022hlt,bahaa2024db,purba2023software,ferrag2023securefalcon,lu2023assessing,bui2023detecting,sun2024enhancing,zhao2025vulnerability,kim2022vuldebert,gujar2024detectbert,peng2023ptlvd,zhao2024python,hin2022linevd,kim2024robust,liu2024pre,nguyen2023mando,gupta2024dl,liang2024source,wu2021peculiar,wu2021self,liu2023software,tanko2025approach,zhang2023vuld,thapa2022transformer,cao2024vulnerability,fu2022linevul,mamede2022transformer,hanif2022vulberta,hou2022vulnerability,zahid2025detectbert,shiaeles2023vuldetect,jianjie2023code,kaanan2024llm,islam2023unbiased,curto2024multivd,le2024software,zhang2024twlog,wang2024scl,almakayeel2024deep,ehrenberg2024python,gong2023gratdet,liu2024automatic,he2025vultr,mechri2025secureqwen,cao2025multi,wang2024extensive,xuan2025large,wu2025information,ni2025abundant,jeon2024design,tian2025efvd,katz2025siexvults,cui2025vulgtda,gunda2025transformer,li2025macd,sun2025hgtjit,oladokun2025effective,wang2025line,shir2025robust,wang2025m2cvd,ferretti2025detecting,perera2025codebert,sultan2025codevul,shang2025cegt,kalouptsoglou2025transfer,tao2025transformer,smaili2025transformer} \\
\hline

MCC & \cite{gujar2024detectbert,hanif2022vulberta,zahid2025detectbert,almakayeel2024deep,ferrag2023securefalcon,sultan2025codevul,smaili2025transformer} \\
\hline

Specificity (TNR) & \cite{saimbhi2024vulnerai,zahid2025detectbert,wu2025information,hanif2022vulberta} \\
\hline

Sensitivity (TPR) & \cite{zahid2025detectbert,li2022software,purba2023software,tanko2025approach,hanif2022vulberta,sun2024gptscan,liu2024automatic,he2025vultr,saimbhi2024vulnerai,ni2025abundant} \\
\hline

FPR & \cite{li2022software,purba2023software,tanko2025approach,cao2024vulnerability,hanif2022vulberta,sun2024gptscan,liu2024automatic,he2025vultr,mamede2022transformer,saimbhi2024vulnerai,bahaa2024db,ni2025abundant,tao2025transformer} \\
\hline

AUC & \cite{bahaa2024db,shiaeles2023vuldetect,cao2025multi,yang2024large,jiang2024haformer,ni2025abundant,sun2025hgtjit,perera2025codebert,sultan2025codevul} \\
\hline

AUROC & \cite{gujar2024detectbert,hin2022linevd,hanif2022vulberta,zahid2025detectbert} \\
\hline

FNR & \cite{purba2023software,tanko2025approach,cao2024vulnerability,mamede2022transformer,hanif2022vulberta,sun2024gptscan,he2025vultr,saimbhi2024vulnerai,bahaa2024db,tao2025transformer} \\
\hline

Macro avg & \cite{ferrag2023securefalcon} \\
\hline

Weighted avg & \cite{ferrag2023securefalcon} \\
\hline

Weighted F1 & \cite{lu2023assessing} \\
\hline

Macro F1 & \cite{lu2023assessing} \\
\hline

MFR & \cite{peng2023ptlvd} \\
\hline

PRAUC & \cite{hin2022linevd} \\
\hline

ROC & \cite{kim2024robust,shiaeles2023vuldetect} \\
\hline

Top-k Accuracy & \cite{liu2023software,yang2024large} \\
\hline

MSE & \cite{gunda2025transformer,sultan2025codevul} \\
\hline

MAE & \cite{gunda2025transformer,sultan2025codevul} \\
\hline

Kappa & \cite{sultan2025codevul} \\
\hline

SP & \cite{sultan2025codevul} \\
\hline

SN & \cite{sultan2025codevul} \\
\hline

MK & \cite{sultan2025codevul} \\
\hline

F2 & \cite{kalouptsoglou2025transfer} \\
\hline

IOU & \cite{tao2025transformer} \\

\bottomrule
\end{tabularx}
\end{table*}
Table \ref{tab:transformer_backbone_frequency} groups the reviewed articles according to the main transformer backbone or family they use. It shows that CodeBERT and its variants are the most commonly used models in software vulnerability detection research, followed by the BERT/RoBERTa/DistilBERT family, GPT/LLM-based models, GraphCodeBERT and graph-aware transformers, other code language models such as CodeT5, PLBART, and UniXcoder, and custom transformer architectures. Overall, the table indicates that recent studies rely heavily on pretrained code-oriented transformer models, while also showing growing interest in graph-aware and LLM-based approaches.

\subsection{RQ4: What are the commonly used evaluation metrics?}

For improved transformer-based approaches, it is important to validate them using various machine learning evaluation metrics. To identify the evaluation metrics commonly used in transformer-based approaches, we examine the metrics adopted in software vulnerability detection research and analyse their frequency of use in measuring vulnerability detection performances. Table \ref{tab:Evaluation} summarizes the evaluation metrics employed in the analysed vulnerability detection studies. The Table \ref{tab:Evaluation} shows that  Accuracy, Precision, Recall, and F1-score are the most widely used measures for assessing model performance, reflecting their suitability for binary and multi-class vulnerability detection tasks. Several studies additionally report specificity (TNR), sensitivity (TPR), false positive rate (FPR) and false negative rate (FNR) to provide a more detailed analysis of detection errors, particularly in security-critical contexts where false alarms and missed vulnerabilities have significant consequences.

To address class imbalance and provide more robust performance evaluation, some research studies employ Matthews Correlation Coefficient (MCC), Area Under the Curve (AUC), and AUROC, which capture overall predictive quality beyond accuracy alone. While fewer studies in the proposed SLR report macro- and weighted averaged metrics, macro and weighted F1-scores and precision/recall-based measures such as PRAUC. Additionally, metrics such as Top-k Accuracy and Receiver Operating Characteristic (ROC) curves are used in a limited number of studies to evaluate ranking-based predictions and threshold independent performance. For Example, Kim et al. \cite{kim2024robust} prepared several labelled ethereum smart contract datasets, addressed class imbalance through over sampling and fine tuning transformer encoders using supervised learning to capture both syntactic and semantic code patterns. They evaluate the model's performance using standard classification metrics such as accuracy, precision, recall, F1-score, and ROC. They use the ROC curve to assess the model's discriminative ability across different decision thresholds. Specifically, ROC plots the true positive rate against the false positive rate, and the corresponding AUC value is used to demonstrate how effectively the fine tuned models distinguish vulnerable contracts from non vulnerable ones, particularly under imbalanced data conditions. On the other hand, Liu et al. \cite{liu2023software} investigates the use of GPT for software vulnerability detection without traditional model fine tuning. They use Top-K accuracy as an evaluation metric because GPT often produces a ranked list of possible vulnerability predictions rather than a single deterministic output. Top-K accuracy measures whether the correct vulnerability appears within the top K predicted candidates e.g Top-1, Top-3, Top-5, which is especially suitable for LLM based and multi class vulnerability detection tasks. This metric reflects realistic usage scenarios where security analysts can review several high confidence predictions and it better captures the practical usefulness of GPT when multiple plausible vulnerability types exist for a given code snippet.

\begin{tcolorbox}[ title={RQ4 Research Finding}]
Most researchers use conventional metrics such as Accuracy, Precision, Recall 
and F1-score to evaluate experimental results. Additionally, some studies employ
measures like FPR, FNR, MCC, AUC, Top-K Accuracy, and ROC to address class imbalance
and provide a more detailed performance analysis. These metrics can help software 
researchers and vendors to evaluate their vulnerability detection approaches for
more generalised results.
\end{tcolorbox}
\subsection{RQ5: What are the fine-grained software vulnerability types identified in the literature?}
In RQ-1, we classified the research papers into binary and fine-grained vulnerability types. However, considering the growing interest of software researchers and vendors in classifying software vulnerabilities into more fine-grained types, we examine how different studies detect various CWE vulnerability types, identify their corresponding CWE numbers and analyse the categories to which these vulnerabilities belong.

\begin{table*}[t]
\scriptsize
\caption{Total number of Vulnerability types detected in each article  }
\label{tab:Vulnerability} 
\begin{tabularx}{\textwidth}{@{} p{2cm} X @{}  }
\toprule
 \textbf{Article} &   \textbf{Vulnerability Types}\\
 \midrule
 Saimbhi and Akpinar \cite{saimbhi2024vulnerai} & Injection (CWE-79, CWE-89, CWE-77, CWE-78),Auth / Access Control(CWE-287, CWE-284, CWE-639),Cryptography(CWE-327, CWE-330, CWE-338, CWE-311),Input Validation	(CWE-22, CWE-434, CWE-502),Web Logic	(CWE-352, CWE-601),Information Disclosure	(CWE-200),Configuration (CWE-16)\\
\hline
 Bahaa et al.\cite{bahaa2024db}& Injection (CWE-78), Buffer/memory corruption (CWE-121, 122, 124, 126, 127, 59), Format string (CWE-134),Numeric/Integer errors (CWE-194, 195, 197), Null pointer / memory management (CWE-690)\\
\hline
 Purba et al.\cite{purba2023software} & Buffer Errors / Memory Corruption	(CWE-120), Injection / SQL Injection (CWE-89)\\\hline
 Ferrag et al.\cite{ferrag2025securefalcon}  & Buffer/memory corruption (CWE-119, 120, 121, 122, 787), Injection: (CWE-78), Numeric errors: (CWE-190), Input validation (CWE-20), Memory management / null dereference (CWE-476, 762)\\\hline
 Lu et al.\cite{lu2023assessing} & Buffer/memory corruption (CWE-119, 125, 787),Injection / XSS:(CWE-79),Input validation (CWE-20) \\\hline
 Sun et al.\cite{sun2024enhancing}  & Buffer/memory corruption (CWE-787) ,Injection (CWE-89 (SQL), 79 (XSS))CSRF (CWE-353),Path / File Traversal (CWE-22)\\\hline
 Kim et al.\cite{kim2022vuldebert}   & Buffer/memory corruption (CWE-119), Resource management / DoS (CWE-399)\\\hline
 Gujar \cite{gujar2024detectbert} &Path / File Traversal (CWE-22),OS / Command Injection (CWE-77),Injection (CWE-79 (XSS), 89 (SQL), 94 (Code Injection)),CSRF (CWE-352),Open Redirect (CWE-601)\\\hline
 Peng et al.\cite{peng2023ptlvd}   &Buffer/memory corruption (CWE-119, 125, 416, 476),Input validation (CWE- 20),Information disclosure (CWE-200), Access control / race conditions (CWE-264, 362),Integer / numeric errors (CWE-189, 190) etc\\\hline
 Zhao et al.\cite{zhao2024python}   & Injection (CWE-78 (OS), 79 (XSS), 89 (SQL), 94 (code injection))CSRF (CWE-352),File upload / RCE (CWE-434)Information disclosure / sensitive data (CWE-200, 319),Cryptography / weak algorithms (CWE-326, 327),Open redirect (CWE-601)\\\hline
 Kim et al.\cite{kim2024robust}  & Arithmetic errors (CWE-189, 190, 191),Logic errors (CWE-840) series,Input / unexpected events (CWE-20, 359), Resource / timing 
(CWE-362, 399)\\\hline
 Nguyen et al.\cite{nguyen2023mando}   & Access control (CWE-284),Integer / numeric (CWE-190),Concurrency / race condition (CWE-362),Input validation / error handling (358, 252),Logic/business errors (CWE-841),Resource / unsafe function usage (CWE-829, 682)\\\hline
 Wu et al.\cite{wu2021peculiar}  & application logic CWE 841\\\hline
 Tanko et al.\cite{tanko2025approach}  & Injection / command (CWE-78),Injection / SQL (CWE-89),Injection / XSS (CWE-79),Path / File Traversal (CWE-22),Cryptography / weak crypto (CWE-327)\\\hline
 Thapa et al.\cite{thapa2022transformer}    & Buffer/memory corruption (CWE-119),Resource management / DoS (CWE-399)\\\hline
 Cao \cite{cao2024vulnerability}  & buffer/memory corruption issue (CWE-120)\\\hline
 Mamedi \cite{mamede2022transformer} &
Resource Management / Denial of Service (CWE-400) others(not specified)\\\hline
 Hanif and Maffeis \cite{hanif2022vulberta}  &\ Integer / Numeric Errors (CWE-190,191)\\
 \hline
 Hou et al.\cite{hou2022vulnerability}   & Buffer / Memory Errors (CWE-120,121) 
\\
\hline
 Zahid \cite{zahid2025detectbert} & Path / File Traversal (CWE-22),Cross-Site Scripting (XSS) (CWE-79),Command / OS Injection (CWE-77), SQL Injection (CWE-89), Code Injection (CWE-94),Cross-Site Request Forgery (CSRF)( CWE-352),Open Redirect / URL Manipulation (CWE-601)\\\hline
kaanan et al.\cite{kaanan2024llm}  & Buffer / Memory Errors( CWE-120) others(not specified)\\\hline
Islam et al.\cite{islam2023unbiased} & Input / Validation Errors( CWE-20,74),Resource Management / Denial of Service (CWE-400,404),Path / File Manipulation / Directory Traversal (CWE-221),Cryptography / Sensitive Data Handling (CWE-311),Integer / Numeric Errors (CWE-190,187),Information Exposure / Security Flaw (CWE-138,467,469)\\\hline
 Sun et al.\cite{sun2024gptscan}  & Access Control / Privileges (CWE-285,284,346),Integer / Numeric Errors / Calculation Errors (CWE-682),Concurrency / Race Conditions(CWE-362),Logic / Business Errors (CWE-841)\\\hline
 Curto et al.\cite{curto2024multivd} & Access Control / Privileges(CWE-732,284, 264),Buffer / Memory Errors ( CWE-787,125,416, 476),Input / Validation Errors(CWE-254,20),Integer / Numeric Errors (CWE-190,189),Concurrency / Race Conditions(CWE-362),Resource Management / DoS (CWE-399),Information Disclosure / Sensitive Data (CWE-200,199)\\\hline
 Alqarni and Azim \cite{alqarni2022low}  & Buffer / Memory Errors ( CWE-120, 476, 805),Resource Management / Memory Leak (CWE-401),Input / Validation Errors (CWE-469)\\\hline
 Gong et al.\cite{gong2023gratdet}  & Logic / Business Errors (CWE-841)\\
\hline
 Liu et al.\cite{liu2024automatic}  & Buffer / Memory Errors (CWE-119), Resource Management / DoS (CWE-399)\\\hline
 Mechri et al.\cite{mechri2025secureqwen}  & Resource management / DoS (CWE-400, 703), Path / File Traversal (CWE-22), Command / OS Injection (CWE-78), Cryptography / Weak Algorithms (CWE-327, 330), SQL Injection (CWE-89), Authentication / Credential Management, CWE-259, Input / Validation Errors(CWE-20), Concurrency / Race Conditions (CWE-377), Certificate / Trust Issues (CWE-295) \\\hline
 Ehrenberg et al.\cite{ehrenberg2024python}  & Buffer / Memory Errors (CWE-787) ,Cross-Site Scripting (XSS)( CWE-79),SQL Injection (CWE-89),Input / Validation Errors (CWE-20),Path / Directory Traversal (CWE-24)\\\hline
 Wang et al.\cite{wang2024extensive}   & Injection (CWE- 78 (OS), 89 (SQL), 94 (code), Web vulnerabilities: 79 (XSS), 352 (CSRF), 601 (Open Redirect),File / resource access: 706\\
\hline
 Shiaeles et al.\cite{shiaeles2023vuldetect}  & Library / API Function Calls (CWE-119, 120, 665, 672, 676),Array Usage	(CWE-125, 126, 127, 787, 20),Pointer Usage (CWE-476, 457, 590, 761, 119/787),Arithmetic Expression (CWE-189, 190, 191, 680, 840/845)  \\
\hline
Katz et al.\cite{katz2025siexvults}  &  Information Disclosure / Sensitive Information Exposure (CWE-200, 201, 203, 204, 208, 209, 214, 215, 532, 535, 536, 537, 538, 550, 598, 615) \\
\hline
Shir et al.\cite{shir2025robust}  &  Buffer / Memory Errors (CWE-121, 122, 124, 126, 127), Integer / Numeric Errors (CWE-190, 191, 194, 195, 197, 680), Resource / Memory Management (CWE-401, 415, 590, 690, 762), Input / Path / Command Issues (CWE-23, 36, 78, 134), Logic / Runtime / Validation Errors (CWE-369, 400, 457) \\
\hline
Ferretti et al.\cite{ferretti2025detecting}  &  Smart Contract Vulnerability Classes: Access Control, Arithmetic, Reentrancy, Unchecked Calls, Other \\

\hline
Sultan et al.\cite{sultan2025codevul}  &  Buffer / Memory Errors (CWE-119, CWE-120), Pointer / Reference Errors (CWE-476), Pointer Subtraction / Addressing Issues (CWE-469), Other / Miscellaneous Vulnerabilities (CWE-other) \\\hline
Shang et al.\cite{shang2025cegt}  &  Reentrancy (CWE-841), Timestamp Dependence (CWE-829), Integer Overflow (CWE-190) \\
\hline
Tao et al.\cite{tao2025transformer}  & Injection Vulnerabilities (CWE-78, 88, 89), Buffer / Memory Errors (CWE-124, 127, 129, 416, 476, 805), Integer / Numeric Errors (CWE-190, 191, 195), Resource Management / DoS (CWE-400, 789, 835), Race / Concurrency / TOCTOU (CWE-363, 367), Resource Lifetime / Handle Issues (CWE-773, 775), Other Memory / Runtime Weaknesses (CWE-401) \\
\hline
Jeon et al.\cite{jeon2024design}  &   Gas exhaustion (CWE-400), Unchecked function call (CWE-252),  balance Equality check point (CWE-697), Incorrect return type (CWE-628 / CWE-704), , Misuse of visibility ( CWE-284), Array length manipulation (CWE-400 / CWE-770), Use of insecure math functions (CWE-190 / CWE-191), Locked ether ( CWE-667 / CWE-703), Data leakage when using private (CWE-200), Misuse of approve function in ERC20 library ( CWE-362), Misuse of var (vCWE-190 / CWE-681), Misuse of multiple return values in internal/private functions ( CWE-393 / CWE-703), Misuse of transfer function in loop ( CWE-400), Misuse of inline assembly (CWE-710), Hardcode of the address ( CWE-798), Deprecated constructions ( CWE-477), False return of ERC20 (CWE-252), Misuse of revert require ( CWE-670) \\

\bottomrule
\end{tabularx}
\end{table*}
Table \ref{tab:Vulnerability} summarizes the diversity of vulnerability types identified in the surveyed articles. For each study, Table \ref{tab:Vulnerability} lists the distinct vulnerability categories identified and maps them to their corresponding CWE \cite{CWESite} identifiers, to the best of our knowledge. For example, Shiaeles et al.  \cite{shiaeles2023vuldetect} present a framework to identify 124+ vulnerabilities. They leverage pre-trained language models to detect multiple CWE vulnerability types in source code. In the first step, they tokenize and normalize code snippets to create textual representations suitable for input to the language model, which generates contextual embeddings capturing both syntax and semantic patterns relevant to vulnerabilities. Model VulDetect treats vulnerability detection as a multi classification task, where each class corresponds to a specific CWE identifier. A classification layer on top of the language model outputs probabilities over these CWE classes, allowing the model to predict one or more vulnerability types for each code snippet. During training, the model is fine tuned on labelled datasets using suitable loss functions for multi class classification, enabling it to learn CWE specific code patterns. By combining contextual embeddings with CWE based classification, VulDetect can identify a wide range of vulnerability types, including subtle or previously unseen weaknesses, in a single unified framework. Similarly, Ni et al. \cite{ni2025abundant} propose a function-level vulnerability detection framework that leverages multiple representations of source code to improve detection accuracy. It integrates three complementary modalities: textual features extracted using a pre-trained code language model UniXcoder to capture semantic patterns, graph-based features from control flow and data flow structures encoded via a Graph Neural Network to capture structural dependencies and optional image or matrix-based representations to learn additional structural patterns. The embeddings from these modalities are fused through a multi-modal network to create a unified function-level representation, which is then fed into a classification layer to predict CWE-based vulnerability types.  Table \ref{tab:GroupedVulnerability} indicates that the literature addresses a wide range of software security issues, with the most common categories including injection vulnerabilities, buffer or memory errors, input and validation problems, access control issues, cryptography-related weaknesses, path or file traversal, numeric and arithmetic errors, resource management problems, race conditions, and information disclosure. The table \ref{tab:GroupedVulnerability} also highlights a separate group of smart-contract-specific vulnerabilities, showing that blockchain and Solidity security form an important part of recent research. Overall, the table provides a high-level view of how vulnerability research is distributed across major weakness categories and which types are most frequently studied.

\begin{tcolorbox}[ title={RQ5 Research Finding}]
Software researchers have shown interest in detecting a wide range of fine-grained vulnerabilities in code, which are mostly
categorized into nine groups based on our knowledge and understanding. Among these, Injection (SQL, OS, XSS, Code), and Buffer/Memory Errors/Corruption are the most dominant. 
\end{tcolorbox}

\twocolumn
\begin{table}[h]
\footnotesize
\caption{Major types of vulnerabilities covered in the reviewed articles}
\label{tab:GroupedVulnerability} 
\begin{tabular}{@{} p{4cm} p{4cm} @{}}
\toprule
 \textbf{Vulnerability Description} & \textbf{Articles}\\
 \midrule
\textbf{Injection (SQL, OS, XSS, Code, Command)} &
\cite{saimbhi2024vulnerai,bahaa2024db,purba2023software,ferrag2025securefalcon,lu2023assessing,sun2024enhancing,gujar2024detectbert,zhao2024python,tanko2025approach,zahid2025detectbert,wang2024extensive,fu2022linevul,peng2023ptlvd,mechri2025secureqwen,tao2025transformer} \\
\hline

\textbf{Buffer / Memory Errors / Corruption} &
\cite{bahaa2024db,purba2023software,ferrag2025securefalcon,lu2023assessing,sun2024enhancing,kim2022vuldebert,thapa2022transformer,cao2024vulnerability,fu2022linevul,alqarni2022low,liu2024automatic,shiaeles2023vuldetect,kaanan2024llm,peng2023ptlvd,curto2024multivd,shir2025robust,sultan2025codevul,tao2025transformer} \\
\hline

\textbf{Input / Validation Errors} &
\cite{saimbhi2024vulnerai,ferrag2025securefalcon,lu2023assessing,peng2023ptlvd,nguyen2023mando,fu2022linevul,ehrenberg2024python,islam2023unbiased,curto2024multivd,alqarni2022low,mechri2025secureqwen} \\
\hline

\textbf{Access Control / Privileges / Authentication} &
\cite{saimbhi2024vulnerai,nguyen2023mando,sun2024gptscan,curto2024multivd,ferretti2025detecting,perera2025codebert} \\
\hline

\textbf{Cryptography / Weak Algorithms / Sensitive Data} &
\cite{saimbhi2024vulnerai,zhao2024python,tanko2025approach,islam2023unbiased,mechri2025secureqwen,katz2025siexvults} \\
\hline

\textbf{Path / File / Directory Traversal} &
\cite{saimbhi2024vulnerai,sun2024enhancing,gujar2024detectbert,tanko2025approach,fu2022linevul,zahid2025detectbert,mechri2025secureqwen,ehrenberg2024python,islam2023unbiased,wang2024extensive,shir2025robust} \\
\hline

\textbf{Cross-Site Request Forgery (CSRF)} &
\cite{saimbhi2024vulnerai,sun2024enhancing,gujar2024detectbert,zhao2024python,zahid2025detectbert,wang2024extensive} \\
\hline

\textbf{Numeric / Arithmetic / Integer Errors} &
\cite{bahaa2024db,ferrag2025securefalcon,kim2024robust,nguyen2023mando,hanif2022vulberta,curto2024multivd,islam2023unbiased,shiaeles2023vuldetect,peng2023ptlvd,shir2025robust,shang2025cegt,tao2025transformer,jeon2024design} \\
\hline

\textbf{Resource Management / Denial of Service} &
\cite{kim2022vuldebert,thapa2022transformer,mamede2022transformer,liu2024automatic,curto2024multivd,islam2023unbiased,mechri2025secureqwen,alqarni2022low,kim2024robust,shir2025robust,tao2025transformer,jeon2024design} \\
\hline

\textbf{Race Conditions / Concurrency / TOCTOU} &
\cite{peng2023ptlvd,kim2024robust,nguyen2023mando,sun2024gptscan,curto2024multivd,mechri2025secureqwen,tao2025transformer,jeon2024design} \\
\hline

\textbf{Information Disclosure / Exposure} &
\cite{saimbhi2024vulnerai,peng2023ptlvd,zhao2024python,islam2023unbiased,curto2024multivd,katz2025siexvults,jeon2024design} \\
\hline

\textbf{Smart Contract Specific Vulnerabilities} &
\cite{ferretti2025detecting,perera2025codebert,shang2025cegt,jeon2024design} \\
\hline

\textbf{Logic / Business / API / Miscellaneous} &
\cite{wu2021peculiar,gong2023gratdet,shiaeles2023vuldetect,mechri2025secureqwen,kim2024robust,nguyen2023mando,sun2024gptscan,jeon2024design} \\
\bottomrule
\end{tabular}
\end{table}

\subsection{RQ6: What Hyper-parameters and environment settings are typically used?}
To optimise existing transformer-based vulnerability approaches for improved software vulnerability detection and classification, software researchers employ different hyperparameters to fine-tune transformer models. Although hyperparameters are algorithm and dataset-dependent. However, in this section, we critically review the hyperparameter configurations employed in the existing transformer-based studies and discuss the corresponding experimental environment settings. These frequently hyperparameter settings can be employed to evaluate the performance of newly developed transformer-based approaches to identify and classify software vulnerabilities, as it is considered one of the important steps in improving their performance. Furthermore, it can help software researchers and vendors to identify frequently used experimental setups required for source-intensive vulnerability detection.
\begin{table*}[t]
\scriptsize
\centering
\caption{Hyperparameters, fine-tuning, pre-training, and platform/environment setup reported in the reviewed articles(Part 1)}
\label{tab:HyperparametersPlatform}

\begin{tabularx}{\textwidth}{@{} p{2cm} p{8cm} X @{}}
\hline
 \textbf{Reference} & \textbf{Hyperparameters / Fine-tuning / Pre-training} & \textbf{Platform / Environment Setup} \\
\hline
Saimbhi and Akpinar \cite{saimbhi2024vulnerai} &
N/A &
Lenovo IdeaPad with AMD Ryzen 5 3500U processor, 8 GB RAM, 256 GB SSD; Python 3.10.4, Pandas library, OpenAI tools \\
\hline

Li et al. \cite{li2022software} &
Adam optimizer, learning rate 0.0001, 5 epochs, batch size 32 &
NVIDIA GeForce RTX 8 GB GPU with 16 GB RAM \\
\hline

Chen and Liu \cite{chen2022hlt} &
RAdam optimizer, learning rate 0.001, 6 encoder layers, 6 multi-head self-attention heads &
Server with Tesla P100 SXM2 16 GB; PyTorch 1.8.0; 8 NVIDIA P100 GPUs \\
\hline

Bahaa et al. \cite{bahaa2024db} &
Adam optimizer, learning rate 3e-5, binary cross-entropy loss, 20 epochs, batch size 20 &
NVIDIA A100 GPU with 34 GB RAM \\
\hline

Purba et al. \cite{purba2023software} &
28 layers, 16 attention heads, batch size 8, 4 epochs &
N/A \\
\hline

Ferrag et al. \cite{ferrag2023securefalcon} &
AdamW optimizer; learning rate 1.85e-4 (warm-up); Z-loss 1e-4; batch size 256; later effective batch size 1152; 100 billion tokens &
32 NVIDIA A100 40 GB GPUs \\
\hline

Lu et al. \cite{lu2023assessing} &
AdamW optimizer, learning rate 5e-5 &
OpenPrompt server with NVIDIA GeForce RTX 4090 \\
\hline

Mahyari \cite{mahyari2024harnessing} &
30 epochs, learning rate 0.01, batch size 64 &
N/A \\
\hline

Sun et al. \cite{sun2024enhancing} &
AdamW optimizer, batch size 16, input length 512, learning rate 2e-5, 12 attention heads &
N/A \\
\hline

Zhao and Liu \cite{zhao2025vulnerability} &
4 encoder layers, dropout 0.2, 4 attention heads, learning rate 1e-3, 50 epochs, batch size 16 &
PyTorch on a server equipped with NVIDIA RTX 3090 GPU \\
\hline

Kim et al. \cite{kim2022vuldebert} &
24 transformer layers, hidden size 1024, 16 self-attention heads &
PyTorch, NVIDIA GeForce Titan \\
\hline

Hanif and Maffeis \cite{hanif2022vulberta} &
Hidden layer with 768 neurons; output layer with 2 or 41 neurons; 10 epochs; learning rate 3e-5 &
PyTorch 1.7 with CUDA 10.2 on Python 3.7. Pretraining: GCP VMs with 48 vCPUs, 240 GB RAM, 2 NVIDIA Tesla A100 40 GB GPUs. Fine-tuning: 48-core Intel Xeon Silver CPU, 292 GB RAM, 2 NVIDIA GTX TITAN Xp GPUs (12 GB each) \\
\hline

Hou et al. \cite{hou2022vulnerability} &
Learning rate \{0.01, 0.001, 0.02\}, batch size \{64, 128, 256\}, embedding size \{64, 128, 256\}, number of heads \{2, 4, 8\}, transformer blocks \{1, 2, 4\} &
ANTLR4 for AST; NVIDIA GeForce GTX 1080 \\
\hline

Zahid \cite{zahid2025detectbert} &
24 attention heads, hidden size 768 &
N/A \\
\hline

Shiaeles et al. \cite{shiaeles2023vuldetect} &
N/A &
ASUS TUF Gaming laptop with Intel Core i7 8th-generation CPU; 6 cores; max frequency 2.2 GHz \\
\hline

Kaanan \cite{kaanan2024llm} &
Maximum input length 512 tokens &
N/A \\
\hline

Jianjie and Le \cite{jianjie2023code} &
Batch size 512, later modified to \(1 \times 1024\) &
N/A \\
\hline

Islam et al. \cite{islam2023unbiased} &
Word embedding size 768, hidden layer 128, learning rate 5e-4, 100 epochs, batch size 512, token length 400 &
8 DGX-A100 NVIDIA GPUs; training and testing took 4--6 hours \\
\hline

Sun et al. \cite{sun2024gptscan} &
4k context token size &
N/A \\
\hline

Curto et al. \cite{curto2024multivd} &
Single linear layer with 15 output neurons, 10 epochs, learning rate 2e-5 &
NVIDIA RTX A6000 GPU \\
\hline

Jeon et al. \cite{jeon2024design} &
Learning rate search \{1e-6, 1e-5, 1e-4, 1e-3\}; batch sizes \{16, 24, 32\}; epochs \{2, 3, 4\}; final setup: learning rate 1e-5, batch size 24, 3 epochs, random seed 2018, Adam optimizer, cross-entropy loss, BERT-base with 12 transformer blocks, hidden size 768, 12 attention heads &
Intel Xeon Silver 4114 2.20 GHz CPU, 256 GB RAM, NVIDIA GeForce Titan RTX, PyTorch \\
\hline

Alqarni and Azim \cite{alqarni2022low} &
Learning rate 2e-5, batch size 64, 10 epochs &
Intel Core i9 processor, Tesla K80 GPU, 12 GB RAM \\
\hline

Le et al. \cite{le2024software} &
10 epochs, learning rate 1e-5, feature embedding size 768 &
N/A \\
\hline
Li\cite{li2025macd} & Batch 16, epochs 10, lr 2e-5, AdamW, cross-entropy & RTX 3070 8GB, Ubuntu 20.04.6, PyTorch \\
\hline
Sun et al.\cite{sun2025hgtjit} & 4 HGT layers, hidden 768, 8 heads, dropout 0.2, batch 64, lr 1e-4, Adam, max 30 epochs, early stop & Ubuntu 18.04, Tesla T4 16GB, PyTorch, DGL \\
\hline
Oladokun \& Rice \cite{oladokun2025effective}& 4 epochs; pretrained on 6 PLs; other hyperparameters not stated & Google Colab, NVIDIA A100, HF Transformers, PyTorch \\
\hline
Wang et al.\cite{wang2025line} & Batch 12, lr 2e-5, seed 123456, 8 Transformer layers, 8 heads, p = 20, k = 100 & Not explicitly stated \\
\hline
Shir et al.\cite{shir2025robust} & Batch 16, lr 1.1e-5, weight decay 3e-4, 30 epochs, early stop, CodeBERT-base, 512-token model limit, 800-token function threshold & Linux x86-64, ARM/x86-64, GCC/Clang, GPU-enabled \\
\hline
Wang et al.\cite{wang2025m2cvd} & 4 epochs, max length 1024, batch 12, lr 2e-5, grad clip 1.0, seed 42, temp 0 & 3$\times$ V100 32GB GPUs \\
\hline
Ferretti et al.\cite{ferretti2025detecting} & Adam, lr 1e-5, batch 8, dropout 0.3, BCEWithLogitsLoss, GridSearchCV & 2-core AMD VM, RTX 2080 Ti 12GB \\
\hline
Perera et al.\cite{perera2025codebert} & Best: 3 layers [256,128,128], 4000 epochs, Sigmoid, Adam & Google Colab, Tesla K80, 12.7GB RAM \\
\hline
Sultan et al.\cite{sultan2025codevul} & Emb dim 768, hidden 512, out 768, 2 GCN layers, MSE alignment & Not explicitly stated \\
\hline
Shang et al.\cite{shang2025cegt}& Adam, cross-entropy, 50 epochs, 5-fold CV & Xeon 3204, RTX 4070, 128 GB RAM \\
\hline
Kalouptsoglou et al.\cite{kalouptsoglou2025transfer} & Fine-tune LR 2e-5 AdamW max len 512; best Big-Vul classifier BiGRU & RTX 4080 Super GPU \\
\hline
Tao et al.\cite{tao2025transformer} & Emb dim 768, dropout 0.1, batch 20, LR 5e-5, Adam & Threadripper 3960X, RTX 4090D, 32GB RAM \\
\hline
Smaili et al.\cite{smaili2025transformer} & LR 1e-3, batch 64, max 50 epochs, early stopping 6, L2 1e-4, Q/K/V 128, hidden 256 & RTX 4070, 32 GB RAM, Ryzen 5700X, PyTorch \\
\hline

Rusinova et al. \cite{rusinova2024explaining} &
12 encoder blocks, hidden size 768, feed-forward/filter size 3072, model sizes 14.5M and 125M parameters &
N/A \\
\hline

Wu et al. \cite{wu2021self} &
Learning rate 0.0015, token dictionary size 20000, 6 encoder blocks &
Ubuntu Linux 18.04; NVIDIA GeForce RTX 2080Ti GPU; Intel Xeon Silver 4214 CPU @ 2.20 GHz \\
\hline

Liu et al. \cite{liu2023software} &
N/A &
GPT-3.5-turbo; 11th Gen Intel Core i7-11800H @ 2.30 GHz; 32 GB RAM; Windows 11 \\

\hline
Zhang et al. \cite{zhang2024twlog} &
N/A &
Linux server; 20-core CPU, 128 GB RAM, NVIDIA Tesla P4 GPU; Python 3.9.18 and PyTorch 1.9.0 \\
\hline

Almakayeel \cite{almakayeel2024deep} &
Learning rate 0.01, ReLU activation, 50 epochs, dropout 0.5, batch size 5 &
Python 3.6.5 on PC with Intel i5-8600K, 250 GB SSD, GeForce 1050Ti 4 GB, 16 GB RAM, 1 TB HDD \\
\hline

Myllari et al. \cite{myllari2025ladle} &
Context length 256; sliding window length \(w=4\) &
Python 3, PyTorch \\
\hline

Gong et al. \cite{gong2023gratdet} &
6 layers, embedding dimension 512, hidden dimension 512, learning rate 0.0001, Adam optimizer, batch size 16 &
Ubuntu 18.04; TITAN RTX GPU; 32 GB RAM; Intel Xeon Gold 5120 CPU @ 2.20 GHz; PyTorch \\
\hline
Liu et al. \cite{liu2024automatic} &
N/A &
GNU GCC 4.8.2 compiler supporting Intel x86, x64, and ARM; Windows GCC (MinGW) and Linux GCC at optimization levels -O0, -O1, -O2, -O3; Windows 10 Enterprise with Intel Xeon W-2133 CPU @ 3.60 GHz and 32 GB memory; Ubuntu 16.04.3 with 16 vCPUs and 32 GB memory \\

\end{tabularx}
\end{table*}
\addtocounter{table}{-1} 
\begin{table*}[t]
\scriptsize
\centering
\caption{Hyperparameters, fine-tuning, pre-training, and platform/environment setup reported in the reviewed articles (Part 2)}
\label{tab:HyperparametersPlateform}
\begin{tabularx}{\textwidth}{@{} p{2cm} p{6.5cm} X @{}}

\hline

He et al. \cite{he2025vultr} &
100 epochs, cross-entropy loss, ReLU activation, AdamW optimizer, batch size 32, learning rate 0.001, dropout 0.5 &
128 GB RAM, 20-core CPU, NVIDIA GeForce RTX 4090 \\
\hline

Mechri et al. \cite{mechri2025secureqwen} &
Pretrained model CodeQwen1.5-7B; BOS token ID 2; EOS token ID 2; hidden activation SiLU; hidden size 4096; intermediate size 13,440; max position embeddings 65,536; model type Qwen2; 32 decoder layers; 4 key-value heads; RMSNorm epsilon 1e-5; RoPE theta 1,000,000; dtype bfloat16; cache enabled; vocab size 92,416; learning rate 1e-4; 15 labels; train batch size 4; eval batch size 8; seed 42; total train batch size 32; total eval batch size 64; Adam optimizer with betas (0.9, 0.999), epsilon 1e-8; 1 epoch &
Multi-GPU distributed setup; Transformers 4.41.1; PyTorch 2.1.0; 16\(\times\) A100 40 GB GPUs \\
\hline

Wang et al. \cite{wang2024scl} &
N/A &
Ubuntu 18.04; Intel Xeon Gold 6130 CPU @ 2.10 GHz; 80 GB RAM; NVIDIA Tesla V100 32 GB GPU; Python 3.8; PyTorch 1.7.0; CUDA 11.0 \\
\hline

Ehrenberg et al. \cite{ehrenberg2024python} &
Learning rate 2e-5, 5 epochs, batch size 8, AdamW optimizer; 2--3 h per training session &
N/A \\
\hline

Cao and Dong \cite{cao2025multi} &
Explored all possible hyperparameter combinations (not specified) &
Two Intel Xeon Gold 6230R CPUs; NVIDIA GeForce RTX 3090 GPU; Python 3.8; PyTorch 1.9.0 \\
\hline

Liu et al. \cite{liu2024making} &
Embedded 768-dimensional vector; FCDS as 1024-dimensional vector; 3 dense layers; dropout 0.5; batch size 8; ReLU; 20 epochs &
N/A \\
\hline

Jiang et al. \cite{jiang2024haformer} &
Embedding dimension 768, input length 512, 12 attention heads, 12 hidden layers, contrastive learning loss 0.05, 10 epochs &
Hugging Face Transformers, PyTorch 2.0; Ubuntu 20.04 server with 2 Intel Xeon 4216 CPUs, 256 GB memory, 8 RTX 3090 GPUs \\
\hline

Wu et al. \cite{wu2025information} &
12 attention heads, learning rate 5e-4, transformer block size 2048 &
Python 3.11, PyTorch 2.2.1, Intel Gold 6148 Skylake @ 2.4 GHz CPUs, NVIDIA V100 GPUs \\
\hline

Yang et al. \cite{yang2024large} &
Max learning rate 5e-6, min learning rate 1e-8, model dimension 256, 8 layers, batch size 64, 2000 epochs &
Intel Xeon 6248R CPU @ 3.00 GHz running Debian GNU/Linux and a single NVIDIA Quadro RTX 8000 GPU; largest model LLMAO with CodeGen-16B \\
\hline

Xuan et al. \cite{xuan2025large} &
Encoder: max length 512, encoding dimension 768, output dimension 512, MLP layers 4; KA: 12 layers, feedforward dimension 768, output dimension 512; training epochs 9, gradient accumulation steps 12, learning rate 3e-5 &
NVIDIA Tesla T4 GPU with 2,560 CUDA cores and 16 GB GDDR6 memory \\
\hline

Wang et al. \cite{wang2024extensive} &
12 layers, 12 self-attention heads, attention head size 64, hidden size 768, intermediate size 3072, 125M parameters &
Linux 5.11.0 on AMD EPYC 7502P 32-Core Processor @ 3.31 GHz with 128 GB RAM and Tesla T4 16 GB GPU \\
\hline

Ni et al. \cite{ni2025abundant} &
Tested hyperparameter values 0, 0.01, 0.02, 0.05, 0.1, 0.2, 0.5, 1, 2, and 5 &
32-core workstation with Intel Xeon Platinum 8358P CPU @ 2.60 GHz, 768 GB RAM, 4\(\times\) NVIDIA GeForce RTX 2080 GPUs, Ubuntu 20.04.6 LTS; Python libraries: Transformers and PyTorch \\
\hline

Do et al. \cite{do2024optimizing} &
Token length set to 32 or 64 &
N/A \\
\hline
Gujar \cite{gujar2024detectbert} &
AdamW optimizer, learning rate 1e-5, 100 epochs &
N/A \\
\hline

Peng et al. \cite{peng2023ptlvd} &
Learning rate 2e-5, AdamW optimizer, cross-entropy loss, batch size 16, 10 epochs, input size 512, hidden dimension 768, output \([0,1]\), 12 attention heads, dropout 0.1 &
Ubuntu 20.04 server with 251 GB memory, 32-core CPU, two Tesla V100 32 GB GPUs; Joern, Python, PyTorch, Transformers, Captum, Scikit-learn, Pandas \\
\hline

Hin et al. \cite{hin2022linevd} &
N/A &
Computing cluster with multiple NVIDIA Tesla V100 GPUs and Xeon E5-2698v3 CPUs @ 2.30 GHz \\
\hline

Zhao et al. \cite{zhao2024python} &
Epochs 1--10, hidden size 768, 12 hidden layers, 12 attention heads, batch size 16 &
N/A \\
\hline

Kim et al. \cite{kim2024robust} &
Input feature 1, padding 256, batch size 8, AdamW optimizer, 10 epochs, learning rate 1e-5 &
NVIDIA Tesla V100 DGX, 32 GB memory; NVIDIA-SMI 450.142.00; CUDA 11.7 \\
\hline

Liu et al. \cite{liu2024pre} &
10 epochs, batch size 128, learning rate 1e-4 &
Pre-trained on 4 NVIDIA RTX 3090 GPUs \\
\hline

Nguyen et al. \cite{nguyen2023mando} &
Embedding size 128; learning rate 0.0005--0.01 for coarse-grained classification and 0.0002--0.005 for fine-grained classification; 8 attention heads; hidden size 128; 100 and 50 epochs &
N/A \\
\hline

Gupta et al. \cite{gupta2024dl} &
Adam optimizer, batch size 32, 50 epochs, learning rate 0.001, input shape (code) (2048,1), input shape (text) (2048,1), dense layer units 64, dropout 0.3 &
Google Colab; Intel Xeon CPU; Tesla T4 GPU with 16 GB memory; 51 GB RAM \\
\hline

Liang et al. \cite{liang2024source} &
Adam optimizer; learning rates 1e-4, 1e-3, 1e-2; hidden channels 64--512; number of heads 1--8 &
Linux server with 128 GB memory, 16-core Intel Xeon processor, NVIDIA RTX A4000 GPU with 16 GB VRAM \\
\hline

Wu et al. \cite{wu2021peculiar} &
N/A &
Ubuntu 18.04 system, 64 GB RAM, Intel i7-9700 CPU, NVIDIA 1080Ti GPU \\
\hline
Tanko \cite{tanko2025approach} &
N/A &
Intel Core i5 CPU, V100 GPU, 32 GB memory, 2 TB disk, Windows 11, Python 3.9.18, Pandas 2.1.3, PyTorch 2.1.0, Scikit-learn 1.3.0, Joern 2.0.201 \\
\hline

Zhang et al. \cite{zhang2023vuld} &
Learning rate 0.001, batch size 64, maximum 50 epochs, 1 transformer encoder layer, word length 512 &
Windows 10, 32 GB RAM, GeForce RTX 3060 GPU, PyTorch \\
\hline

Thapa et al. \cite{thapa2022transformer} &
110M parameters, 12 layers, hidden size 768, 12 attention heads &
GPU internal RAM e.g.\ 16 GB; GPT-2 with 1.5B parameters also discussed \\
\hline

\hline
Cao \cite{cao2024vulnerability} &
N/A &
PyTorch 1.9.0, NVIDIA Quadro RTX 6000 GPU, Intel Xeon Gold 6126 CPU \\
\hline

Fu and Tantithamthavorn \cite{fu2022linevul} &
12 Transformer encoder blocks, hidden size 768, 12 attention heads &
NVIDIA RTX 3090 GPU \\
\hline

Mamede \cite{mamede2022transformer} &
BERT maximum length 512 tokens &
VDet for Java, a VS Code extension \\
\hline
Tian \& Zhang\cite{tian2025efvd} & Pre-trained CodeBERT; 256-dim CodeBERT output; 128-dim Word2Vec; 10 attention heads; 256 hidden dim; lr 0.0001; early stopping & RTX 3090, CUDA 12.7, PyTorch 1.8.1 \\
\hline
Katz et al.\cite{katz2025siexvults} & LR \{1e-5, 1e-4, 1e-3, 1e-2\}, dropout \{0.2, 0.3\}, activations \{ReLU, ELU, Sigmoid\}, batch \{32, 64\}, epochs \{50, 60\}, 50-trial random search, early stopping, 5-fold CV, pretrained transformer models & N/A \\

\end{tabularx}
\end{table*}
\addtocounter{table}{-1} 
\begin{table*}[t]
\scriptsize
\centering
\caption{Hyperparameters, fine-tuning, pre-training, and platform/environment setup reported in the reviewed articles (Part 3)}
\label{tab:HyperparametersPlateform}
\begin{tabularx}{\textwidth}{@{} p{2cm} p{7cm} X @{}}
\hline
Cui et al.\cite{cui2025vulgtda} & Batch 128, lr 0.001, dropout 0.2, BCE + CosineSimilarity + BCE loss & Linux for Joern; hardware not stated \\
\hline
Vanam et al.\cite{vanam2025software} & SFOA tuning, cross-entropy loss, 100-epoch training curves; numeric hyperparameters not given & N/A \\
\hline
Gunda et al.\cite{gunda2025transformer} & Pretrained CodeBERT, BCE + MSE, Adam, threshold 0.4, Gaussian noise, 10 epochs & N/A \\

\end{tabularx}
\end{table*}

Table \ref{tab:HyperparametersPlatform} presents the hyperparameter and experimental environment setups used in the surveyed articles. For each study, we list the key training hyperparameters along with the hardware and software environments used in the experiments.
Hyperparameters are critically important in ML and DL because they directly influence how a model learns from data, how well it generalizes and how efficiently it trains. For Example, Mechri et al.  \cite{mechri2025secureqwen} employ hyperparameters extensively to optimize LLM for identifying CWE based vulnerabilities. They use pre trained CodeQwen models and configures model specific hyperparameter such as 32 decoder layers, 4 attention heads, a hidden size of 4096, an intermediate MLP size of 13,440, and maximum position embeddings of 65,536 to define the model’s capacity. Training hyperparameters include a learning rate of 0.0001, the Adam optimizer with standard betas, batch sizes of 4 for training and 8 for evaluation and one epoch. The model predicts 15 CWE vulnerability classes using a classification head, and distributed multi GPU training with 16 NVIDIA A100 40GB GPUs is employed to handle memory and computational demands. These hyperparameters, spanning model architecture, training, and distributed execution, are crucial for ensuring high accuracy, F1-score, and reliable multi class vulnerability detection in Python codebases. Similarly, Chen and Liu \cite{chen2022hlt}  divide code into statements, functions, files, and then encodes each level using transformer encoders to capture both local and global contextual information. These embeddings are aggregated through a hierarchical attention mechanism, allowing the model to combine fine grained and coarse grained features for vulnerability prediction. For better performance, stable convergence and high detection accuracy across different code structures; they tuned hyperparameters such as learning rate, batch size, number of transformer layers, hidden size, attention heads, and training epochs. 
\begin{table*}[t]
\footnotesize
\caption{Most important transformer hyperparameters and their observed ranges in the reviewed articles }
\label{tab:Hyperparameterall} 
\begin{tabularx}{\textwidth}{@{} p{3cm} p{7cm} p{2cm} X @{}}
\toprule
 \textbf{Hyperparameter} & \textbf{Description} & \textbf{Common Value} & \textbf{Observed Range} \\
 \midrule
d model / Hidden Size & Dimensionality of hidden representations or embeddings used in transformer or related encoder modules. & 768 & 128 -- 4096 \\
\hline

Epochs & Number of complete passes over the training set. & 10 & 1 -- 2000 \\
\hline

Batch Size & Number of samples processed in one training step. & 16--32 & 4 -- 1152 \\
\hline

Learning Rate & Step size used by the optimizer to update model weights. & 2e-5 to 1e-4 & 1e-6 -- 0.02 \\
\hline

Vocabulary Size & Size of tokenizer vocabulary; often inherited from pretrained models and not always reported. & Pretrained / not reported & 20{,}000 -- 92{,}416 \\
\hline

Dropout & Regularization rate used to reduce overfitting during training. & 0.2 -- 0.3 & 0.1 -- 0.5 \\
\hline

Number of Attention Heads & Number of parallel self-attention heads in transformer layers. & 12 & 1 -- 24 \\
\hline

Number of Layers & Depth of the network, usually transformer encoder/decoder blocks or related stacked layers. & 12 & 1 -- 32 \\
\hline

Maximum Input Length / Token Length & Maximum number of tokens or context size accepted by the model. & 512 & 32 -- 65{,}536 \\
\hline

Weight Decay / L2 Regularization & Regularization strength applied to model weights during training. & 1e-4 to 3e-4 & 1e-4 -- 3e-4 \\
\hline

Gradient Accumulation Steps & Number of steps accumulated before one optimizer update; used to simulate larger batches. & 12 & 12 \\
\hline

Gradient Clipping & Maximum norm/value used to stabilize training updates. & 1.0 & 1.0 \\
\hline

Random Seed & Fixed seed used for reproducibility across runs. & 42 / 2018 & 42 -- 123456 \\
\hline

Embedding Dimension & Size of token or feature embeddings before deeper processing. & 768 & 64 -- 768 \\
\hline

Feed-forward / Intermediate Size & Size of inner feed-forward network in transformer blocks or related modules. & 3072 & 768 -- 13{,}440 \\
\hline

Output Dimension & Final projected feature dimension used before classification or fusion. & 512 & 512 -- 768 \\
\hline

Optimizer & Algorithm used to update model parameters during training. & Adam / AdamW & Adam, AdamW, RAdam \\
\hline

Loss Function & Objective function optimized during training. & Cross-entropy & Cross-entropy, Binary Cross-Entropy, BCEWithLogitsLoss, BCE+MSE, BCE+CosineSimilarity
+BCE, MSE alignment, Contrastive loss, Z-loss \\
\hline

Activation Function & Non-linear function used in hidden/output layers. & ReLU & ReLU, ELU, Sigmoid, SiLU \\
\hline

Early Stopping & Stops training when validation performance stops improving. & Used in several studies & Present / not reported \\
\hline

Cross-validation & Repeated train-validation splitting for robust evaluation and tuning. & 5-fold CV & 5-fold CV \\

\bottomrule
\end{tabularx}
\end{table*}

Table \ref{tab:Hyperparameterall} lists the most important hyperparameters for transformer models and their typical ranges. For each hyperparameter, the table provides a brief description and the values commonly used in experiments. Overall, the Table \ref{tab:Hyperparameterall} provides a concise overview of the configuration ranges for transformer models, helping guide experimental setup and reproducibility. We believe these values are not standard and can vary across approaches, depending on the nature of the software vulnerability data. However, these values can serve as a baseline, as they are frequently reported in the literature for software vulnerability detection and classification. Moreover, the experimental environment can help software researchers and developers identify in advance the resources required for vulnerability detection and classification by considering transformer-based algorithms and datasets before actually running the experiments.  

\begin{tcolorbox}[ title={RQ6 Research Finding}]
Researchers explore a wide range of hyperparameters for fine-tuning their models
and environment settings. The most important hyperparameters are the learning rate, 
batch size, number of transformer layers, hidden size, number of attention heads 
and training epochs.
\end{tcolorbox}
\subsection{RQ7: At what level of granularity is the vulnerability detection performed?}
For more fine-grained analysis of software vulnerability detection, this RQ explores the granularity levels at which researchers identify vulnerabilities in their experiments.
\begin{table}
\footnotesize
\caption{Vulnerability granularity levels in the reviewed articles }
\label{tab:Granularity} 
\begin{tabular}{@{} p{3cm} p{1cm} p{3cm} @{}}
\toprule
 \textbf{Category} & \textbf{Count} & \textbf{Articles}\\
 \midrule

Function Level & 40 & \cite{chen2022hlt,purba2023software,ferrag2023securefalcon,lu2023assessing,bui2023detecting,sun2024enhancing,zhao2025vulnerability,zhao2024python,kim2024robust,gupta2024dl,wu2021peculiar,tanko2025approach,cao2024vulnerability,fu2022linevul,hanif2022vulberta,shiaeles2023vuldetect,jianjie2023code,sun2024gptscan,do2024optimizing,jeon2024design,alqarni2022low,le2024software,gong2023gratdet,he2025vultr,mechri2025secureqwen,wang2024scl,cao2025multi,liu2024making,xuan2025large,wang2024extensive,ni2025abundant,cui2025vulgtda,gunda2025transformer,li2025macd,shir2025robust,wang2025m2cvd,sultan2025codevul,kalouptsoglou2025transfer,smaili2025transformer} \\
\hline

Statement Level & 9 & \cite{gujar2024detectbert,hin2022linevd,liu2024pre,mamede2022transformer,zahid2025detectbert,kaanan2024llm,islam2023unbiased,sun2024gptscan,tao2025transformer} \\
\hline

Line-Level & 13 & \cite{mahyari2024harnessing,peng2023ptlvd,nguyen2023mando,rusinova2024explaining,liu2023software,fu2022linevul,curto2024multivd,alqarni2022low,le2024software,liu2024making,yang2024large,xuan2025large,wang2025line} \\
\hline

Task Workflow Level & 1 & \cite{zhang2024twlog} \\
\hline

Binary Code & 2 & \cite{liu2024automatic,jiang2024haformer} \\
\hline

File/Program Level & 3 & \cite{li2022software,mahyari2024harnessing,vanam2025software} \\
\hline

Slice Level & 2 & \cite{bahaa2024db,zhang2023vuld} \\
\hline

Code Gadget & 2 & \cite{kim2022vuldebert,thapa2022transformer} \\
\hline

Contract-Level & 5 & \cite{liang2024source,nguyen2023mando,ferretti2025detecting,perera2025codebert,shang2025cegt} \\
\hline

Code Snippet-Level & 4 & \cite{hou2022vulnerability,ehrenberg2024python,wang2024extensive,tian2025efvd} \\
\hline

Feature Level & 2 & \cite{almakayeel2024deep,saimbhi2024vulnerai} \\
\hline

Log Entry Level & 1 & \cite{myllari2025ladle} \\
\hline

Log Sequence-Level & 1 & \cite{wu2025information} \\
\hline

Commit-Level & 1 & \cite{sun2025hgtjit} \\
\hline

Flow/Program Analysis Level & 1 & \cite{katz2025siexvults} \\
\hline

App Level & 1 & \cite{oladokun2025effective} \\
\bottomrule
\end{tabular}
\end{table}

Table \ref{tab:Granularity} presents the \textbf{levels of granularity} at which previous research has identified vulnerabilities in datasets. Granularity of vulnerability refers to the degree of detail or precision in presenting and analysing a security vulnerability. We compiled this information based on details from research articles and our own understanding. Most of the studies focus on a single granularity level; however, a few of them, such as \cite{mahyari2024harnessing,nguyen2023mando,liang2024source,fu2022linevul,mamede2022transformer,jianjie2023code,sun2024gptscan,alqarni2022low,le2024software,liu2024making,wang2024extensive,xuan2025large}, identify vulnerabilities at multiple levels. For example, Nguyen et al. \cite{nguyen2023mando} propose a framework that models smart contracts as heterogeneous graphs, in which nodes represent entities such as functions, variables, statements, and contracts, and edges capture relations such as function calls, data dependencies, and control flow. Each node and edge is embedded with features encoding its type and semantic role, allowing the model to distinguish different interactions within the contract. A heterogeneous graph transformer (HGT) is applied to these graphs, using type specific multi head attention to aggregate contextual information from neighbours while respecting node and edge types. By leveraging these node embeddings and a readout function, MANDO-HGT can detect vulnerabilities at multiple granularity levels, including function level and contract level, enabling fine grained identification of unsafe operations as well as coarse grained assessment of vulnerable functions or contracts. Their model is trained with multi level supervision on labelled smart contract datasets, allowing it to capture complex structural and semantic patterns and achieve robust vulnerability detection across various levels of granularity. Similarly, Xuan et al. \cite{xuan2025large} propose a methodology that leverages pretrained LLMs to detect software vulnerabilities at multiple levels of granularity, including line- and function-level. The approach treats vulnerability detection as a hierarchical classification problem, enabling the model to capture both fine grained and coarse grained vulnerability patterns.

\begin{figure*}[!ht]
    \centering
    \includegraphics[width=0.80\textwidth]{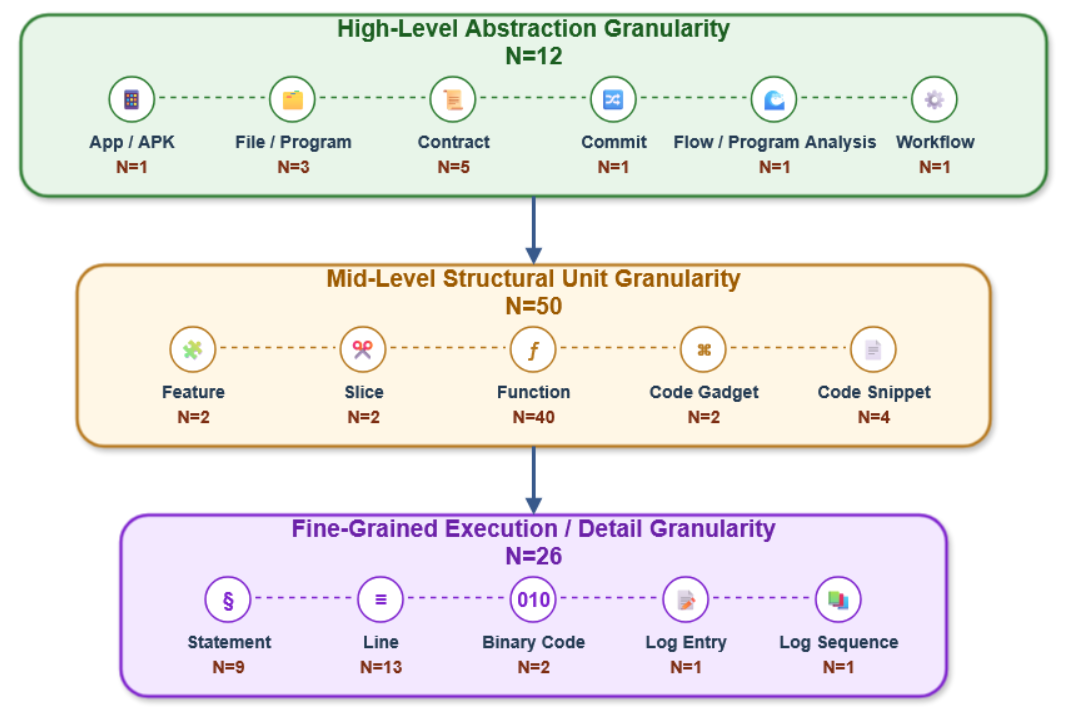}
    \caption{Granularity Level }
    \label{fig:granularity}
\end{figure*}

Figure \ref{fig:granularity} presents a top-down breakdown of software granularity, along with the number of research articles reporting vulnerabilities at each level. According to the best of our knowledge, levels progress from conceptual and process level views (applications and workflows), to the functional structure (features and functions) and finally to low level execution details (statements and binary code).
\begin{tcolorbox}[ title={RQ7 Research Finding}]
Researchers detect vulnerabilities at fine grained and coarse grained levels. 
While over 12 studies have explored multi granularity detection, most focus 
on single granularity detection, with function level vulnerability detection being 
the most dominant.
\end{tcolorbox}
\subsection{RQ8: What baseline models are used for comparison in these studies?}
For software researchers, it is always challenging to identify baseline approaches to compare against their proposed transformer-based approach. For this purpose, we evaluated existing transformer-based approaches to identify the baseline methods commonly used by researchers to evaluate and compare their methods.
\begin{table*}[t]
\scriptsize
\centering
\caption{The baseline Approaches Used to Evaluate Vulnerability Approaches}
\label{tab:baseline}

\begin{tabularx}{\textwidth}{@{} p{3.1cm} X @{}}
\hline
\textbf{Article} & \textbf{Baseline} \\
\hline
 Saimbhi and Akpinar \cite{saimbhi2024vulnerai} &  GPT-4 vs GPT-3.5, GPT-4 vs GPT-4 Turbo \\
\hline
 li et al \cite{li2022software} &  CNN, LSTM, Bi-LSTM, and CNN-LSTM.  \\
\hline
  Chen and liu \cite{chen2022hlt}  &  Transformer and
Embedding-Transformer    \\
\hline
 Bahaa et al. \cite{bahaa2024db} & SOTA approaches: VulDeePecker,   \\\hline
Purba et al. \cite{purba2023software}  &  Flawfinder, Static Analysis,RATS ,Checkmarx 
VulDeePecker,CodeGen,Davinci   \\\hline
 Ferrag et al. \cite{ferrag2023securefalcon}  & RoBERTa, BERT,CodeBERT etc    \\\hline
 lu ey al. \cite{lu2023assessing}  & CodeBERT, CodeT5, and CodeGPT. (Each Model fine tunning campare with prompt tunning)  \\\hline
 Mahyari \cite{mahyari2024harnessing}  & LSTM, VulDeeLocator    \\\hline
 Bui et al.  &  Russell, VulDeePecker,SySeVR,Devign    \\\hline
 Sun et al. \cite{sun2024enhancing}  & LineVUL,Reveal,Devign,SySeVR  \\\hline
  Zhao and liu  \cite{zhao2025vulnerability}  &  Siamese, TokenCNN,VulDeePecker,CodeBERT    \\\hline
 Kim et al.\cite{kim2022vuldebert}  & BiLSTM    \\\hline
 Hanif and Maffeis \cite{hanif2022vulberta}  &  Baseline-BiLSTM, Baseline-TextCNN   \\\hline
 Hou et al. \cite{hou2022vulnerability}  &  VulDeepecker 
AE-KNN   \\\hline
 Zahid \cite{zahid2025detectbert}  &  LineVD  \\\hline
Shiaeles at al.\cite{shiaeles2023vuldetect}  &  LSTM, CodeBERT, yseVR and VulDeBERT   \\\hline
 Kannan \cite{kaanan2024llm}  & RoBERTa,CodeBERT,EnsembleLLM,GPT-2   \\\hline
 Jianjie and le \cite{jianjie2023code}  &  BiLSTM, TextCNN,GCNN,Devign    \\\hline
 Islam et al.\cite{islam2023unbiased}  & BiLSTM,TextCNN,RoBERTa,CodeBERT,Devign,
VulDeePecker,VELVET   \\\hline
 Sun et al.\cite{sun2024gptscan}  & GPT   \\\hline
Curto et al.\cite{curto2024multivd}  &  LineVul  \\\hline
 Jeon et al.\cite{jeon2024design}  & SVM ,Eth2Vec ,DR-GCN   \\\hline
 Alqarni and Azim \cite{alqarni2022low}  &  RNN+LSTM, Bi-LSTM   \\\hline
Le et al. \cite{le2024software}  &  C/C++ language results  \\\hline
 Zhang et al. \cite{zhang2024twlog}  &  PCA,SVM, deeplog  \\\hline
Almakayeel \cite{almakayeel2024deep}  & RHSODL-AMD , AAMD-OELAC ,GBWODL-AMC , DBN , J48 ,Naïve–Bayes ,Linear-SVM,LSTM classifier , KNN , RF, CNN encoder-RNN , DNN , CNN classifier, Dense model   \\\hline
 Gong et al.\cite{gong2023gratdet} &  Mythril, Smartcheck , Oyente,Slither, Securify, Manticore, Peculiar,BLSTM-ATT, TextCNN, CGE    \\
\hline Liu et al. \cite{liu2024automatic} &  Ins2vec-TCNN, VDiscover,A-BiLSTM,SAFE,SAFE+ROS  \\
\hline He et al.\cite{he2025vultr}  &  VulDeePecker,SySeVR,Devign, VulCNN, IVDetect, and mVulPreter  \\
\hline Mechri et al.\cite{mechri2025secureqwen} &  VulDeePecker,,Devign,SecureFalcon etc  \\
\hline
 Cao \cite{cao2024vulnerability}  &  Cppcheck ,SyseVR,Deepwukong   \\
\hline
 Wang et al. \cite{wang2024scl} &  ReGVD,CodeBERT ContraBERT ,GraphCodeBERT  \\
\hline
 Ehrenberg et al. \cite{ehrenberg2024python}  &  compare results of models with each other   \\
\hline
 Cao and  Dong \cite{cao2025multi}  &  SySeVR, devign,CodeBERT, CodeBERT-mmd, CodeBERT-dann   \\

\hline
 Liu et al. \cite{liu2024making} &  Devign , SySeVR,REVEAL  \\
\hline
 Jiang et al. \cite{jiang2024haformer} &  Gemini, Asm2Vec , SAFE,jTrans ,Trex ,Asteria-Pro   \\
\hline
 Wu et al. \cite{wu2025information}  &   K-Nearest Neighbor (KNN), Decision Tree (DT), and Multi-layer Perception (MLP)   \\
\hline
 Yang et al. \cite{yang2024large}  &  DeepFL , DeepRL4FL , and TRANSFER-FL    \\\hline
 Xuan et al. \cite{xuan2025large} &  GRD,AMPLE,Reveal,MAGNET,LIVABLE,SySeVR,DeepWukong,
LineVul,ReVeal (re), IVDetect  \\\hline
 Wang et al. \cite{wang2024extensive}  &  Compare models with each other
\\\hline Ni et al.\cite{ni2025abundant}  &  UniXcoder,IVDetect,Devign,Reveal,LineVD,LineVul,VulCNN,
VulDeeLocator,Cppcheck   \\
\hline Do et al.\cite{do2024optimizing}  &  CodeT5,Bert \\
\hline Gujar \cite{gujar2024detectbert}  &  GCN,GAT  \\
\hline
 Peng et al. \cite{peng2023ptlvd}  &  SAC,IG,Saliency,Deeplift,Deeplift Shap,GuidedBackprop,InputxGradient   \\
\hline
 Hin et al. \cite{hin2022linevd}  &  IVDetect 
\\\hline
 Kim et al. \cite{kim2024robust}  &  Different dataset    \\
\hline
 Liu et al.\cite{liu2024pre}   &  Bi-LSTM, Transformer,VulDeePecker ,Devign, SySeVR, ReVeal etc  \\
\hline
 Nguyen et al.\cite{nguyen2023mando}  &  GNNs etc   \\\hline
 Gupta et al.\cite{gupta2024dl}  &  Bert,CodeBert,GraphCodeBert,TextCNN,TextGCN,
Devign(AST),DL-VulBERT   \\\hline
 Liang et al.\cite{liang2024source}  &  Checkmark,Flawfinder,RATS,LineVul,LineVD,DeeplineDP
,VulSniper  \\\hline
 Wu et al.\cite{wu2021peculiar}  &  Honeybadger,Manticore, MythrilOsiris ,Oyente, Securify, 
Slither ,Smartcheck, DR-GCN, TMP   \\
\hline
 Rusinova et al. \cite{rusinova2024explaining}   &  LIME and SHAP   \\
\hline
 Wu et al.\cite{wu2021self}   &  BLSTM, BGRU    \\
\hline
 Liu et al.\cite{liu2023software}   & VulDeePeckker,SySeVR,Original GPT    \\
\hline
 Tanko \cite{tanko2025approach}  &  Different Graph representation techniques  \\
\hline
 Zhang et al.\cite{zhang2023vuld}   & VulDeePecker,SySeVR-BGRU,SySeVR-ABGRU,Russell   \\
\hline
 Thapa et al.\cite{thapa2022transformer}   &  VulDeePecker
Original, BiLSTM, BiGRU, BERTBase, GPT-2 Base, CodeBERT, DistilBERT, RoBERTa etc  \\
\hline

 Fu \& Tantithamthavorn \cite{fu2022linevul}   &  CodeBERT,IVDect,Reveal,SyseVR,Devign,BoW+RF,
VulDeePecker etc  \\
\hline
Tian \& Zhang\cite{tian2025efvd} & VulDeePecker, SySeVR, Devign, Reveal, GGNN, GCN, R-GCN, SVM, RF, MLP, BGNN4VD \\
\hline
Katz et al.\cite{katz2025siexvults} & ChatGPT-4o, CodeBERT, CodeT5; discussed DeepWukong, VulDeePecker, SonarQube, LineVul \\
\hline
Cui et al.\cite{cui2025vulgtda} & Devign, VulCNN, VulDeePecker \\
\hline
Gunda et al. \cite{gunda2025transformer} & Logistic Regression \\
\hline
Li \cite{li2025macd} & IVDetect, ReVeal, Devign, SySeVR, VulDeePecker \\
\hline
Sun et al.\cite{sun2025hgtjit} & VCCFinder, DeepJIT, CC2Vec, CCT5, CodeJIT \\
\hline
Oladokun \& Rice \cite{oladokun2025effective} & AutoML (RF, SVC, MLP) \\
\hline
Wang et al. \cite{wang2025line} & ChatGPT 3.5/4o, Devign, ReGVD, CodeBERT, UniXcoder-base, CodeT5+, UniXcoder-nine, TRACED, DeepDFA, PDBERT \\
\hline
Shir et al. \cite{shir2025robust} & RNN, LSTM, GRU, Longformer, prior LLVM-IR/assembly works \\
\hline
Wang et al. \cite{wang2025m2cvd} & ChatGPT 3.5/4o, Devign, ReGVD, CodeBERT, CodeT5, UniXcoder-base, UniXcoder-nine, TRACED \\
Ferretti et al. \cite{ferretti2025detecting} & BERT, DistilBERT, CodeBERT, Gemini, stacking classifiers \\
\hline
Perera et al. \cite{perera2025codebert} & CodeBERT variants; prior smart-contract tools in related comparison \\
\hline
Sultan et al. \cite{sultan2025codevul} & CodeBERT, GraphCodeBERT, CodeT5, SantaCoder, Devign, ReVeal, IVDetect, BGNN4VD, ContraFlow, $\mu$VulDeePecker, SySeVR, MANDO \\
\hline

Shang et al. \cite{shang2025cegt} & Oyente, Mythril, Slither, LSTM, RNN, GCN, BiLSTM, ReChecker, DR-GCN, TMP, CGE, DL4SC \\
\hline
Kalouptsoglou et al. \cite{kalouptsoglou2025transfer} & VulDeePecker, SySeVR, Devign, ReVeal, LineVul, traditional embeddings \\
\hline
Tao et al. \cite{tao2025transformer} & Flawfinder, Checkmarx, SySeVR, IVDetect, VDTC, VulDeeLocator, LineVul \\
\hline
Smaili et al. \cite{smaili2025transformer} & VulDeePecker, Devign, Reveal, GCL4SVD, GSVD, MGVD, DGVD, CodeT5 \\
\end{tabularx}
\end{table*}

Table \ref{tab:baseline} summarizes the baseline models and approaches used in the surveyed articles for software vulnerability detection. Baseline is a reference method or model used for comparison to evaluate the performance of a new approach. It provides a point of reference so researchers can judge whether their approach is actually improving performance in software vulnerability detection or classification. For example, Islam et al \cite{islam2023unbiased}  combines transformer based source code representation with a semantic vulnerability graph (SVG) to detect software vulnerabilities more accurately. The approach constructs a graph capturing semantic relationships between code elements, such as function calls, variable dependencies and control flow interactions, which is then integrated with a transformer encoder to generate contextual embeddings of code snippets. To evaluate their method, they use baseline comparison techniques including BiLSTM, TextCNN, RoBERTa, CodeBERT, Devign, VulDeeP-ecker, and VELVET. By comparing against these baselines, they demonstrate that combining semantic graph information with transformer embeddings significantly improves performance metrics such as accuracy, precision, recall and F1-score, highlighting the effectiveness of their approach in capturing subtle and complex vulnerability patterns in source code.  Similarly, Cao and Dong \cite{cao2025multi} use pretrained code models to detect vulnerabilities across different programming languages and domains. In the first step, they pretrain a language model on multiple source code repositories to capture general code semantics and syntactic patterns. After that, source code from various domains is tokenized and represented using the pretrained embeddings, which are fed into a classification layer to predict vulnerabilities. To enable cross domain detection, the model incorporates domain adaptation techniques, aligning features from source and target domains to handle variations in coding styles and domain specific constructs. For evaluation, the study uses baseline methods such as single domain pretrained models, standard transformer models without cross domain adaptation and classical machine learning classifiers like SVM or LSTM-based models. Comparisons against these baselines demonstrate that the multi source pretrained model significantly improves detection accuracy, F1-score, and robustness across unseen domains.

\begin{table}
\footnotesize
\caption{Methodological Taxonomy of Baseline Used in Vulnerability Detection }
\label{tab:baseline_taxonomy} 
\begin{tabular}{@{} p{3cm} p{5cm} }
\toprule
 \textbf{Category} & \textbf{Representative Methods / Tools} \\
 \midrule
Traditional Machine Learning 
& SVM, KNN, Decision Trees, Na\"ive Bayes, Random Forest \\
\hline
Deep Learning (Sequence-based) 
& LSTM, BiLSTM, GRU, CNN, CNN--LSTM \\
\hline
Transformer-based Models 
& BERT, RoBERTa, CodeBERT, GraphCodeBERT, UniXcoder, CodeT5 \\
\hline
Graph-based Models 
& GCN, GAT, DR-GCN, AST-based Devign \\
\hline
Static Analysis Tools 
& Flawfinder, RATS, Cppcheck, Checkmarx \\
\hline
Symbolic / Formal Analysis Tools 
& Mythril, Oyente, Slither, Manticore \\
\hline
Large Language Models (LLMs) 
& GPT-2, GPT-3.5, GPT-4, GPT-4 Turbo \\
\hline
State-of-Art frameworks
& VulDeePecker, SySeVR, Devign, LineVul etc \\

\bottomrule
\end{tabular}
\end{table}
Table \ref{tab:baseline_taxonomy} shows the summary of baseline categories and associated values. Table \ref{tab:baseline_taxonomy} highlights the diversity of baselines across conceptual, functional and low level software analysis techniques. These baselines include widely adopted deep learning frameworks such as Devign, ReVeal, VulDeePecker, SecureFalcon, SySeVR, IVDetect and LineVul, which serve as benchmarks for evaluating the effectiveness of new vulnerability detection approaches.
\begin{figure*}[h!]
    \centering
    \includegraphics[width=0.80\textwidth]{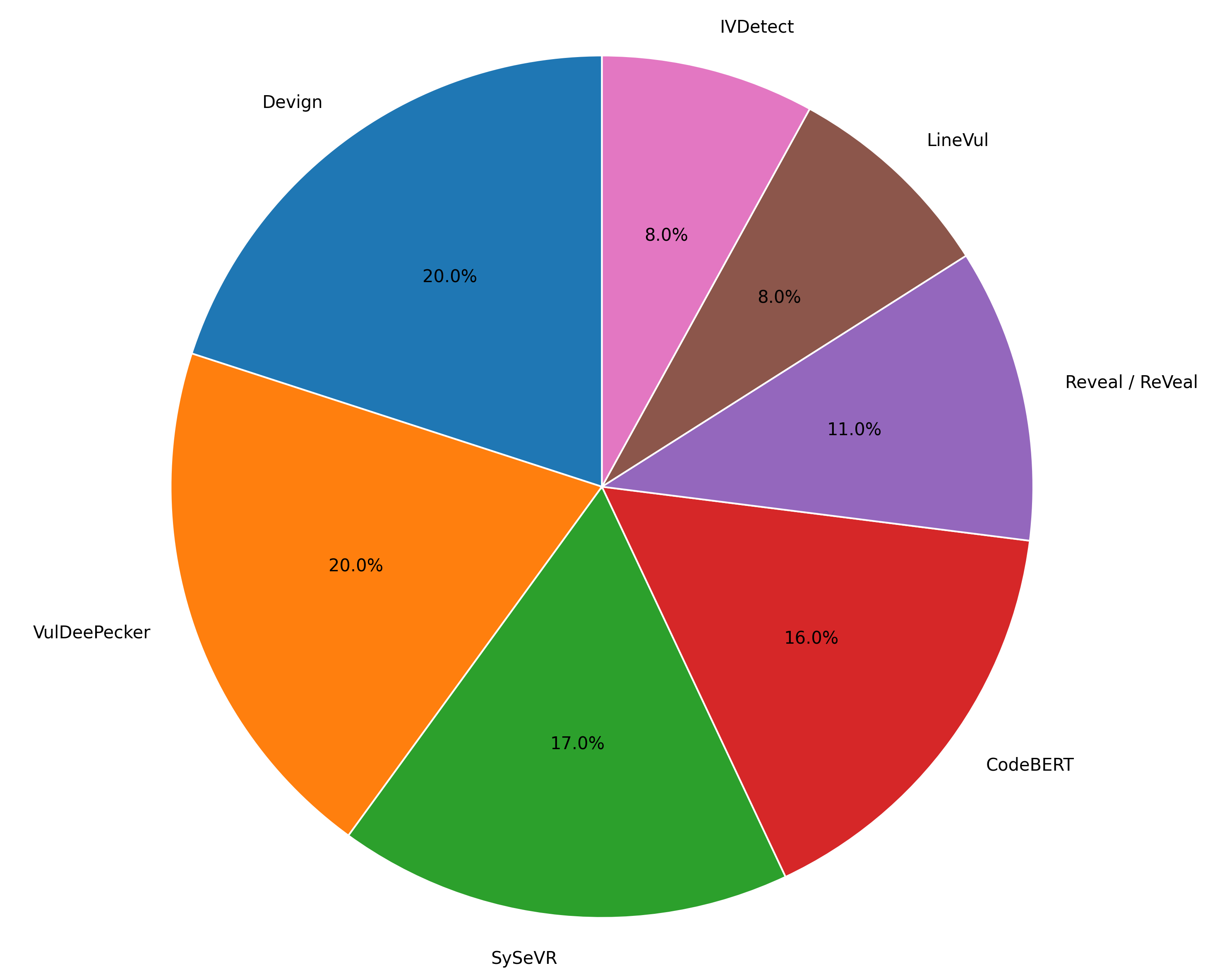}
    \caption{Popular Baselines }
    \label{fig:popular_baseline}
\end{figure*}

Figure \ref{fig:popular_baseline} shows the popular baseline methods and their proportion of use in studies and research articles. VulDeePecker and Devign  holds the largest share at 20.0\%, followed by SySeVR at 17.0\% and CodeBERT at 16.0\%, Reveal at 11.0\%, LineVul and IVDetect 8.0\%  indicating that these approaches contribute the most among the models compared. 

\begin{tcolorbox}[ title={RQ8 Research Finding}]
Researchers mostly use seven categories of models and tools as baselines for 
their experimental results. Among these, VulDeePecker, Devign, SySeVR are the most 
frequently used baselines.
\end{tcolorbox}
 \subsection{RQ9: How many existing approaches focus on multi-lingual software vulnerability detection using transformers?}
Recently, researchers have started exploring multi-lingual-based vulnerability detection using a transformer. For example, Le et al. \cite {le2024software} compare two transformer models, CodeBERT and ChatGPT, for
vulnerability prediction in low-resource programming languages. In the First step, they prepare
a labelled dataset of vulnerable and non-vulnerable code snippets and apply the same samples
to both models. Then, CodeBERT is fine-tuned on the dataset, while ChatGPT is evaluated
using prompt-based in-context learning. 

\begin{table}
\footnotesize
\caption{Papers addressing multi-language vulnerability detection }
\label{tab:multilanguagepapers} 
\begin{tabular}{@{} p{1cm} p{4.3cm} p{2cm}  @{} }
\toprule

\textbf{Articles} & \textbf{Multi-language Detection Methodology}  & \textbf{Languages} \\
 \midrule

Mahyari \cite{mahyari2024harnessing} &
Uses multi-language vulnerability data rather than restricting the task to one programming language. &
C/C++, Java, PHP, C\# \\
\hline

Jianjie and Le \cite{jianjie2023code} &
Uses datasets such as CodeXGLUE / Juliet, which cover more than one programming language. &
C/C++, Java \\
\hline

Le et al. \cite{le2024software} &
The study is explicitly centered on vulnerability prediction in low-resource languages using CodeBERT and ChatGPT &
Rust,Kotlin, Swift \\
\hline

He et al. \cite{he2025vultr} &
Uses mixed-source datasets, such as SARD + NVD, that include vulnerabilities across multiple programming languages. &
C/C++, Java, Python, PHP, and others \\
\hline

Gujar \cite{gujar2024detectbert} &
Uses CVEfixes and related sources, so the detection setting is not restricted to a single language. &
50+ languages through CVEfixes \\
\hline
Gunda et al. \cite{gunda2025transformer} &
Uses a custom dataset containing both Python and C code, making it clearly multi-language. &
Python, C \\
\bottomrule
\end{tabular}
\end{table}
 Table \ref{tab:multilanguagepapers} identifies the studies in the review that address multi-language vulnerability detection rather than focusing on a single programming language. It shows that only a small number of papers explicitly work across multiple languages, using either multi-language benchmark datasets or custom datasets containing code from more than one language. The covered languages include C/C++, Java, PHP, C\#, Python and C, and low-resource languages such as Rust, Kotlin, and Swift, while some studies rely on broader datasets like CVEfixes, which contain vulnerabilities from more than 50 programming languages. Overall, Table \ref{tab:multilanguagepapers} highlights that multi-language vulnerability detection is still less common in the literature compared with single-language or domain-specific studies.

\begin{tcolorbox}[ title={RQ9 Research Finding}]
A fewer number of reviewed papers work across multiple programming languages. 
Most existing studies still focus on single-language or domain-specific vulnerability 
detection. This encourages researchers to develop software vulnerability approaches 
\end{tcolorbox}
\section{Threats to Validity}
This review may not include all relevant studies because the selection of papers depended on the databases searched, the search strings and keywords used, and the inclusion and exclusion criteria applied during screening. As a result, some relevant studies may have been missed, especially if they described transformer models, vulnerability detection, or software security using different terminology. Another possible threat is publication bias, since the review mainly includes published papers, which are more likely to report positive or stronger results, while negative findings, replication failures, workshop drafts, and unpublished studies may be underrepresented. In addition, several parts of the review required manual classification, including transformer family, granularity level, vulnerability type coverage, single-language versus multi-language scope, and baseline grouping. Because these categories involve interpretation, some studies may reasonably fit more than one group, and a few borderline cases may have been simplified. A further limitation is that many papers did not clearly report important details such as exact baselines, hyperparameters, granularity, language scope, vulnerability type, experimental setup, or dataset composition. In such cases, some entries had to be inferred from the datasets used, the experiments described, or the overall problem setting, which introduces a risk of misinterpretation. There is also a dataset-based inference threat, since some papers used datasets that contained multiple CWE types or programming languages, whereas the model itself only performed binary vulnerability detection; interpreting dataset diversity as model capability may therefore overestimate the study's actual contribution. Finally, granularity was not consistently reported across papers, and levels such as function-level, line-level, statement-level, contract-level, and flow-level had to be grouped into broader categories. Because some studies span multiple granularity levels, assigning them to one or more categories may also require subjective judgment.
\section{Open Directions}
Future research on transformer-based software vulnerability detection should move beyond benchmark-oriented classification and focus more on practical and fine-grained security solutions. In particular, future studies should give equal importance to both vulnerability detection and vulnerable code localization. Although many existing approaches aim to determine whether a code sample is vulnerable, such predictions are of limited practical value unless the model can also identify the exact vulnerable region, such as the function, statement, line, contract fragment, or execution path responsible for the weakness. Therefore, future systems should be designed to jointly support classification and localization so that they can provide more actionable outputs for developers. This would improve their usefulness in debugging, remediation, patch generation, and vulnerability triage, and would make transformer-based methods more relevant for real-world software engineering practice \cite{li2022software,fu2022linevul,hin2022linevd,tao2025transformer}.

Another important direction is the development of richer code representations and more generalized detection models. The findings of this review indicate that the way code is represented has a significant impact on detection performance. Sequence-based representations are effective for capturing lexical and contextual patterns, while graph-based representations better encode structural and dependency information. Future work should therefore explore hybrid representations that combine token sequences with abstract syntax trees, control-flow graphs, data-flow graphs, and program dependence information in order to capture both local and global vulnerability patterns. At the same time, future models should move beyond narrow settings limited to a single language or a small number of vulnerability categories. More generalized approaches are needed to support multiple programming languages and broader vulnerability coverage, which would improve the scalability, transferability, and practical relevance of vulnerability detection systems \cite{chen2022hlt,liang2024source,nguyen2023mando,bahaa2024db,mahyari2024harnessing,kim2022vuldebert}.

A further promising research direction is the integration of transformer models and large language models with traditional program analysis techniques. Although transformer-based methods have demonstrated strong contextual understanding, they still face limitations in semantic precision and may produce predictions based on statistical patterns rather than verifiable security reasoning. In contrast, traditional static analysis, dynamic analysis, symbolic execution, and taint analysis provide program-level evidence that can validate or refine model outputs. Future work should therefore investigate hybrid frameworks that combine the learning ability of transformers and LLMs with the precision of traditional analysis methods. In addition, explainability should become a central goal of future research. Vulnerability detection systems should not only identify security flaws but also explain why the code is vulnerable, highlight the suspicious region, and help developers understand the root cause. Improved explainability would enhance trust, usability, and adoption in real-world software security workflows \cite{purba2023software,sun2024gptscan,rusinova2024explaining,shiaeles2023vuldetect}.

Dataset quality and evaluation design also require substantial improvement. This review shows that many studies rely on a limited number of benchmark datasets, some of which may contain noisy labels, duplicated samples, synthetic examples, class imbalance, or hidden data leakage. These limitations can lead to overly optimistic performance results and weak real-world generalization. Future research should focus on constructing balanced, diverse, and high-quality datasets with transparent collection procedures, accurate labels, realistic code contexts, and broader vulnerability coverage. More challenging and realistic evaluation settings are also needed, including cross-project, cross-version, cross-language, and temporally separated testing. In addition, future studies should more clearly distinguish between dataset diversity and actual model capability, since a dataset may include multiple languages or vulnerability types even when the model itself only performs limited binary detection \cite{lu2023assessing,gujar2024detectbert,kim2024robust,kalouptsoglou2025transfer}.

Finally, future work should focus on improving robustness, adaptability, and real-world usability. Current models often remain sensitive to tokenization strategies, preprocessing choices, hyperparameter settings, and dataset composition, which means that further research is needed on security-aware pretraining, effective fine-tuning, expressive embeddings, and systematic ablation studies. Future systems should also be better equipped to detect previously unseen or zero-day vulnerabilities through approaches such as self-supervised learning, anomaly detection, continual learning, and transfer learning. In addition, research should explore cross-representation learning across source code, binaries, commits, logs, and runtime traces, as well as stronger integration into practical developer workflows such as IDEs, code review systems, and CI/CD pipelines. Overall, the next generation of transformer-based vulnerability detection systems should aim not only for high predictive accuracy, but also for localization capability, explainability, robustness, reproducibility, and deployment readiness \cite{kim2022vuldebert,hin2022linevd,gujar2024detectbert,kim2024robust,shir2025robust,ferretti2025detecting,yang2024large}.

\section{Conclusion}

Software vulnerabilities continue to pose a serious threat to modern software systems as applications grow in size, complexity, and interconnectivity. In recent years, transformer-based models have emerged as a powerful direction for software vulnerability detection due to their ability to capture semantic, structural, and contextual information from source code more effectively than many traditional machine learning and deep learning techniques. Despite this rapid growth, prior systematic literature reviews have mainly focused on conventional vulnerability detection approaches, with limited attention to transformer-based methods as a distinct research area. To address this gap, this study presented a systematic literature review of transformer-based software vulnerability detection research, guided by eight research questions.

Following Kitchenham’s SLR guidelines, we reviewed 80 primary studies published between 2021 and 2025 and analyzed them across multiple dimensions, including task type, datasets, vulnerability types, granularity levels, baselines, evaluation metrics, hyperparameter, and language coverage. The findings show that transformer-based approaches are now widely applied to binary vulnerability detection and vulnerability type detection. The review also reveals that Devign and Big-Vul are the most frequently used datasets, while Devign and VulDeePecker appear as the most common baseline models in comparative experiments. Most studies operate at the function level, although recent work increasingly explores finer granularities such as line-level, statement-level, and commit-level detection. The review further shows that the literature covers a broad range of vulnerability categories, yet only a limited number of studies explicitly address multi-language vulnerability detection.

Overall, this survey provides a focused and comprehensive synthesis of transformer-based vulnerability detection research and highlights several important research gaps. These include the lack of standardized benchmarks, limited support for multi-language detection, inconsistent reporting of hyperparameter and baselines, insufficient attention to explainability, and limited evaluation in real-world development settings. The results of this study provide a useful reference for researchers and practitioners by consolidating current knowledge and identifying the most promising directions for future work. In particular, future progress is likely to depend on stronger benchmark standardization, broader multi-language datasets, better reporting practices, explainable transformer architectures, and deeper integration of transformer models and large language models with graph-based analysis and program analysis techniques. Such advances will be essential for building more accurate, robust, and practically deployable software vulnerability detection systems.

\printcredits

\textbf{Acknowledgements:} The authors acknowledge the use of the writing assistance tool (Grammarly and ChatGPT) to improve the writing quality of this paper. Following its use, the authors thoroughly reviewed and revised the content, and they take full responsibility for the final version of the paper.

\textbf{Conflict of Interest:}The authors declare no conflicts of interest.

\bibliographystyle{cas-model2-names}

\bibliography{cas-refs}

@inproceedings{thapa2022transformer,
  title={Transformer-based language models for software vulnerability detection},
  author={Thapa, Chandra and Jang, Seung Ick and Ahmed, Muhammad Ejaz and Camtepe, Seyit and Pieprzyk, Josef and Nepal, Surya},
  booktitle={Proceedings of the 38th Annual Computer Security Applications Conference},
  pages={481--496},
  year={2022}

}

@inproceedings{zhang2023vuld,
  title={VulD-Transformer: source code vulnerability detection via transformer},
  author={Zhang, Xuejun and Zhang, Fenghe and Zhao, Bo and Zhou, Bo and Xiao, Boyang},
  booktitle={Proceedings of the 14th Asia-Pacific Symposium on Internetware},
  pages={185--193},
  year={2023}
}

@article{tanko2025approach,
  title={AN APPROACH FOR VULNERABILITY DETECTION IN WEB APPLICATIONS USING GRAPH NEURAL NETWORKS AND TRANSFORMERS},
  author={TANKO, MOHAMMED YAHAYA and SULTAN, ABU BAKAR MD and OSMAN, MOHD HAFEEZ and ZULZALIL, HAZURA},
  journal={Journal of Theoretical and Applied Information Technology},
  volume={103},
  number={1},
  year={2025}
}

@article{zahid2025detectbert,
  title={DetectBERT: A Transformer-Based Approach for Statement-Level Vulnerability Detection in Python Code},
  author={Zahid, Mohammed Junead Sheriff},
  journal={Authorea Preprints},
  year={2025},
  publisher={Authorea}
}

@article{kim2024robust,
  title={Robust vulnerability detection in solidity-based ethereum smart contracts using fine-tuned transformer encoder models},
  author={Kim, Jaehyun and Lee, Sangmyeong and Kim, Howon and others},
  journal={IEEE Access},
  year={2024},
  publisher={IEEE}
}

@article{bahaa2024db,
  title={DB-CBIL: A DistilBert-Based Transformer Hybrid Model using CNN and BiLSTM for Software Vulnerability Detection},
  author={Bahaa, Ahmed and Kamal, Aya El-Rahman and Fahmy, Hanan and Ghoneim, Amr S},
  journal={IEEE Access},
  year={2024},
  publisher={IEEE}
}

@inproceedings{hou2022vulnerability,
  title={A vulnerability detection algorithm based on transformer model},
  author={Hou, Fujin and Zhou, Kun and Li, Longbin and Tian, Yuan and Li, Jie and Li, Jian},
  booktitle={International Conference on Artificial Intelligence and Security},
  pages={43--55},
  year={2022},
  organization={Springer}
}

@inproceedings{kim2022vuldebert,
  title={Vuldebert: A vulnerability detection system using bert},
  author={Kim, Soolin and Choi, Jusop and Ahmed, Muhammad Ejaz and Nepal, Surya and Kim, Hyoungshick},
  booktitle={2022 IEEE International Symposium on Software Reliability Engineering Workshops (ISSREW)},
  pages={69--74},
  year={2022},
  organization={IEEE}
}

@inproceedings{hanif2022vulberta,
  title={Vulberta: Simplified source code pre-training for vulnerability detection},
  author={Hanif, Hazim and Maffeis, Sergio},
  booktitle={2022 International joint conference on neural networks (IJCNN)},
  pages={1--8},
  year={2022},
  organization={IEEE}
}

@inproceedings{fu2022linevul,
  title={Linevul: A transformer-based line-level vulnerability prediction},
  author={Fu, Michael and Tantithamthavorn, Chakkrit},
  booktitle={Proceedings of the 19th International Conference on Mining Software Repositories},
  pages={608--620},
  year={2022}
}

@inproceedings{mamede2022transformer,
  title={A transformer-based ide plugin for vulnerability detection},
  author={Mamede, Cl{\'a}udia and Pinconschi, Eduard and Abreu, Rui},
  booktitle={Proceedings of the 37th IEEE/ACM International Conference on Automated Software Engineering},
  pages={1--4},
  year={2022}
}

@article{chan2023transformer,
  title={Transformer-based vulnerability detection in code at EditTime: Zero-shot, few-shot, or fine-tuning?},
  author={Chan, Aaron and Kharkar, Anant and Moghaddam, Roshanak Zilouchian and Mohylevskyy, Yevhen and Helyar, Alec and Kamal, Eslam and Elkamhawy, Mohamed and Sundaresan, Neel},
  journal={arXiv preprint arXiv:2306.01754},
  year={2023}
}

@article{cao2024vulnerability,
  title={Vulnerability detection based on transformer and high-quality number embedding},
  author={Cao, Yang and Dong, Yunwei and Peng, Jiajie},
  journal={Concurrency and Computation: Practice and Experience},
  volume={36},
  number={28},
  pages={e8292},
  year={2024},
  publisher={Wiley Online Library}
}

@article{harzevili2025systematic,
  title={A Systematic Literature Review on Automated Software Vulnerability Detection Using Machine Learning},
  author={Harzevili, Nima shiri and Belle, Alvine boaye and Wang, Junjie and Wang, Song and Jiang, Zhen ming (jack) and Nagappan, Nachiappan},
  journal={ACM COMPUTING SURVEYS},
  volume={57},
  number={3},
  year={2025},
  publisher={ASSOC COMPUTING MACHINERY 1601 Broadway, 10th Floor, NEW YORK, NY USA}
}

@inproceedings{hin2022linevd,
  title={LineVD: Statement-level vulnerability detection using graph neural networks},
  author={Hin, David and Kan, Andrey and Chen, Huaming and Babar, M Ali},
  booktitle={Proceedings of the 19th international conference on mining software repositories},
  pages={596--607},
  year={2022}
}

@article{ni2025abundant,
  title={Abundant Modalities Offer More Nutrients: Multi-Modal-Based Function-level Vulnerability Detection},
  author={Ni, Chao and Yin, Xin and Li, Xinrui and Xu, Xiaodan and Yu, Zhi},
  journal={ACM Transactions on Software Engineering and Methodology},
  year={2025},
  publisher={ACM New York, NY}
}

@inproceedings{yang2024large,
  title={Large language models for test-free fault localization},
  author={Yang, Aidan ZH and Le Goues, Claire and Martins, Ruben and Hellendoorn, Vincent},
  booktitle={Proceedings of the 46th IEEE/ACM International Conference on Software Engineering},
  pages={1--12},
  year={2024}
}

@inproceedings{li2022software,
  title={Software Vulnerability Detection Based on Anomaly-Attention},
  author={Li, Shanshan and Chen, Deng and Zhang, Jun and Wang, Haoyu and Li, Lei and Qian, Yuyang and Liu, Hailun},
  booktitle={2022 4th International Conference on Robotics and Computer Vision (ICRCV)},
  pages={261--265},
  year={2022},
  organization={IEEE}
}

@inproceedings{purba2023software,
  title={Software vulnerability detection using large language models},
  author={Purba, Moumita Das and Ghosh, Arpita and Radford, Benjamin J and Chu, Bill},
  booktitle={2023 IEEE 34th International Symposium on Software Reliability Engineering Workshops (ISSREW)},
  pages={112--119},
  year={2023},
  organization={IEEE}
}

@article{ghaffarian2017software,
  title={Software vulnerability analysis and discovery using machine-learning and data-mining techniques: A survey},
  author={Ghaffarian, Seyed Mohammad and Shahriari, Hamid Reza},
  journal={ACM computing surveys (CSUR)},
  volume={50},
  number={4},
  pages={1--36},
  year={2017},
  publisher={ACM New York, NY, USA}
}

@article{lin2020software,
  title={Software vulnerability detection using deep neural networks: a survey},
  author={Lin, Guanjun and Wen, Sheng and Han, Qing-Long and Zhang, Jun and Xiang, Yang},
  journal={Proceedings of the IEEE},
  volume={108},
  number={10},
  pages={1825--1848},
  year={2020},
  publisher={IEEE}
}

@article{nong2022open,
  title={Open science in software engineering: A study on deep learning-based vulnerability detection},
  author={Nong, Yu and Sharma, Rainy and Hamou-Lhadj, Abdelwahab and Luo, Xiapu and Cai, Haipeng},
  journal={IEEE Transactions on Software Engineering},
  volume={49},
  number={4},
  pages={1983--2005},
  year={2022},
  publisher={IEEE}
}

@article{shiri2024systematic,
  title={A systematic literature review on automated software vulnerability detection using machine learning},
  author={Shiri Harzevili, Nima and Boaye Belle, Alvine and Wang, Junjie and Wang, Song and Jiang, Zhen Ming and Nagappan, Nachiappan},
  journal={ACM Computing Surveys},
  volume={57},
  number={3},
  pages={1--36},
  year={2024},
  publisher={ACM New York, NY}
}

@article{eberendu2022systematic,
  title={A systematic literature review of software vulnerability detection},
  author={Eberendu, Adanma Cecilia and Udegbe, Valentine Ikechukwu and Ezennorom, Edmond Onwubiko and Ibegbulam, Anita Chinonso and Chinebu, Titus Ifeanyi and others},
  journal={European Journal of Computer Science and Information Technology},
  volume={10},
  number={1},
  pages={23--37},
  year={2022}
}

@article{croft2022data,
  title={Data preparation for software vulnerability prediction: A systematic literature review},
  author={Croft, Roland and Xie, Yongzheng and Babar, Muhammad Ali},
  journal={IEEE Transactions on Software Engineering},
  volume={49},
  number={3},
  pages={1044--1063},
  year={2022},
  publisher={IEEE}
}

@inproceedings{peng2023ptlvd,
  title={PTLVD: Program Slicing and Transformer-based Line-level Vulnerability Detection System},
  author={Peng, Tao and Chen, Shixu and Zhu, Fei and Tang, Junwei and Liu, Junping and Hu, Xinrong},
  booktitle={2023 IEEE 23rd International Working Conference on Source Code Analysis and Manipulation (SCAM)},
  pages={162--173},
  year={2023},
  organization={IEEE}
}

@article{senanayake2023android,
  title={Android source code vulnerability detection: a systematic literature review},
  author={Senanayake, Janaka and Kalutarage, Harsha and Al-Kadri, Mhd Omar and Petrovski, Andrei and Piras, Luca},
  journal={ACM Computing Surveys},
  volume={55},
  number={9},
  pages={1--37},
  year={2023},
  publisher={ACM New York, NY}
}

@article{le2022survey,
  title={A survey on data-driven software vulnerability assessment and prioritization},
  author={Le, Triet HM and Chen, Huaming and Babar, M Ali},
  journal={ACM Computing Surveys},
  volume={55},
  number={5},
  pages={1--39},
  year={2022},
  publisher={ACM New York, NY}
}

@article{bassi2023systematic,
  title={A systematic literature review on software vulnerability prediction models},
  author={Bassi, Deepali and Singh, Hardeep},
  journal={IEEE Access},
  volume={11},
  pages={110289--110311},
  year={2023},
  publisher={IEEE}
}

@article{sohan2020systematic,
  title={A systematic literature review and quality analysis of Javascript malware detection},
  author={Sohan, Md Fahimuzzman and Basalamah, Anas},
  journal={IEEE Access},
  volume={8},
  pages={190539--190552},
  year={2020},
  publisher={IEEE}
}

@inproceedings{chen2022hlt,
  title={HLT: A Hierarchical Vulnerability Detection Model Based on Transformer},
  author={Chen, Yupan and Liu, Zhihong},
  booktitle={2022 4th International Conference on Data Intelligence and Security (ICDIS)},
  pages={50--54},
  year={2022},
  organization={IEEE}
}

@inproceedings{lu2023assessing,
  title={Assessing the effectiveness of vulnerability detection via prompt tuning: An empirical study},
  author={Lu, Guilong and Ju, Xiaolin and Chen, Xiang and Yang, Shaoyu and Chen, Liang and Shen, Hao},
  booktitle={2023 30th Asia-Pacific Software Engineering Conference (APSEC)},
  pages={415--424},
  year={2023},
  organization={IEEE}
}

@inproceedings{mahyari2024harnessing,
  title={Harnessing the power of llms in source code vulnerability detection},
  author={Mahyari, Andrew A},
  booktitle={MILCOM 2024-2024 IEEE Military Communications Conference (MILCOM)},
  pages={251--256},
  year={2024},
  organization={IEEE}
}

@inproceedings{bui2023detecting,
  title={Detecting software vulnerabilities based on source code analysis using GCN transformer},
  author={Bui, Van-Cong and Do, Xuan-Cho},
  booktitle={2023 RIVF International Conference on Computing and Communication Technologies (RIVF)},
  pages={112--117},
  year={2023},
  organization={IEEE}
}

@inproceedings{sun2024enhancing,
  title={Enhancing Source Code Vulnerability Detection Using Flattened Code Graph Structures},
  author={Sun, Zhequ and Liu, Ke and Yang, Yu},
  booktitle={2024 6th International Conference on Frontier Technologies of Information and Computer (ICFTIC)},
  pages={209--213},
  year={2024},
  organization={IEEE}
}

@inproceedings{zhao2025vulnerability,
  title={Vulnerability Code Similarity Detection Method Based on Transformer},
  author={Zhao, Jialin and Liu, Weiwei},
  booktitle={2025 5th International Conference on Sensors and Information Technology},
  pages={851--854},
  year={2025},
  organization={IEEE}
}

@inproceedings{gujar2024detectbert,
  title={DetectBERT: Code Vulnerability Detection},
  author={Gujar, Shantanu Sudhir},
  booktitle={2024 Global Conference on Communications and Information Technologies (GCCIT)},
  pages={1--21},
  year={2024},
  organization={IEEE}
}

@inproceedings{zhao2024python,
  title={Python Source Code Vulnerability Detection Based on CodeBERT Language Model},
  author={Zhao, Kunpeng and Duan, Shuya and Qiu, Ge and Zhai, Jinyuan and Li, Mingze and Liu, Long},
  booktitle={2024 7th International Conference on Algorithms, Computing and Artificial Intelligence (ACAI)},
  pages={1--6},
  year={2024},
  organization={IEEE}
}

@inproceedings{liu2024pre,
  title={Pre-training by predicting program dependencies for vulnerability analysis tasks},
  author={Liu, Zhongxin and Tang, Zhijie and Zhang, Junwei and Xia, Xin and Yang, Xiaohu},
  booktitle={Proceedings of the IEEE/ACM 46th International Conference on Software Engineering},
  pages={1--13},
  year={2024}
}

@inproceedings{nguyen2023mando,
  title={Mando-hgt: Heterogeneous graph transformers for smart contract vulnerability detection},
  author={Nguyen, Hoang H and Nguyen, Nhat-Minh and Xie, Chunyao and Ahmadi, Zahra and Kudendo, Daniel and Doan, Thanh-Nam and Jiang, Lingxiao},
  booktitle={2023 IEEE/ACM 20th International Conference on Mining Software Repositories (MSR)},
  pages={334--346},
  year={2023},
  organization={IEEE}
}

@inproceedings{gupta2024dl,
  title={DL-VulBERT: A Deep Learning Classifier for the Identification of Software Vulnerabilities},
  author={Gupta, Aditya Raj and Tomar, Deepak Singh and Shekhar, Raj},
  booktitle={2024 15th International Conference on Computing Communication and Networking Technologies (ICCCNT)},
  pages={1--7},
  year={2024},
  organization={IEEE}
}

@inproceedings{liang2024source,
  title={A source code vulnerability detection method based on adaptive graph neural networks},
  author={Liang, Chen and Wei, Qiang and Jiang, Zirui and Wang, Yisen and Du, Jiang},
  booktitle={Proceedings of the 39th IEEE/ACM International Conference on Automated Software Engineering Workshops},
  pages={187--196},
  year={2024}
}

@inproceedings{wu2021peculiar,
  title={Peculiar: Smart contract vulnerability detection based on crucial data flow graph and pre-training techniques},
  author={Wu, Hongjun and Zhang, Zhuo and Wang, Shangwen and Lei, Yan and Lin, Bo and Qin, Yihao and Zhang, Haoyu and Mao, Xiaoguang},
  booktitle={2021 IEEE 32nd International Symposium on Software Reliability Engineering (ISSRE)},
  pages={378--389},
  year={2021},
  organization={IEEE}
}

@inproceedings{rusinova2024explaining,
  title={Explaining of Transformer-based Models for Vulnerable Function Detection},
  author={Rusinova, Zalina and Chernyshov, Yury and Dolganov, Anton},
  booktitle={2024 IEEE Ural-Siberian Conference on Biomedical Engineering, Radioelectronics and Information Technology (USBEREIT)},
  pages={304--307},
  year={2024},
  organization={IEEE}
}

@inproceedings{wu2021self,
  title={Self-attention based automated vulnerability detection with effective data representation},
  author={Wu, Tongshuai and Chen, Liwei and Du, Gewangzi and Zhu, Chenguang and Shi, Gang},
  booktitle={2021 IEEE Intl Conf on Parallel \& Distributed Processing with Applications, Big Data \& Cloud Computing, Sustainable Computing \& Communications, Social Computing \& Networking (ISPA/BDCloud/SocialCom/SustainCom)},
  pages={892--899},
  year={2021},
  organization={IEEE}
}

@inproceedings{liu2023software,
  title={Software vulnerability detection with gpt and in-context learning},
  author={Liu, Zhihong and Liao, Qing and Gu, Wenchao and Gao, Cuiyun},
  booktitle={2023 8th International Conference on Data Science in Cyberspace (DSC)},
  pages={229--236},
  year={2023},
  organization={IEEE}
}

@inproceedings{saimbhi2024vulnerai,
  title={VulnerAI: GPT Based Web Application Vulnerability Detection},
  author={Saimbhi, Sohan Singh and Akpinar, Kevser Ovaz},
  booktitle={2024 International Conference on Artificial Intelligence, Metaverse and Cybersecurity (ICAMAC)},
  pages={1--6},
  year={2024},
  organization={IEEE}
}

@inproceedings{kaanan2024llm,
  title={LLM-Based Approach for Buffer Overflow Detection in Source Code},
  author={Kaanan, Emran and Karim, Tasmin and Shaon, Md Shazzad Hossain and Sultan, Md Fahim and Cuzzocrea, Alfredo and Akter, Mst Shapna},
  booktitle={2024 27th International Conference on Computer and Information Technology (ICCIT)},
  pages={1898--1902},
  year={2024},
  organization={IEEE}
}

@inproceedings{jianjie2023code,
  title={Code Defect Detection Method Based on BERT and Ensemble},
  author={JianJie, Ye and Le, Wei},
  booktitle={2023 9th International Conference on Computer and Communications (ICCC)},
  pages={2130--2138},
  year={2023},
  organization={IEEE}
}

@inproceedings{islam2023unbiased,
  title={An unbiased transformer source code learning with semantic vulnerability graph},
  author={Islam, Nafis Tanveer and Parra, Gonzalo De La Torre and Manuel, Dylan and Bou-Harb, Elias and Najafirad, Peyman},
  booktitle={2023 IEEE 8th European Symposium on Security and Privacy (EuroS\&P)},
  pages={144--159},
  year={2023},
  organization={IEEE}
}

@inproceedings{sun2024gptscan,
  title={Gptscan: Detecting logic vulnerabilities in smart contracts by combining gpt with program analysis},
  author={Sun, Yuqiang and Wu, Daoyuan and Xue, Yue and Liu, Han and Wang, Haijun and Xu, Zhengzi and Xie, Xiaofei and Liu, Yang},
  booktitle={Proceedings of the IEEE/ACM 46th International Conference on Software Engineering},
  pages={1--13},
  year={2024}
}

@misc{shiaeles2023vuldetect,
  title={VulDetect: A novel technique for detecting software vulnerabilities using Language Models},
  author={Shiaeles, MOA},
  year={2023}
}

@article{do2024optimizing,
  title={Optimizing software vulnerability detection using RoBERTa and machine learning},
  author={Do, Cho Xuan and Luu, Nguyen Trong and Nguyen, Phuong Thi Lan},
  journal={Automated Software Engineering},
  volume={31},
  number={2},
  pages={40},
  year={2024},
  publisher={Springer}
}

@inproceedings{curto2024multivd,
  title={MultiVD: A Transformer-based Multitask Approach for Software Vulnerability Detection},
  author={Curto, Claudio and Giordano, Daniela and Palazzo, Simone and Indelicato, D},
  booktitle={Proceedings of the 21st International Conference on Security and Cryptography},
  pages={416--423},
  year={2024}
}

@article{jeon2024design,
  title={Design and evaluation of highly accurate smart contract code vulnerability detection framework},
  author={Jeon, Sowon and Lee, Gilhee and Kim, Hyoungshick and Woo, Simon S},
  journal={Data Mining and Knowledge Discovery},
  volume={38},
  number={3},
  pages={888--912},
  year={2024},
  publisher={Springer}
}

@inproceedings{le2024software,
  title={Software vulnerability prediction in low-resource languages: An empirical study of codebert and chatgpt},
  author={Le, Triet Huynh Minh and Babar, M Ali and Thai, Tung Hoang},
  booktitle={Proceedings of the 28th International Conference on Evaluation and Assessment in Software Engineering},
  pages={679--685},
  year={2024}
}

@inproceedings{zhang2024twlog,
  title={TWLog: Task Workflow-Based Log Anomaly Detection},
  author={Zhang, Suqiong and Fan, Dongyi and He, Lili and Liu, Yi and Chen, Deng},
  booktitle={Asia-Pacific Web (APWeb) and Web-Age Information Management (WAIM) Joint International Conference on Web and Big Data},
  pages={3--16},
  year={2024},
  organization={Springer}
}

@inproceedings{alqarni2022low,
  title={Low Level Source Code Vulnerability Detection Using Advanced BERT Language Model.},
  author={Alqarni, Mansour and Azim, Akramul},
  booktitle={Canadian AI},
  year={2022}
}

@article{almakayeel2024deep,
  title={Deep learning-based improved transformer model on android malware detection and classification in internet of vehicles},
  author={Almakayeel, Naif},
  journal={Scientific Reports},
  volume={14},
  number={1},
  pages={25175},
  year={2024},
  publisher={Nature Publishing Group UK London}
}

@article{myllari2025ladle,
  title={Ladle: a method for unsupervised anomaly detection across log types},
  author={Myll{\"a}ri, Juha and Aalto, Tatu and Nurminen, Jukka K},
  journal={Automated Software Engineering},
  volume={32},
  number={2},
  pages={34},
  year={2025},
  publisher={Springer}
}

@article{gong2023gratdet,
  title={GRATDet: Smart Contract Vulnerability Detector Based on Graph Representation and Transformer.},
  author={Gong, Peng and Yang, Wenzhong and Wang, Liejun and Wei, Fuyuan and HaiLaTi, KeZiErBieKe and Liao, Yuanyuan},
  journal={Computers, Materials \& Continua},
  volume={76},
  number={2},
  year={2023}
}

@inproceedings{liu2024automatic,
  title={Automatic Software Vulnerability Detection in Binary Code},
  author={Liu, Shigang and Li, Lin and Ban, Xinbo and Chen, Chao and Zhang, Jun and Camtepe, Seyit and Xiang, Yang},
  booktitle={International Conference on Machine Learning for Cyber Security},
  pages={148--166},
  year={2024},
  organization={Springer}
}

@article{he2025vultr,
  title={VulTR: Software vulnerability detection model based on multi-layer key feature enhancement},
  author={He, Haitao and Wang, Sheng and Wang, Yanmin and Liu, Ke and Yu, Lu},
  journal={Computers \& Security},
  volume={148},
  pages={104139},
  year={2025},
  publisher={Elsevier}
}

@article{mechri2025secureqwen,
  title={SecureQwen: Leveraging LLMs for vulnerability detection in python codebases},
  author={Mechri, Abdechakour and Ferrag, Mohamed Amine and Debbah, Merouane},
  journal={Computers \& Security},
  volume={148},
  pages={104151},
  year={2025},
  publisher={Elsevier}
}

@article{wang2024scl,
  title={SCL-CVD: Supervised contrastive learning for code vulnerability detection via GraphCodeBERT},
  author={Wang, Rongcun and Xu, Senlei and Tian, Yuan and Ji, Xingyu and Sun, Xiaobing and Jiang, Shujuang},
  journal={Computers \& Security},
  volume={145},
  pages={103994},
  year={2024},
  publisher={Elsevier}
}

@article{ehrenberg2024python,
  title={Python source code vulnerability detection with named entity recognition},
  author={Ehrenberg, Melanie and Sarkani, Shahram and Mazzuchi, Thomas A},
  journal={Computers \& Security},
  volume={140},
  pages={103802},
  year={2024},
  publisher={Elsevier}
}

@article{cao2025multi,
  title={Multi-source cross-domain vulnerability detection based on code pre-trained model},
  author={Cao, Yang and Dong, Yunwei},
  journal={Information and Software Technology},
  pages={107764},
  year={2025},
  publisher={Elsevier}
}

@article{liu2024making,
  title={Making vulnerability prediction more practical: Prediction, categorization, and localization},
  author={Liu, Chongyang and Chen, Xiang and Li, Xiangwei and Xue, Yinxing},
  journal={Information and Software Technology},
  volume={171},
  pages={107458},
  year={2024},
  publisher={Elsevier}
}

@article{jiang2024haformer,
  title={HAformer: Semantic fusion of hex machine code and assembly code for cross-architecture binary vulnerability detection},
  author={Jiang, Xunzhi and Wang, Shen and Gong, Yuxin and Yu, Tingyue and Liu, Li and Yu, Xiangzhan},
  journal={Computers \& Security},
  volume={145},
  pages={104029},
  year={2024},
  publisher={Elsevier}
}

@article{ferrag2025securefalcon,
  title={Securefalcon: Are we there yet in automated software vulnerability detection with llms?},
  author={Ferrag, Mohamed Amine and Battah, Ammar and Tihanyi, Norbert and Jain, Ridhi and Maimu{\c{t}}, Diana and Alwahedi, Fatima and Lestable, Thierry and Thandi, Narinderjit Singh and Mechri, Abdechakour and Debbah, Merouane and others},
  journal={IEEE Transactions on Software Engineering},
  year={2025},
  publisher={IEEE}
}

@article{wu2025information,
  title={What information contributes to log-based anomaly detection? Insights from a configurable transformer-based approach},
  author={Wu, Xingfang and Li, Heng and Khomh, Foutse},
  journal={Automated Software Engineering},
  volume={32},
  number={2},
  pages={58},
  year={2025},
  publisher={Springer}
}

@inproceedings{devlin2019bert,
  title={Bert: Pre-training of deep bidirectional transformers for language understanding},
  author={Devlin, Jacob and Chang, Ming-Wei and Lee, Kenton and Toutanova, Kristina},
  booktitle={Proceedings of the 2019 conference of the North American chapter of the association for computational linguistics: human language technologies, volume 1 (long and short papers)},
  pages={4171--4186},
  year={2019}
}

@article{feng2020codebert,
  title={Codebert: A pre-trained model for programming and natural languages},
  author={Feng, Z},
  journal={arXiv preprint arXiv:2002.08155},
  year={2020}
}

@article{xuan2025large,
  title={Large language models based vulnerability detection: How does it enhance performance?},
  author={Xuan, Cho Do and Quang, Dat Bui and Quang, Vinh Dang},
  journal={International Journal of Information Security},
  volume={24},
  number={1},
  pages={69},
  year={2025},
  publisher={Springer}
}

@article{wang2024extensive,
  title={An extensive study of the effects of different deep learning models on code vulnerability detection in Python code},
  author={Wang, Rongcun and Xu, Senlei and Ji, Xingyu and Tian, Yuan and Gong, Lina and Wang, Ke},
  journal={Automated Software Engineering},
  volume={31},
  number={1},
  pages={15},
  year={2024},
  publisher={Springer}
}

@article{vaswani2017attention,
  title={Attention is all you need},
  author={Vaswani, Ashish and Shazeer, Noam and Parmar, Niki and Uszkoreit, Jakob and Jones, Llion and Gomez, Aidan N and Kaiser, {\L}ukasz and Polosukhin, Illia},
  journal={Advances in neural information processing systems},
  volume={30},
  year={2017}
}

@inproceedings{perl2015vccfinder,
  title={Vccfinder: Finding potential vulnerabilities in open-source projects to assist code audits},
  author={Perl, Henning and Dechand, Sergej and Smith, Matthew and Arp, Daniel and Yamaguchi, Fabian and Rieck, Konrad and Fahl, Sascha and Acar, Yasemin},
  booktitle={Proceedings of the 22nd ACM SIGSAC conference on computer and communications security},
  pages={426--437},
  year={2015}
}

@misc{CWESite,
  title        = {CWE Vulnerabilities },
author = {CWE}, 
  howpublished = {https://cwe.mitre.org/},
  note         = {Accessed: 2025-11-18}
}

@article{wu2022code,
  title={Code vulnerability detection based on deep sequence and graph models: A survey},
  author={Wu, Bolun and Zou, Futai},
  journal={Security and Communication Networks},
  volume={2022},
  number={1},
  pages={1176898},
  year={2022},
  publisher={Wiley Online Library}
}

@inproceedings{chernis2018machine,
  title={Machine learning methods for software vulnerability detection},
  author={Chernis, Boris and Verma, Rakesh},
  booktitle={Proceedings of the fourth ACM international workshop on security and privacy analytics},
  pages={31--39},
  year={2018}
}

@misc{Example,
  title        = {Real-World Examples of Application Security Breaches},
author = {Brian Pavicic}, 
  howpublished = {https://true-positives.com/appsec-blog/cybersecurity-breaches-real-world-examples-lessons-learned},
  note         = {Accessed: 2025-11-25}
}

@article{alaoui2022deep,
  title={Deep learning for vulnerability and attack detection on web applications: A systematic literature review},
  author={Alaoui, Rokia Lamrani and Nfaoui, El Habib},
  journal={Future Internet},
  volume={14},
  number={4},
  pages={118},
  year={2022},
  publisher={MDPI}
}

@article{ameh2025c3,
  title={C3-VULMAP: A Dataset for Privacy-Aware Vulnerability Detection in Healthcare Systems},
  author={Ameh, Jude Enenche and Otebolaku, Abayomi and Shenfield, Alex and Ikpehai, Augustine},
  journal={Electronics},
  volume={14},
  number={13},
  pages={2703},
  year={2025},
  publisher={MDPI}
}

@article{matloob2025healthcare,
  title={Healthcare fraud detection using adaptive learning and deep learning techniques},
  author={Matloob, Irum and Khan, Shoab and Rukaiya, Rukaiya and Alfraihi, Hessa and Ali Khan, Javed},
  journal={Evolving Systems},
  volume={16},
  number={2},
  pages={72},
  year={2025},
  publisher={Springer}
}

@article{gao2024sguard,
  title={sGuard+: Machine learning guided rule-based automated vulnerability repair on smart contracts},
  author={Gao, Cuifeng and Yang, Wenzhang and Ye, Jiaming and Xue, Yinxing and Sun, Jun},
  journal={ACM Transactions on Software Engineering and Methodology},
  volume={33},
  number={5},
  pages={1--55},
  year={2024},
  publisher={ACM New York, NY}
}

@techreport{kitchenham2007guidelines,
  title={Guidelines for performing systematic literature reviews in software engineering},
  author={Kitchenham, Barbara and Charters, Stuart and others},
  year={2007},
  institution={Technical report, ver. 2.3 ebse technical report. ebse},
  publisher={Keele, UK}
}

@inproceedings{wang2025line,
  title={Line-level semantic structure learning for code vulnerability detection},
  author={Wang, Ziliang and Li, Ge and Li, Jia and Dong, Yihong and Xiong, Yingfei and Jin, Zhi},
  booktitle={Proceedings of the 16th International Conference on Internetware},
  pages={269--280},
  year={2025}
}

@article{shir2025robust,
  title={Robust Vulnerability Detection across Compilations: LLVM-IR vs. Assembly with Transformer Model},
  author={Shir, Rony and Surve, Priyanka and Elovici, Yuval and Shabtai, Asaf},
  journal={Proceedings of the ACM on Software Engineering},
  volume={2},
  number={ISSTA},
  pages={618--639},
  year={2025},
  publisher={ACM New York, NY, USA}
}

@article{su2026source,
  title={Source code vulnerability detection based on deep learning: a review},
  author={Su, Huading and Xu, Zhen and Zhang, Yan and Tan, Qian},
  journal={Cybersecurity},
  volume={9},
  number={1},
  pages={2},
  year={2026},
  publisher={Springer}
}

@inproceedings{khan2026leveraging,
  title={Leveraging Transformers to Discover Software Vulnerabilities based on Source Code Slices},
  author={Khan, Abdur Rehman and Xu, Yue and Li, Yuefeng},
  booktitle={Proceedings of the 2026 Australasian Information Security Conference},
  pages={1--9},
  year={2026}
}

@inproceedings{tian2025efvd,
  title={EFVD: A framework of source code vulnerability detection via fusion of enhanced graph representation learning and pre-trained transformer-based model},
  author={Tian, Lei and Zhang, Cheng},
  booktitle={Proceedings of the 2025 5th International Conference on Computer Network Security and Software Engineering},
  pages={316--320},
  year={2025}
}

@inproceedings{katz2025siexvults,
  title={SIExVulTS: Sensitive Information Exposure Vulnerability Detection System using Transformer Models and Static Analysis},
  author={Katz, Kyler and Moshtari, Sara and Mujhid, Ibrahim and Mirakhorli, Mehdi and Garcia, Derek},
  booktitle={2025 ACM/IEEE International Symposium on Empirical Software Engineering and Measurement (ESEM)},
  pages={230--241},
  year={2025},
  organization={IEEE}
}

@inproceedings{cui2025vulgtda,
  title={VulGTDA: A Software Vulnerability Detection Method via Graph Transformer and Domain Adaptation},
  author={Cui, Haoqiang and Zhang, Cheng and Cai, Feifan},
  booktitle={2025 5th International Conference on Neural Networks, Information and Communication Engineering (NNICE)},
  pages={1053--1056},
  year={2025},
  organization={IEEE}
}

@inproceedings{vanam2025software,
  title={Software Vulnerability Detection in Source Code Using Superb Fairy-Wren Deep Transformer Guided Model for Next Generation Software Security},
  author={Vanam, Raghavender Reddy and Yadavali, Vinaysimha Varma and Elumalai, Saravanan and others},
  booktitle={2025 6th International Conference on Inventive Research in Computing Applications (ICIRCA)},
  pages={101--106},
  year={2025},
  organization={IEEE}
}

@inproceedings{li2025macd,
  title={MACD: Source Code Vulnerability Detection Method Integrating Mamba and Attention},
  author={Li, Jingen},
  booktitle={2025 7th International Conference on Electronics and Communication, Network and Computer Technology (ECNCT)},
  pages={376--380},
  year={2025},
  organization={IEEE}
}

@inproceedings{gunda2025transformer,
  title={Transformer-Based Semantic Embeddings and Hybrid Neural Networks for Robust Software Vulnerability Detection},
  author={Gunda, Brahma Sagar and Krishna, G Bala and Rawat, Sandeep Singh},
  booktitle={2025 Innovations in Power and Advanced Computing Technologies (i-PACT)},
  pages={1--9},
  year={2025},
  organization={IEEE}
}

@inproceedings{reza2026empirical,
  title={An Empirical Analysis of Transformer-Based Models with LIME Explainability for JavaScript Vulnerability Detection},
  author={Reza, SM Imran and Moon, Iffath Tanjim and Fahim, Md Forhan Shahriar and Alam, Adrita and Sheikh, Md Tanjid},
  booktitle={2026 5th International Conference on Electrical, Computer \& Telecommunication Engineering (ICECTE)},
  pages={1--6},
  year={2026},
  organization={IEEE}
}

@inproceedings{oladokun2025effective,
  title={How Effective are pretrained Programming Language-based Language Models (PLLMs) in the Detection of Android Vulnerabilities?},
  author={Oladokun, Olanrewaju and Rice, Jacqueline},
  booktitle={2025 IEEE Canadian Conference on Electrical and Computer Engineering (CCECE)},
  pages={150--154},
  year={2025},
  organization={IEEE}
}

@article{sun2025hgtjit,
  title={HgtJIT: Just-in-Time Vulnerability Detection Based on Heterogeneous Graph Transformer},
  author={Sun, Xiaobing and Zhou, Mingxuan and Cao, Sicong and Wu, Xiaoxue and Bo, Lili and Wu, Di and Li, Bin and Xiang, Yang},
  journal={IEEE Transactions on Dependable and Secure Computing},
  year={2025},
  publisher={IEEE}
}

@article{wang2025m2cvd,
  title={M2CVD: Enhancing Vulnerability Understanding through Multi-Model Collaboration for Code Vulnerability Detection},
  author={Wang, Ziliang and Li, Ge and Li, Jia and Li, Jia and Yan, Meng and Xiong, Yingfei and Jin, Zhi},
  journal={ACM Transactions on Software Engineering and Methodology},
  year={2025},
  publisher={ACM New York, NY}
}

@ARTICLE{ferrag2023securefalcon,
  author={Ferrag, Mohamed Amine and Battah, Ammar and Tihanyi, Norbert and Jain, Ridhi and Maimuţ, Diana and Alwahedi, Fatima and Lestable, Thierry and Thandi, Narinderjit Singh and Mechri, Abdechakour and Debbah, Merouane and Cordeiro, Lucas C.},
  journal={IEEE Transactions on Software Engineering}, 
  title={SecureFalcon: Are We There Yet in Automated Software Vulnerability Detection With LLMs?}, 
  year={2025},
  volume={51},
  number={4},
  pages={1248-1265},
  doi={10.1109/TSE.2025.3548168}}

@inproceedings{ferretti2025detecting,
  title={Detecting Smart Contract Vulnerabilities using Transformers and LLMs},
  author={Ferretti, Stefano and D’Angelo, Gabriele and Ghini, Vittorio and Tomasone, Marco B},
  booktitle={2025 IEEE International Conference on Pervasive Computing and Communications Workshops and other Affiliated Events (PerCom Workshops)},
  pages={7--12},
  year={2025},
  organization={IEEE}
}

@inproceedings{perera2025codebert,
  title={CodeBERT-Based Embeddings for Detecting Vulnerable Smart Contracts},
  author={Perera, Awarjana and Pillai, Babu and Tharani, Jeyakumar Samantha and Rao, Aravinda S and Muthukkumarasamy, Vallipuram},
  booktitle={2025 IEEE 50th Conference on Local Computer Networks (LCN)},
  pages={1--9},
  year={2025},
  organization={IEEE}
}

@inproceedings{sultan2025codevul,
  title={CodeVul+: A Structure-Aware Framework for Cross-Repository Vulnerability Detection},
  author={Sultan, Md Fahim and Akter, Mst Shapna and Cuzzocrea, Alfredo},
  booktitle={2025 IEEE International Conference on Big Data (BigData)},
  pages={4406--4415},
  year={2025},
  organization={IEEE}
}

@article{shang2025cegt,
  title={CEGT: Smart contract vulnerability detection via connectivity-enhanced GCN-transformer},
  author={Shang, Jiandong and Li, Jiaru and Sui, Yizhe and Guo, Hengliang and Gao, Xu and Zhang, Dujuan and Guo, Yang and Wu, Gang},
  journal={Journal of Systems and Software},
  volume={227},
  pages={112454},
  year={2025},
  publisher={Elsevier}
}

@article{kalouptsoglou2025transfer,
  title={Transfer learning for software vulnerability prediction using Transformer models},
  author={Kalouptsoglou, Ilias and Siavvas, Miltiadis and Ampatzoglou, Apostolos and Kehagias, Dionysios and Chatzigeorgiou, Alexander},
  journal={Journal of Systems and Software},
  volume={227},
  pages={112448},
  year={2025},
  publisher={Elsevier}
}

@article{tao2025transformer,
  title={Transformer-based statement level vulnerability detection by cross-modal fine-grained features capture},
  author={Tao, Wenxin and Su, Xiaohong and Ke, Yekun and Han, Yi and Zheng, Yu and Wei, Hongwei},
  journal={Knowledge-Based Systems},
  volume={316},
  pages={113341},
  year={2025},
  publisher={Elsevier}
}

@article{smaili2025transformer,
  title={A transformer-based framework for software vulnerability detection using attention-driven convolutional neural networks},
  author={Smaili, Abdelkarim and Zhang, Yanan and Mekkaoui, Djamel Eddine and Midoun, Mohamed Amine and Talhaoui, Mohamed Zakariya and Hamidaoui, Meryem and Kong, Weiqiang},
  journal={Engineering Applications of Artificial Intelligence},
  volume={160},
  pages={111859},
  year={2025},
  publisher={Elsevier}
}

@article{do2026novel,
  title={A novel approach for software vulnerability detection based on ensemble learning model},
  author={Do Xuan, Cho and Quang, Dat Bui and Quang, Vinh Dang},
  journal={Computers and Electrical Engineering},
  volume={130},
  pages={110848},
  year={2026},
  publisher={Elsevier}
}


\end{document}